\documentclass[preprint,12pt]{elsarticle}

\usepackage{graphicx}
\usepackage[margin=0.8in]{geometry}
\usepackage[figuresright]{rotating}
\usepackage{bm}
\usepackage{booktabs}
\usepackage{xcolor,colortbl}
\usepackage{color}

\definecolor{Gray}{gray}{0.85}
\definecolor{LightCyan}{rgb}{0.88,1,1}

\newcolumntype{a}{>{\columncolor{Gray}}c}
\newcolumntype{b}{>{\columncolor{white}}c}

\usepackage{amsmath}
\usepackage{amsfonts}
\usepackage{amssymb}
 
\usepackage{lineno}
 \usepackage {hyperref}
\hypersetup{
    colorlinks,
    linkcolor={red!50!black},
    citecolor={blue!50!black},
    urlcolor={blue!80!black}
}
\usepackage{subfig}

\usepackage{algorithm}
\usepackage{algpseudocode} %\usepackage[noend]{algpseudocode}
\makeatletter
\def\BState{\State\hskip-\ALG@thistlm}
\makeatother
\floatstyle{plain}
\newfloat{myalgo}{tbhp}{mya}

\journal{Journal Name}

\begin{document}

\begin{frontmatter}



%% e.g. \dochead{17th International Conference on Dynamical Processes in Excited States of Solids}

\title{A high order hybridizable discontinuous Galerkin method for incompressible miscible displacement in heterogeneous media}


\author{M. S. Fabien$^1$, M. G. Knepley$^2$, and B. M. Rivi\`{e}re$^1$}

\address{1 Department of Computational and Applied Mathematics, Rice University, Houston, TX 7005, USA}
\address{2 Department of Computer Science and Engineering, State University of New York at Buffalo, Buffalo, New York, 14260, USA}
 
\begin{abstract}
We present a new method for approximating solutions to the incompressible miscible displacement problem in porous media.  At the discrete level, the coupled nonlinear system has been split into two linear systems that are solved sequentially.  The method is based on a hybridizable discontinuous Galerkin method for the Darcy flow, which produces a mass--conservative flux approximation, and a hybridizable discontinuous Galerkin method for the transport equation.  The resulting method is high order accurate.  Due to the implicit treatment of the system of partial differential equations, we observe computationally that no slope limiters are needed.  Numerical experiments are provided that show that the method converges optimally and is robust for highly heterogeneous porous media in 2D and 3D.
\end{abstract}


\end{frontmatter}
\section{Introduction}
\label{}
Miscible displacement is a fundamental concept in geophysics, and it is applicable as a model for groundwater movement and enhanced oil recovery \cite{lantz1970rigorous,killough1987fifth,todd1972development,homsy1987viscous,douglas1983approximation,todd1972development}.  The equations that govern miscible displacement form a system of coupled nonlinear partial differential equations.  Numerous techniques have been proposed to resolve the nonlinearity in the miscible displacement system, for instance, implicit-explicit, semi-implicit, and fully implicit methods (see \cite{li2015high} for a literature review of these approaches).  Moreover, the choice of discretization is also a critical decision in the solution process.  Accuracy, mass balance, and efficiency of implementation are all valid concerns.  With respect to the Darcy flow system, incorrect approximations to the velocity can cause oscillations and instability when used in the convection-dominated transport system.  Utilizing compatible discreizations (as defined in \cite{dawson2004compatible}) for flow and transport maintains local/global mass conservation, which provides stability and accuracy in the numerical methods.  
\\ \indent
Mixed finite element (MFE) methods have the compatibility property when the velocity space is taken to be $H(\text{div},\Omega)$ conforming.  However, MFE results in a semi-definite saddle-point system, which needs specialized block solvers, especially in the high order regime.  Through hybridization \cite{cockburn2004characterization,arnold1985mixed}, the MFE method is called a hybrid mixed finite element technique (HMFE), and becomes more practical in some regards.  Namely, one is able to significantly reduce the number of degrees of freedom, as well as generate a symmetric positive definite system.  This is possible by introducing a certain Lagrange multiplier such that the degrees of freedom associated with the velocity $\bm {u}$ and pressure $p$ can be eliminated to obtain a globally coupled system for the multiplier only.  A combined HMFE-discontinuous Galerkin method for miscible displacement was examined in~\cite{zhang2017combined}.%In this work, for the Darcy flow we consider a hybridized mixed (HMFE) method~\cite{cockburn2004characterization}.
\\ \indent 
Discontinuous Galerkin (DG) methods are popular methods, in part because they have a number of attractive features, e.g. high accuracy, local mass conservation, completely discontinuous approximations that expose parallelism and allow for $hp$--adaptation, and they are able to handle nonconforming meshes \cite{cockburn2000development}.  On the other hand, DG methods in general have more degrees of freedom than their continuous counterparts, and this causes major challenges for linear solvers.  The hybridizable discontinuous Galerkin method (HDG) addresses this issue \cite{CockburnDGRS09,CockburnGL09,CockburnDG08}.  Similar to the HMFE method, a global system solely in terms of the approximate trace of the concentration variable can be obtained.  To do this, we prescribe a specific numerical flux for the approximate concentration variable.  The numerical flux is defined such that we can express it and the approximation to the concentration, in terms of an additional unknown defined on the skeleton of the mesh.  To ensure that the numerical trace is single valued, we require that the normal component of the numerical flux across the element boundaries is continuous.  In this paper we consider an HDG method for both Darcy flow and transport.  An approach related to the HDG scheme called the hybrid high order method (HHO) was applied to the miscible displacement problem in~\cite{anderson2018arbitrary}.  Here they consider a variety of meshes in 2D, such as triangular, Cartesian, Kershaw, and hexagonal-dominant meshes.  In their framework upto degree three polynomials are considered, and it is demonstrated that a piecewise linear basis provides the best balance between computational efficient and accuracy.  Our work examines polynomials upto degree sixteen, and several challenging benchmarks are considered in heterogeneous porous media.  The connection between HHO and HDG is not fully understood at this time~\cite{cockburn2016bridging}.
\\ \indent
For higher orders, hybridization can significantly reduce the number of degrees of freedom, as well as the total number of nonzeros in the discretization matrix \cite{huerta2013efficiency,samii2016parallel}.  The reduction in degrees of freedom (and total nonzeros in discretization operators) is of great importance, as we use high order accurate approximations for the simulation of complex flow-transport systems.  Furthermore, since our algorithm decouples flow and transport, at every time step, multiple linear systems are to be solved.  In the most simple situation only two linear solves are needed per time step, one for obtaining the pressure and velocity and the other for concentration.  In some situations one could lag the pressure/velocity update and use it for multiple time steps before obtaining a new profile.  However, iterative coupling and high order time stepping may be utilized to enhance the solution, which increases the number of linear solves that are needed per time step.   Efficient linear solvers are required for this class of problems~\cite{fabien1}, as the dominant cost occurs during this phase of the simulation.
\\ \indent
The HDG method with polynomial degree $k$ boasts optimal order of accuracy $k+1$ in the $L^2$ norm for all approximate variables, possesses a local postprocessing that can enhance the accuracy of the scalar variable (with a order of accuracy $k+2$), and retains favorable aspects of DG methods (e.g. local mass conservation, ability to handle unstructured meshes, etc.).  The HDG method for flow and for transport are compatible in the sense that is defined in \cite{dawson2004compatible}; which means that stronger discrete analogs of global conservation for flow, and local conservation for transport are satisfied.  Further more, if more accuracy is desired, one can resort to a simple element by element postprocessing that projects the flow velocity into an $H(\text{div},\Omega)$ conforming subspace.  This postprocessing is available since the scalar and flux unknowns converge optimally and the normal component of the numerical flux for the HDG method is single valued \cite{CockburnDGRS09}.  To the best of our knowledge, there are very few papers on HDG for complex porous media flows.  Recently, we applied the HDG method to two-phase flows in~\cite{fabien2}%A HDG-HDG method for two-phase flow in porous media explored in \cite{fabien2}.
\\ \indent
Examples of standard DG methods for miscible displacement can be found in \cite{riviere2002discontinuous,li2015high}.  Classical primal non-compatible DG methods that are used for the Darcy system require special attention.  The totally discontinuous Darcy velocity must be constructed by taking the gradient of pressure (which reduces the accuracy of the Darcy velocity a full order).  Further, a non-compatible Darcy velocity can cause oscillations and instability when used in the convection-dominated transport given by equation~(\ref{eq:modelB}).  As such, weighted average stabilization \cite{ern2008discontinuous} and velocity reconstructions must be utilized \cite{bastian2003superconvergence,ern2007accurate}.  In comparison, the HDG method gives optimal convergence rates of $k+1$ for both pressure and Darcy velocity approximations, and if an $H(\text{div},\Omega)$ conforming velocity is required, projections with optimal convergence rates exist \cite{CockburnDGRS09}.  The scheme presented in this paper is high order accurate.  It also admits a discrete local mass balance, has a velocity that has a numerical trace with a continuous normal component, and allows for hybridization, which significantly reduces the total number of degrees of freedom.
\\\indent
An outline of the paper is is given.  Section~\ref{sec:model} contains the model problem and section~\ref{sec:disc} the numerical scheme.  Simulations are shown in section~\ref{sec:numerical}.  Conclusions follow.
\section{Model problem}
\label{sec:model}
The displacement of one incompressible fluid by another in the domain $\Omega \subset \mathbb{R}^d$ (for $d=2,3$) over the time interval $(0,T)$ is governed by the following three coupled equations:
\begin{align} 
&\nabla \cdot {\bm u} = q^I-q^P,\quad {\bm u} = -\frac{ { K} }{\mu(c)} \nabla p, && \textrm{in}~~\Omega \times (0,T],
\label{eq:modelA}
\\
&\phi \frac{\partial c}{\partial t} +
 q^Pc
+
   \nabla   \cdot ({\bm u}c   -   {\bm D }({\bm u}) \nabla c) = q^I \bar{c}, && \textrm{in}~~\Omega \times (0,T].
\label{eq:modelB}
\end{align}
The primary unknowns are the pressure of the fluid mixture denoted by $p$, the concentration of the solvent in the fluid mixture denoted by $c$, and the velocity denoted by $\bm u$.  The dispersion-diffusion tensor is denoted by $ {\bm D }({\bm u})$, $\phi$ is the porosity of the medium, $\mu$ is the viscosity of the fluid mixture, and ${  K} $ is the permeability of the porous medium.  For simplicity, we assume that on each element, $K$ is a scalar.  We note that $ K$ can vary spatially.  The functions $q^I$ and $q^P$ are the flow rates at injection and production wells respectively, and $\bar{c}$ is the fluid concentration  prescribed at the injection wells.  We assume that the dispersion-diffusion tensor depends on the velocity:
$$
{\bm D}({\bm u})
=
(d_m + \alpha_t\| {\bm u} \|) {\bm I}
+
(\alpha_l-\alpha_t)
\frac{{\bm u} {\bm u}^T}{\| {\bm u} \|},
$$
where $\alpha_t,$ and $\alpha_l$ are the tangential and longitudinal dispersivities, respectively.  The molecular diffusivity is denoted by $d_m$.  For the viscosity, we assume the common quarter-power mixing law~\cite{koval1963method}
$$
{\mu(c)}= (c (\mu_s)^{-0.25} + (1-c) (\mu_o)^{-0.25})^{-4},
$$
where $\mu_s$ (respectively, $\mu_o$) is the viscosity of the solvent (respectively, resident fluid).
\\ \indent
The system is completed by no flow boundary conditions and an initial condition for the concentration:
\begin{align*} 
{\bm u} \cdot {\bm n} &= 0,&&\textrm{on}~~ \partial\Omega \times (0,T],
\\
{\bm D}({\bm u}) \nabla c \cdot {\bm n} &= 0,&&\textrm{on}~~ \partial\Omega \times (0,T],
\\
c({\bm x},t) &= c^0({\bm x}),&& \textrm{in}~~ \Omega \times \{0\},
\end{align*}
where $\bm n$ denotes the outward unit normal vector to $\Omega$.  %In order to guarantee uniqueness for the Darcy system, an additional constraint on the pressure variable is applied.

\section{Discretization}
\label{sec:disc}
Here we describe the spatial discretization for the miscible displacement system.  Sections~\ref{pv_sec} and \ref{cn_sec} require some notation that we clarify.  We assume that the domain $\Omega$ has been partitioned into a non-overlapping set of elements, $ \mathcal{E}_h$.  The skeleton of the mesh, denoted by $\Gamma_h$, consists of all unique edges (or faces in 3D) of the mesh.  The collection of all element boundaries is denoted by $\partial \mathcal{E}_h$, and is distinct from $\Gamma_h$, as interior edges (or faces) are duplicated.

 The set $\mathbb{Q}_{k}$ is the typical tensor product finite element space; whose members are tensor products of polynomials of degree $k$ in each coordinate direction.  The symbols denoting inner products have a special distinction depending on its arguments and the underlying domain of integration.  Given $E\in \mathcal{E}_h$, and $e\in \Gamma_h$, we have
\begin{alignat*}{2}
( {\bm q} ,{\bm r} )_E &= \int_E {\bm q} \cdot {\bm r} ,
 \quad \quad
&&{\bm q} ,{\bm r} \in L^2(E)\times L^2(E),
\\
(u,v)_E &= \int_E u v,
 \quad \quad
&&u,v\in L^2(E),
\\
\langle u ,v\rangle_e &= \int_e u v,  
 \quad \quad
&&u,v\in L^2(e).
%%\\
%%\langle \mu , \xi \rangle_e &= \int_e \mu  \xi ,  
%% \quad \quad
%%&& \mu , \xi \in L^2(e).
\end{alignat*} 
 
To facilitate high order approximation we use a nodal basis, which is nodal at Gauss-Lobatto-Legendre points, and we use Gauss-Legendre points and weights for high order quadrature.  Invoking a change of variables we evaluate our basis on the reference element (or edge/face) using barycentric interpolation of the second kind \cite{berrut2004barycentric}.  This approach obviates usage of generalized Vandermond matrices, or their inversion.
 
 \subsection{Pressure and velocity approximation}
 \label{pv_sec}

\subsubsection{Hybridizable discontinuous Galerkin}
 \label{sec:darcy_hdg}
A HDG method is used to discretize equations~(\ref{eq:modelA}).  The following discrete spaces are needed:
\begin{equation}
\begin{split}
W_h^{\text{DG}} &= \{ w\in L^2(\Omega): w|_E \in \mathbb{Q}_{k}(E),~~~\forall E \in \mathcal{E}_h \},
\\
{\bm V}_h^{DG} &= W_h^{\text{DG}} \times W_h^{\text{DG}},
\\
M_h^{DG} &= \{ \zeta \in L^2(\Gamma_h): \zeta|_e \in \mathbb{Q}_k(e),~~~\forall e \in \Gamma_h \}.
\end{split}
\label{eq:bdm_spaces}
\end{equation}
The HDG method seeks $({\bm u}_h , p_h,\widehat{p}_h)\in {\bm V}_h^{\text{DG}} \times W_h^{\text{DG}} \times M_h^{\text{DG}}$ such that
\begin{align}
( \mu(c_h){  K}^{-1} {\bf u}_h , {\bf v})_{\mathcal{E}_h}
-
(   p_h , \nabla\cdot {\bf v} )_{\mathcal{E}_h}
+
\langle \widehat{p}_h, {\bf v}\cdot {\bf n} \rangle_{\partial\mathcal{E}_h}
&=
0
 \label{eq:hdg_darcy1}
\\
-( {\bf u}_h , \nabla w)_{\mathcal{E}_h}
+
\langle
\widehat{{\bf u} }_h
\cdot
{\bm n}
,
w
\rangle_{\partial\mathcal{E}_h}
  &= (q^I-q^P,w)_{\mathcal{E}_h},
   \label{eq:hdg_darcy2}
\\
\langle  
  \widehat{\bf u}_h  \cdot {\bm n}
  , \zeta \rangle_{\partial \mathcal{E}_h } 
 &= 0,
 \label{eq:hdg_darcy3}
\end{align}
for all $({\bm v}, w,\zeta)\in {\bm V}_h^{\text{DG}} \times W_h^{\text{DG}}\times M_h^{\text{DG}}$.  The numerical traces are given as follows:
\begin{align*}
\tilde{p}_h
&=
\widehat{p}_h  ,
\\
\widehat{\bf u}_h
&=
{\bf u}_h
+
( p_h  - \tilde{p}_h ){\bm n}.
\end{align*}
The HDG system written in matrix form can be expressed as
\begin{equation}
\def\arraystretch{1.5}
\begin{bmatrix}
{\bf A} & -\textbf{B}^T & \textbf{C}^T
\\
\textbf{B} & \textbf{D} & \textbf{E}
\\
\textbf{C} & \textbf{G} & \textbf{H}
\end{bmatrix}
\begin{bmatrix}
\textbf{U}
\\
\textbf{P}
\\
\widehat{\textbf{P}}
\end{bmatrix}
=
\begin{bmatrix}
\textbf{R}_u
\\
\textbf{R}_p
\\
\textbf{R}_{\widehat{p}}
\end{bmatrix}
 \notag
,
\end{equation}
and isolating interior unknowns gives
\begin{equation}
\def\arraystretch{1.5}
\begin{bmatrix}
\textbf{U}
\\
\textbf{P}
\end{bmatrix}
=
\begin{bmatrix}
\textbf{A} & -\textbf{B}^T
\\
\textbf{B}  & \textbf{D}
\end{bmatrix}^{-1}
\Bigg(
\begin{bmatrix}
\textbf{R}_u
\\
\textbf{R}_p
\end{bmatrix}
-
\begin{bmatrix}
\textbf{C}^T
\\
\textbf{E}
\end{bmatrix}
\widehat{\textbf{P}}
\Bigg)
 \label{hdg_invert}
 .
\end{equation}
Due to the discontinuous nature of HDG, the inverted matrix in equation~\eqref{hdg_invert} can be applied in an element by element manner.  The equation that enforces continuity of the normal component of the numerical trace of the Darcy velocity is
\begin{equation}
\textbf{C} \textbf{U} + \textbf{G}\textbf{P} + \textbf{H} \widehat{\textbf{P}} = \textbf{R}_{\widehat{p}}.
\end{equation}
We can condense the interior unknowns to obtain a globally coupled system only defined in terms of $\widehat{P}$, the pressure on the mesh skeleton, 
$$
\mathbb{H}\widehat{\textbf{P}} = \mathbb{F},
$$
where
\begin{equation}
\begin{split}
\mathbb{H}
& =
\textbf{H} - [\textbf{C}~\textbf{G}]
\begin{bmatrix}
\textbf{A} & -\textbf{B}^T
\\
\textbf{B}  & \textbf{D} 
\end{bmatrix}^{-1}
\begin{bmatrix}
\textbf{C}^T
\\
\textbf{E}
\end{bmatrix},
\\
\mathbb{F} &=
\textbf{R}_{\widehat{p}}
-
[\textbf{C}~\textbf{G}]
\begin{bmatrix}
\textbf{A} & -\textbf{B}^T
\\
\textbf{B}  & \textbf{D} 
\end{bmatrix}^{-1}
\begin{bmatrix}
\textbf{R}_u
\\
\textbf{R}_p
\end{bmatrix}.
\end{split}
\end{equation}
We note that the expressions for $\mathbb{H}$ and $\mathbb{F}$ can be obtained at the element level.  The HDG method for Darcy flow has a number of appealing features, notably:
\begin{itemize}
\item Static condensation reduces the total number of degrees of freedom.  This is especially important for discontinuous Galerkin methods which give rise to a large number of unknowns.  The plethora of unknowns is increased even further by problems in high dimensions and higher order polynomial approximation spaces.
\item HDG allows for flexibility in the selection of approximation spaces in comparison to MFE.
\item HDG possesses a local mass balance property, which can be crucial for coupled flow-transport problems~\cite{dawson2004compatible}.
\item The approximations for ${\bm u}_h$, $p_h$, and $\widehat{p}_h$ all converge at the optimal rate of $k+1$ in the $L^2$ norm..
%\item Postprocessing for $p $ results in a new approximation $p^*$ that superconverges at the rate of $k+2$.
\item The numerical trace of ${\bm u}_h$ has its normal component continuous, which renders the HDG method a compatible discretization~\cite{dawson2004compatible}.
\end{itemize} 
%  \subsection{Relation between HMFE and HDG for Darcy flow} 
% The HMFE and HDG methods are closely related, especially for second order elliptic problems.  We note that both methods can use the same computational framework.  In subsection~\ref{sec:darcy_hdg}, if we set $\tau \equiv 0$, and replace the pressure space $W_h^{\text{DG}}$ with $W_h^{\text{BDM}}$, one recovers the hybridized BDM (HMFE) method as defined in~\cite{cockburn2004characterization}.  As such, it is straightforward to modify HDG method to obtain the hybridized BDM method.
% 
%It should be noted that the HMFE method has the benefit of a Darcy velocity with continuous normal component.  For the HDG method, only the numerical trace of the Darcy velocity has its normal component continuous.  The HDG method requires a stabilization parameter, whereas the HMFE method does not.  Both methods result in a symmetric positive definite system, with comparable computational costs as the pressure trace space is $M_h^{\text{DG}}$.  The pressure on the volume space ($p_h$) converges at the rate of $k-1$ for HMFE ($k>0$) and $k$ for HDG ($k\ge0$) in the $L^2$ norm.  In section~\ref{sec:numerical} we compare the HDG-HDG and HMFE-HDG methods for the miscible displacement problem.

 \subsection{Concentration approximation}
  \label{cn_sec}  
   
 \subsubsection{Hybridizable discontinuous Galerkin} 
 In this section we describe the HDG method used to discretize the convection-diffusion equation~(\ref{eq:modelB}).  
 %In order to define the method, the following discrete spaces are needed:
%\begin{equation}
%\begin{split}
%W_h&= \{ w\in L^2(\Omega): w|_E \in \mathbb{Q}_{k}(E),~~~\forall E \in \mathcal{E}_h \},
%\\
%{\bm V}_h &= W_h \times W_h ,
%\\
%M_h &= \{ \zeta \in L^2(\Gamma_h): \zeta|_e \in \mathbb{Q}_k(e),~~~\forall e \in \Gamma_h \}
%%\\
%%M_h(0) &= \{ \zeta \in M_h: \mu=0,~~~\text{on } \Gamma_{p_D}\}
%.
%\end{split}
%\label{hdg_spaces}
%\end{equation}
Let 
$$
{\bm q}_h= - {\bm D}({\bm u}_h) \nabla c.
$$ 
 It is assumed that the velocity ${\bm u}_h$ has been computed.  We utilize the discontinuous finite element spaces as stated in equation~(\ref{eq:bdm_spaces}).  The HDG method seeks $({\bm q}_h , c_h , \widehat{c}_h )\in {\bm V}_h^{DG} \times W_h^{DG}\times M_h^{DG} $ such that
\begin{align}
(  ({\bm D}({\bm u_h}))^{-1} {\bm q}_h , {\bm v})_{\mathcal{E}_h}
-
(   c_h , \nabla\cdot {\bm v} )_{\mathcal{E}_h}
+
\langle \widehat{c}_h, {\bm v}\cdot {\bm n} \rangle_{\partial\mathcal{E}_h}
&=
0,
\label{eq:concentration1}
\\
\bigg(\phi \frac{\partial c_h}{\partial t},w\bigg)_{\mathcal{E}_h}
+
(q^P c_h,w)_{\mathcal{E}_h}
 -
 ( {\bm u}_h c_h +{\bm q}_h , \nabla w)_{\mathcal{E}_h}
+
\langle
(\widehat{{\bm q}_h } 
+
{\bm u}_h \widehat{c}_h
)
\cdot
{\bm n}
,
w
\rangle_{\partial\mathcal{E}_h}
 &= (q^I \bar{c},w)_{\mathcal{E}_h},
 \label{eq:concentration2}
\\
\langle  
  (\widehat{\bm q}_h + {\bm u}_h \widehat{c}_h)  \cdot {\bm n}
  , 
  \zeta 
  \rangle_{ \partial \mathcal{E}_h } 
 &= 0,
 \label{eq:concentration3}
\end{align}
for all $({\bm v}, w,\zeta)\in {\bm V}_h^{DG} \times W_h^{DG}\times M_h^{DG} $.  We use explicit formulas for the dispersion-diffusion tensor (see subsection~\ref{subsec_manu}), which means that we can analytically precompute its inverse.  The numerical traces take the following form
\begin{align*}
\tilde{c}_h
&=
\widehat{c}_h ,
\\
\widehat{\bm q}_h 
&=
{\bm q}_h 
+
\tau ( c_h  - \tilde{c}_h ){\bm n},
\end{align*}
 where $\tau$ is a stabilization term that is piecewise constant on element boundaries.  Following \cite{nguyen2009implicit,chen2015robust}, given $e\in \Gamma_h$, we set
 $$
 \tau ({\bm x})  = |{\bm u}({\bm x}) \cdot {\bm n}| +  \max{ (\| {\bm D}({\bm u}({\bm x}))\|_{\infty} ,1) },
 \forall {\bm x} \in e.
 $$
Further details on the selection of $\tau$ can be found in~\cite{CockburnDGRS09}.  The choice of $\tau$ becomes less important as the polynomial order increases, since the numerical dissipation is on the order of $\mathcal{O}(h^{k+1})$.  HDG methods for the convection-diffusion problem have been studied by numerous authors, including the case of convection-dominated diffusion, and small diffusion coefficients \cite{fu2015analysis,CockburnDGRS09,nguyen2009implicit,qiu2016hdg,chen2014analysis}.
 
Equations~\eqref{eq:concentration1},~\eqref{eq:concentration2},and~\eqref{eq:concentration3} give rise to the following matrix system
\begin{equation}
\def\arraystretch{1.5}
\begin{bmatrix}
\textbf{A} & -\textbf{B}^T & \textbf{J}^T
\\
\textbf{B} & \textbf{D} & \textbf{E}
\\
\textbf{J} & \textbf{G} & \textbf{H}
\end{bmatrix}
\begin{bmatrix}
\textbf{Q}
\\
\textbf{C}
\\
\widehat{\textbf{C}}
\end{bmatrix}
=
\begin{bmatrix}
\textbf{R}_q
\\
\textbf{R}_c
\\
\textbf{R}_{\widehat{c}}
\end{bmatrix}
 \notag
,
\end{equation}
and isolating interior unknowns gives
\begin{equation}
\def\arraystretch{1.5}
\begin{bmatrix}
\textbf{Q}
\\
\textbf{C}
\end{bmatrix}
=
\begin{bmatrix}
\textbf{A} & -\textbf{B}^T
\\
\textbf{B}  & \textbf{D} 
\end{bmatrix}^{-1}
\Bigg(
\begin{bmatrix}
\textbf{R}_q
\\
\textbf{R}_c
\end{bmatrix}
-
\begin{bmatrix}
\textbf{C}^T
\\
\textbf{E}
\end{bmatrix}
\widehat{\textbf{C}}
\Bigg)
 \label{eq:hdg_invert}
 .
\end{equation}
Due to the discontinuous nature of HDG, the inverted matrix in equation~(\ref{eq:hdg_invert}) can be applied in an element by element manner.  The equation that enforces continuity of the normal component of the numerical trace is
\begin{equation*}
\textbf{J} \textbf{Q} + \textbf{G}\textbf{C} + \textbf{H} \widehat{\textbf{C}} = \textbf{R}_{\widehat{c}}.
\end{equation*}
We can condense the interior unknowns to obtain a globally coupled system only defined in terms of $\widehat{\textbf{C}}$, the concentration on the mesh skeleton, $\mathbb{H}\widehat{\textbf{C}} = \mathbb{F}$:
\begin{equation*}
\begin{split}
\mathbb{H}
& =
\textbf{H} - [\textbf{J}~\textbf{G}]
\begin{bmatrix}
\textbf{A} & -\textbf{B}^T
\\
\textbf{B}  & \textbf{D} 
\end{bmatrix}^{-1}
\begin{bmatrix}
\textbf{J}^T
\\
\textbf{E}
\end{bmatrix},
\\
\mathbb{F} &=
\textbf{R}_{\widehat{c}}
-
[\textbf{J}~\textbf{G}]
\begin{bmatrix}
\textbf{A} & -\textbf{B}^T
\\
\textbf{B}  & \textbf{D} 
\end{bmatrix}^{-1}
\begin{bmatrix}
\textbf{R}_q
\\
\textbf{R}_c
\end{bmatrix}.
\end{split}
\end{equation*}
We note that the expressions for $\mathbb{H}$ and $\mathbb{F}$ can be obtained at the element level.  The HDG method for transport has a number of appealing features, notably:
\begin{itemize}
\item  It is locally conservative.

\item  Static condensation reduces the total number of degrees of freedom, as well as total number of nonzero entries in the discretization matrix.  This is especially important for discontinuous Galerkin methods, as they give rise to a large number of unknowns compared to continuous Galerkin.  
 
\item  The normal component of the numerical flux $\widehat{ {\bm q}}_h$ is continuous. 
 
\item  The approximations for ${\bm q}_h$, $c_h$, and $\widehat{c}_h$ all converge at the optimal rate of $k+1$ in the $L^2$ norm.

%\item Postprocessing for $c $ results in a new approximation $c^*$ that superconverges at the rate of $k+2$.
\end{itemize}
    
 \subsection{Semi-implicit algorithm} 
 The semi-implicit algorithm is described here.  The Darcy system~\eqref{eq:modelA} is split from the transport system~\eqref{eq:modelB}.  We first solve the Darcy problem, given a concentration profile.  The HDG method simultaneously recovers both the pressure and velocity.  After the velocity is obtained, we insert it into the transport system to generate an updated concentration profile.  The HDG scheme is formulated in Algorithm~\ref{alg:misc_disp2}.  Let $t_n$ denote the time at step $n$, and $n_{\textrm{nsteps}}$ be the number of time steps to be taken.  The time step size is given by $\Delta t$.  A superscript of $n$ denotes the $n$th time step, so that 
 $$
 c^n_h({\bm x},t_n) : = c^{n}_h,
 \quad \widehat{c}^n_h({\bm x},t_n) : = \widehat{c}^{n}_h,
 \quad p^n_h({\bm x},t_n) : = p^{n}_h, 
 \quad\widehat{p}^n_h({\bm x},t_n) : = \widehat{p}^{n}_h,
 \quad {\bm q}^n_h({\bm x},t_n) : = {\bm q}^{n}_h,
 \quad {\bm u}^n_h({\bm x},t_n) : = {\bm u}^{n}_h.
 $$
Initial conditions correspond to $n=0$.
%%\newenvironment{Algorithm}[1][tbh]%
%%{\begin{myalgo}[#1]
%%\centering
%%\begin{minipage}[t]{5cm}
%%\vspace{0pt} 
%%\begin{algorithm}[H]}%
%%{\end{algorithm}
%%\end{minipage}
%%\end{myalgo}}
\begin{center}
%%\begin{minipage}[t]{8cm}
%%  \vspace*{-5ex}  
%%  \begin{algorithm}[H] 
%%%\begin{Algorithm}[H]
%%\caption{Semi-implicit HMFE-HDG method for the miscible displacement problem.}
%%\label{alg:misc_disp1}
%% \begin{algorithmic}[1]
%%%\Procedure{MyProcedure}{}
%%      \For{$n=0$ to $n_{\textrm{nsteps}}-1$ }%\Comment{We have the answer if r is 0}
%%        \State Using $c_h^n$ and ${\bm q}_h^n$, solve system~\eqref{eq:hmfe_darcy1},~\eqref{eq:hmfe_darcy2}, and~\eqref{eq:hmfe_darcy3} for $\widehat{p}_h^n$.  Recover ${\bm u}_h^n$ and $p_h^n,$ element by element via equations~\eqref{eq:static_evap_darcy1}-\eqref{eq:static_evap_darcy2}.
%%        \State Using $c_h^n$ and ${\bm u}_h^n$, solve system~\eqref{eq:modelB} for $\widehat{  c}_h^{n+1}$.  Recover ${\bm q}_h^{n+1}$ and ${  c}_h^{n+1},$  element by element via equations~\eqref{eq:hdg_invert}.
%%        
%%		\State ${  \bm q}_h^{n} \gets {\bm q}_h^{n+1}$.        
%%		\State ${  c}_h^{n} \gets {  c}_h^{n+1}$.
%%        
%%        \State $t_{n+1} \gets t_n + \Delta t$.
%%      \EndFor %\label{euclidendwhile}
%%%\EndProcedure
%%\end{algorithmic}
%%%\end{Algorithm}
%%\end{algorithm}
%%\end{minipage}%
%%\hspace*{1ex}
\begin{minipage}[t]{8cm}
  \vspace*{-5ex} 
%\begin{Algorithm}[H]
  \begin{algorithm}[H] 
\caption{Semi-implicit HDG method for the miscible displacement problem.}
\label{alg:misc_disp2}
 \begin{algorithmic}[1]
%\Procedure{MyProcedure}{} eq:hdg_darcy3
      \For{$n=0$ to $n_{\textrm{nsteps}}-1$ }%\Comment{We have the answer if r is 0}
        \State Using $c_h^n$ and ${\bm q}_h^n$, solve system~\eqref{eq:hdg_darcy1}~\eqref{eq:hdg_darcy2}, and~\eqref{eq:hdg_darcy3} for $\widehat{p}_h^n$.  Recover ${\bm u}_h^n$ and $p_h^n,$ element by element via equations~\eqref{hdg_invert}.
        \State Using $c_h^n$ and ${\bm u}_h^n$, solve system~\eqref{eq:concentration1},~\eqref{eq:concentration2}, and~\eqref{eq:concentration3} for $\widehat{  c}_h^{n+1}$.  Recover ${\bm q}_h^{n+1}$ and ${  c}_h^{n+1},$  element by element via equations~\eqref{eq:hdg_invert}.
        
		\State ${  \bm q}_h^{n} \gets {\bm q}_h^{n+1}$.        
		\State ${  c}_h^{n} \gets {  c}_h^{n+1}$.
        
        \State $t_{n+1} \gets t_n + \Delta t$.
      \EndFor %\label{euclidendwhile}
%\EndProcedure
\end{algorithmic}
%\end{Algorithm}
\end{algorithm}
\end{minipage} 
 \end{center}
 
\subsection{Reduced computational cost of hybridization} 
A key feature of hybridization is that it reduces the total number of degrees of freedom (DOFs) compared to their classical counterparts~\cite{cockburn2016static}.  Specific information regarding total degrees of freedom and total nonzero entries in the discretization matrix for different element types can be found in~\cite{huerta2013efficiency,samii2016parallel,bui2016construction}.  For clarity we consider some quantitative examples to illustrate this point.  
\subsubsection{Example 1 (2D quadrilateral mesh)} 
\begin{figure}[ht!]
%\hspace*{-5ex}
\subfloat[Ratio vs $k$]{\includegraphics[trim = 10mm 80mm 20mm 85mm, clip, scale = 0.5]{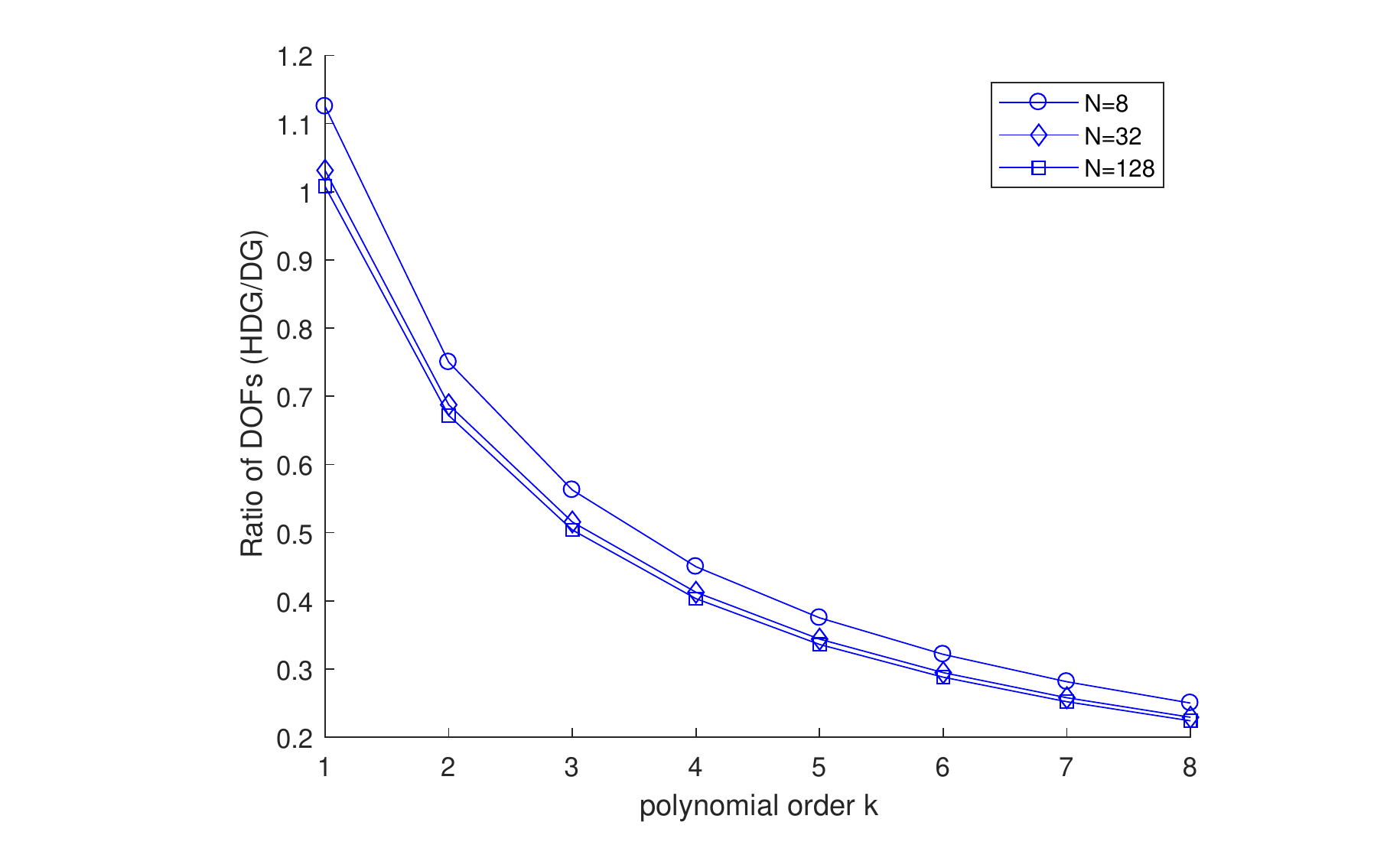}\label{fig:dof_1}}
\hspace*{-5ex}
\subfloat[Ratio vs $N$]{\includegraphics[trim = 10mm 80mm 20mm 85mm, clip, scale = 0.5]{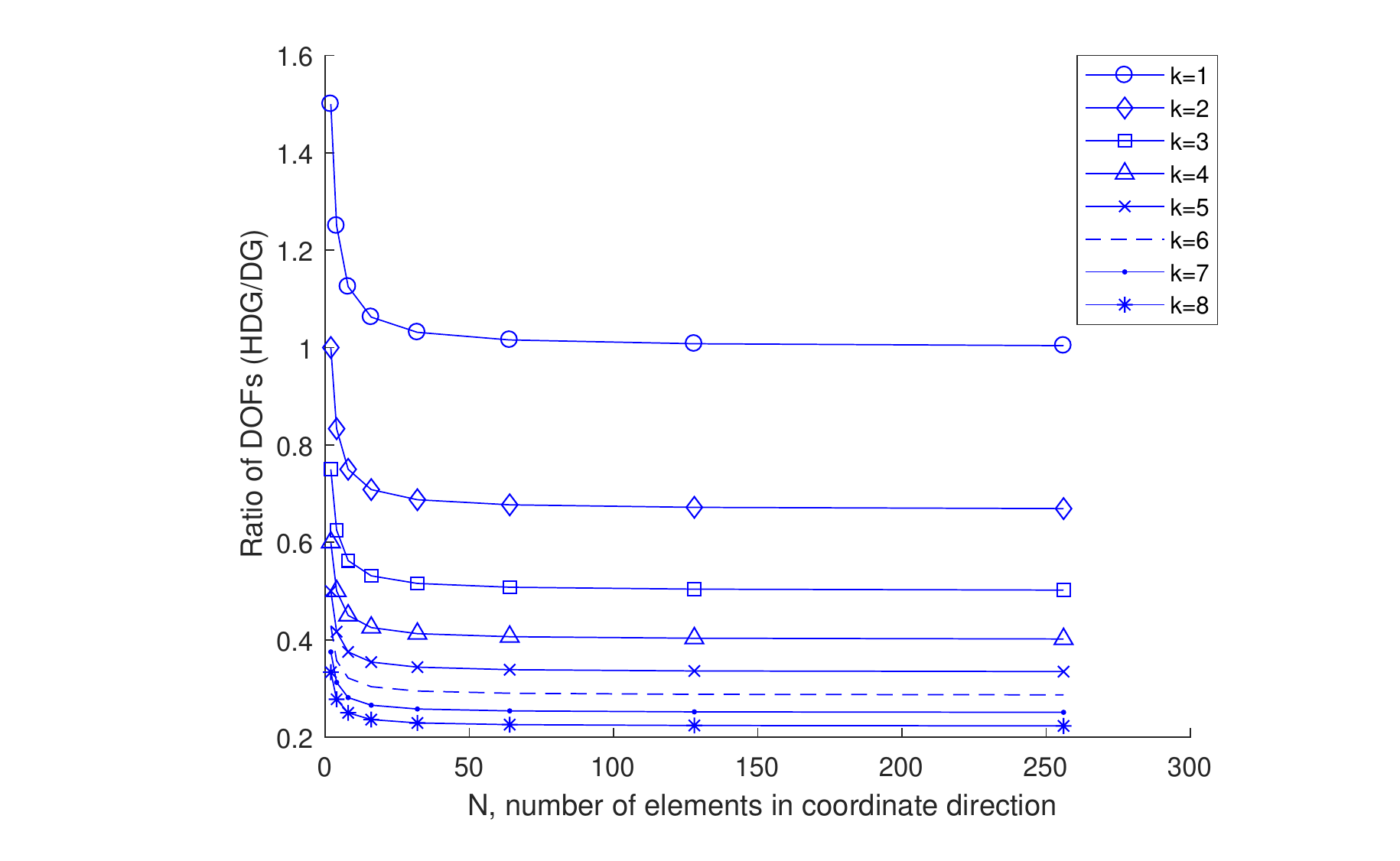}
\label{fig:dof_2}}
\caption{Ratio of DOFs, HDG to DG (2D quadrilateral meshes).  The ratio vs $k$ (left) and the ratio vs $N$ (right) shows that the HDG method benefits more for large $k$.}
\label{fig:dof_3}
\end{figure}

We consider a 2D uniform mesh of the unit square with $N\times N$ quadrilateral elements ($N$ elements in each coordinate direction).  The DOFs for classical DG methods are $(k+1)^2 N^2$, since there are $N^2$ elements and $(k+1)^2=\dim{ \mathbb{Q}_k(E)}$.  For HDG we have $(k+1)(2N^2+2N)$ DOFs, since there are $2N^2+2N$ faces and $k+1 =\dim{ \mathbb{Q}_k(e)}$.  It then follows that
$$
\tilde{r}
:=
\frac{ 
\textrm{HDG}_{\textrm{DOFs}}
}
{
\textrm{DG}_{\textrm{DOFs}}
}
<1,
$$
whenever $k > 1 + 2N^{-1}$.  Fig.~\ref{fig:dof_3} visualizes this ratio, for sample $k$ and $N$.  From Fig~\ref{fig:dof_2}, it is evident that for a given polynomial order, increasing the number of elements decreases the ratio $\tilde{r}$.  However, much more substantial reductions for $\tilde{r}$ are obtained if the polynomial degree is increased, which is clear from Figs~\ref{fig:dof_1} and~\ref{fig:dof_2}.
For $N>2$ and $k>1$, the HDG method has fewer DOFs than classical DG.  For $k=3$ and large $N$, classical DG has almost twice the number of DOFs compared to HDG.
\subsubsection{Example 2 (3D tetrahedral mesh)} 
In 3D the situation is similar.  Here we consider the unit cube, which is partitioned into $N\times N \times N$ hexahedra ($N$ hexahedra in each coordinate direction).  Each hexahedron is then divided into 5 tetrahedra.  Let $\mathbb{P}_k(E)$ denote the space of polynomials of degree at most $k$ on the domain $E$.  Here the DOFs for classical DG methods are $5(k+3)(k+2)(k+1) N^3/6$, since there are $5N^3$ elements and $(k+3)(k+2)(k+1)/6 =\dim{ \mathbb{P}_k(E)}$.  For HDG we have $(k+2)(k+1)(6N^3+2N^2)/2$ DOFs, since there are $6N^3+2N^2$ faces and $(k+2)(k+1)/2 =\dim{ \mathbb{P}_k(e)}$.  We then have
$$
\tilde{r}
:=
\frac{ 
\textrm{HDG}_{\textrm{DOFs}}
}
{
\textrm{DG}_{\textrm{DOFs}}
}
<1,
$$
whenever $k > (3 + 6N^{-1})/5$.
\begin{figure}[ht!]
%\hspace*{-5ex}
\subfloat[Ratio vs $k$]{\includegraphics[trim = 10mm 80mm 20mm 85mm, clip, scale = 0.5]{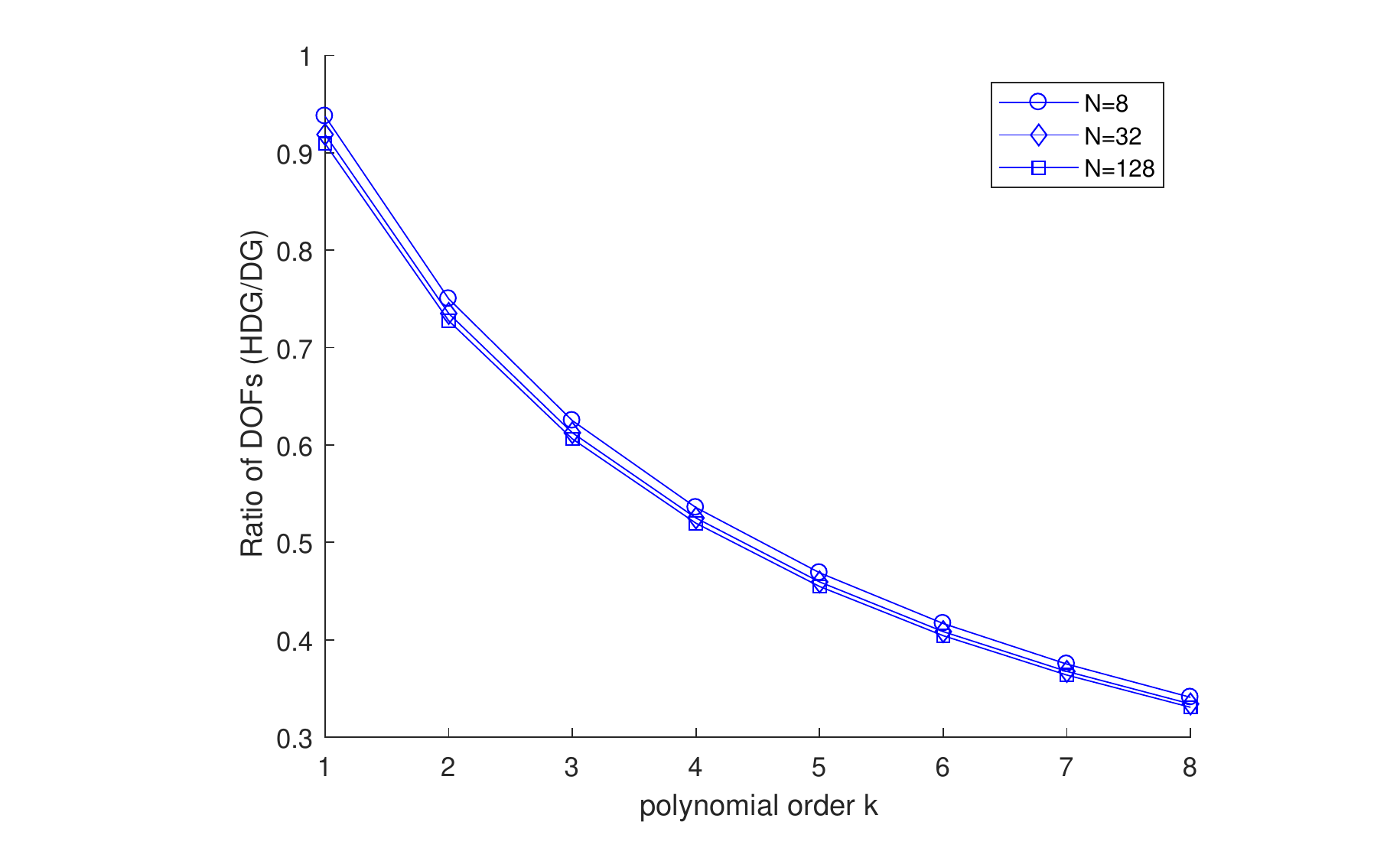}\label{fig:dof_4}}
\hspace*{-5ex}
\subfloat[Ratio vs $N$]{\includegraphics[trim = 10mm 80mm 20mm 85mm, clip, scale = 0.5]{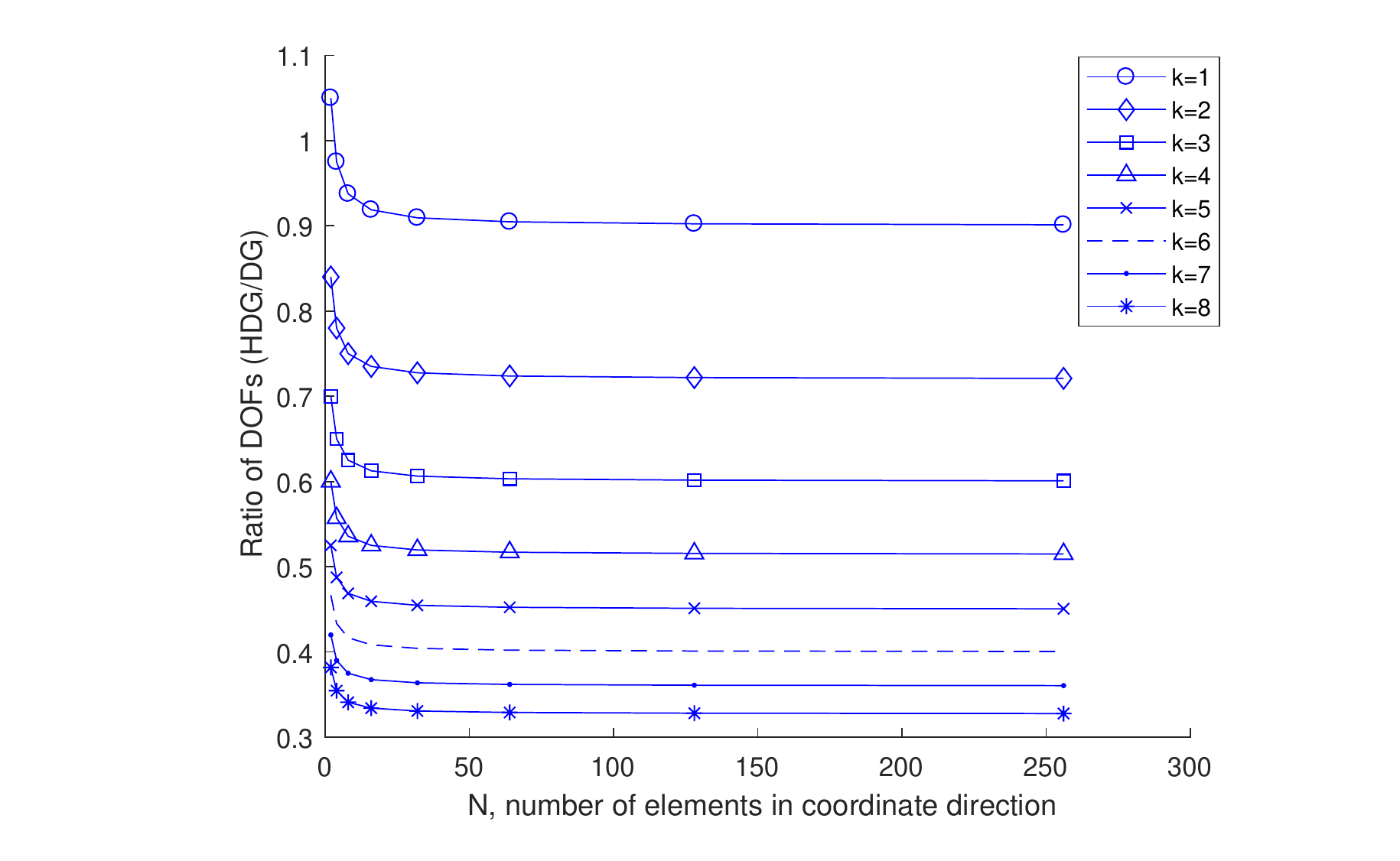}
\label{fig:dof_5}}
\caption{Ratio of DOFs, HDG to DG (3D tetrahedral meshes).  The ratio vs $k$ (left) and the ratio vs $N$ (right) shows that the HDG method benefits more for large $k$.}
\label{fig:dof_6}
\end{figure} 
From Fig.~\ref{fig:dof_6}, we observe that the same conclusions can be drawn as the previous example in 2D.  The HDG method benefits much more from large $k$, this is depicted in Figs.~\ref{fig:dof_4} and~\ref{fig:dof_5}.  For $k>0$ and $N>3$, the HDG method will have fewer DOFs than classical DG.
 
\section{Numerical experiments} 
\label{sec:numerical}
Four numerical experiments are given in this section to test our algorithm.  We use implicit Euler time stepping for all of the following experiments.  The initial concentration is zero for all numerical experiments; except for subsection~\ref{subsec_manu}, where it is determined from the manufactured solution.%  In subsections~\ref{subsec_hom1},~\ref{sec:spe2d}, and~\ref{sec:spe3d} we use the HMFE-HDG scheme (Algorithm~\ref{alg:misc_disp1}).  Subsections~\ref{subsec_manu} and~\ref{sec:lens} compare the HMFE-HDG scheme and the HDG-HDG scheme (Algorithm~\ref{alg:misc_disp2}).
 
\subsection{Manufactured solution in 2D}
\label{subsec_manu}
To test the convergence properties of our method, we use the method of manufactured solutions.  The domain is set to be the unit square, and is partitioned into $N\times N$ quadrilateral elements.  For simplicity, we use Dirichlet boundary conditions, which are obtained through the following prescribed solutions:
\begin{align*}
            p(x,y,t)  &= 1 + xy\tanh(1-x) \tanh(1-y)\exp(-t),
            \\
            c(x,y,t) &= \cos(t) \sin(\pi x)   \sin(\pi y)/(2\pi)^2 .
\end{align*}
Initial conditions are easily acquired from the above expressions.  We set $\Delta t = 0.1/( (k+1) N^k)$, and $T=0.5$.  For ease of generating the source and sink functions, a homogeneous permeability taken.  Also, the standard quarter power mixing law is used:
$$
 { K}   =  9.44\cdot 10^{-3} ,
  ~~~
 {\mu(c)}= (c (\mu_s)^{-0.25} + (1-c) (\mu_o)^{-0.25})^{-4}   %mu_o/mu_s
,
$$
with mobility ratio $\mu_o/\mu_s=2$.  We remark that ${\bm D}( \cdot )$ is assumed to be symmetric and uniformly positive definite.  Moreover, the dispersion-diffusion tensor is of size $d\times d$, where $d$ is the underlying dimension.  We set $d_m=1.0$, $\alpha_t = (1.8) \cdot 10^{-6}$ and $\alpha_l = (1.8) \cdot 10^{-5}$.  The molecular diffusion coefficient, time step, and permeability are selected such that proper convergence rates may be extracted.  The porosity is fixed constant, $\phi\equiv 0.2$.  Table~\ref{tab_convergence_rates_hdg} shows the results of the HDG scheme.  We observe that the expected convergence rates are met.  The HDG scheme results in optimal rates of $k+1$ in the $L^2$ norm for all approximate variables.

 \begin{table}[htb!]
\centering
\begin{tabular}{l b a b a b a b a b}
\hline
 &    \multicolumn{3}{c}{$\|p_h-p\|_{L^2(\Omega)}$} & \multicolumn{2}{c}{$\|  {\bm u}_h-{\bm u}\|_{L^2(\Omega)}$} 
&  \multicolumn{2}{c}{$\|  c_h-  c\|_{L^2(\Omega)}$} & \multicolumn{2}{c}{$\| {\bm q}_h - {\bm q}\|_{L^2(\Omega)}$} 
\\ 
\cline{3-4} \cline{5-6}  \cline{7-8}  \cline{9-10}
$k$  &$N$& Error & Rate & Error & Rate &  Error& Rate & Error& Rate
\\ 
\hline
\vspace*{-2.6ex}
\\
  
1& 4  &  4.5289e-03 & -&9.9657e-05 &-&9.9657e-04 &-&4.5389e-05 &-\\
 & 8  &  1.3640e-03 &1.7313 &3.0906e-05 &1.6891 &3.0906e-04 &1.6891 &1.3671e-05 &1.7312 \\
 & 16 &  3.7910e-04 &1.8472 &8.6748e-06 &1.8330 &8.6748e-05 &1.8330 &3.7997e-06 &1.8471 \\
 & 32 &  1.0028e-04 &1.9186 &2.3091e-06 &1.9095 &2.3091e-05 &1.9095 &1.0051e-06 &1.9185 \\
 & 64 &  2.5814e-05 &1.9578 &5.9693e-07 &1.9517 &5.9693e-06 &1.9517 &2.5874e-07 &1.9578 \\
\\
2& 4  & 4.5027e-05 &-&1.0323e-07 &-&1.0323e-05 &-&4.5027e-06 &-\\
 & 8  & 6.4664e-06 &2.7998 &1.4819e-08 &2.8004 &1.4819e-06 &2.8004 &6.4664e-07 &2.7998 \\
 & 16 & 8.6804e-07 &2.8971 &1.9975e-09 &2.8912 &1.9975e-07 &2.8912 &8.6804e-08 &2.8971 \\
 & 32 & 1.1252e-07 &2.9476 &2.5998e-10 &2.9417 &2.5998e-08 &2.9417 &1.1252e-08 &2.9476 \\
\\
 3& 4   &3.5178e-06 &-&8.0583e-09 &-&8.0583e-07 &-&3.5178e-07 &-\\
  & 8   &2.4617e-07 &3.8369 &5.6509e-10 &3.8339 &5.6509e-08 &3.8339 &2.4617e-08 &3.8369 \\
  & 16  &1.6248e-08 &3.9213 &3.7459e-11 &3.9151 &3.7459e-09 &3.9151 &1.6248e-09 &3.9213 \\
  & 32  &1.0434e-09 &3.9609 &2.4137e-12 &3.9560 &2.4137e-10 &3.9560 &1.0434e-10 &3.9609 \\
 \\
 4& 4 &   3.8959e-06 &-&1.2033e-10 &-&7.7980e-07 &-&3.8959e-07 &-\\
  & 8 &   1.3339e-07 &4.8682 &4.1143e-12 &4.8703 &3.0924e-08 &4.6563 &1.3339e-08 &4.8682 \\
  & 16&   4.3950e-09 &4.9237 &1.3666e-13 &4.9120 &1.0972e-09 &4.8168 &4.3950e-10 &4.9237 \\
  & 32&   1.4138e-10 &4.9582 &4.4401e-15 &4.9438 &3.6669e-11 &4.9032 &1.4138e-11 &4.9582 \\
  \\
 5&4 &   1.4760e-07 &-&4.1182e-11 &-&4.9140e-10 &-&1.4760e-08 &-\\
  &8 &   2.5930e-09 &5.8309 &7.1927e-13 &5.8393 &9.1152e-12 &5.7525 &2.5930e-10 &5.8309 \\
  &16 &  4.3030e-11 &5.9131 &1.1943e-14 &5.9123 &1.5738e-13 &5.8560 &4.3030e-12 &5.9131 \\
  &32 &  6.9392e-13 &5.9544 &1.9357e-16 &5.9471 &2.5974e-15 &5.9210 &6.9392e-14 &5.9544 \\
\hline
 \end{tabular}
\caption{HDG method.  Errors and convergence rates for $p_h,{\bm u}_h,c_h,$ and ${\bm q}_h$, on a Cartesian mesh of $N\times N$ elements.  The variables ${\bm u}_h,c_h,p_h$ and ${\bm q}_h$ converge at the rate of $k+1$ in the $L^2$ norm.}
\label{tab_convergence_rates_hdg}
\end{table}
 
\subsection{Homogeneous permeability in 2D}
\label{subsec_hom1}
\begin{figure}[ht!]
%\hspace*{-5ex}
\subfloat[$t=2.5$]{\includegraphics[trim = 10mm 80mm 20mm 85mm, clip, scale = 0.5]{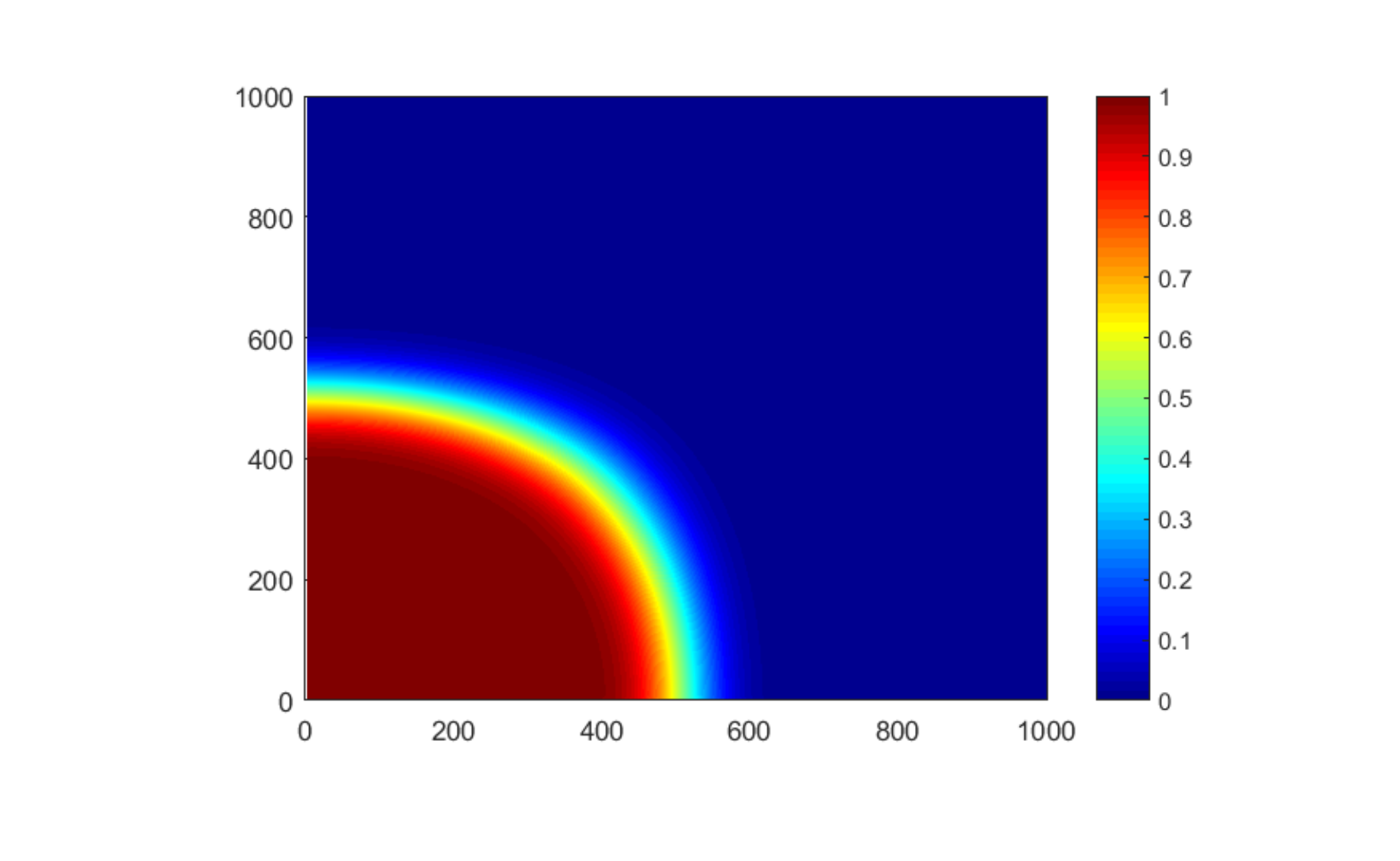}}
\hspace*{-5ex}
\subfloat[$t=5$]{\includegraphics[trim = 10mm 80mm 20mm 85mm, clip, scale = 0.5]{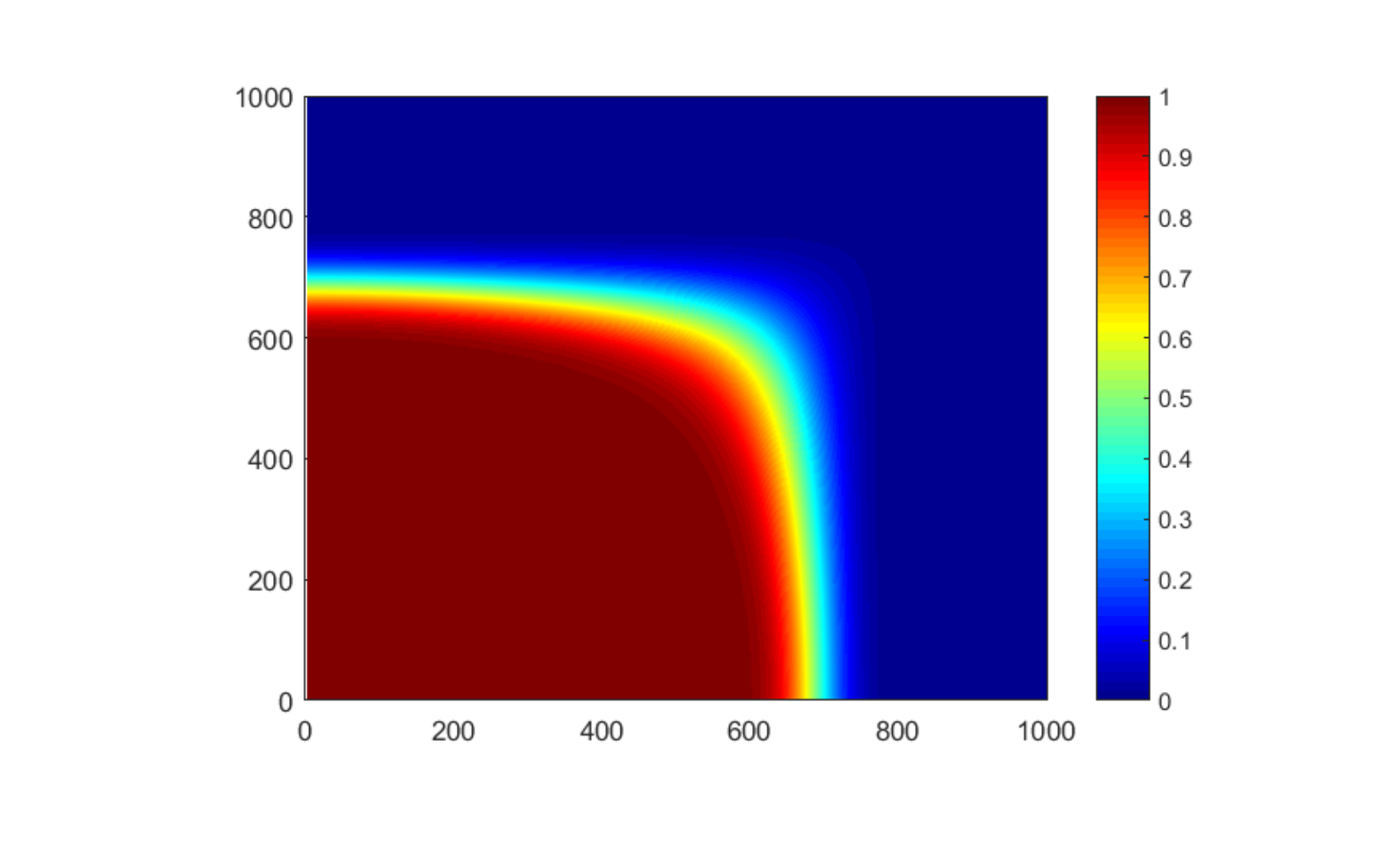}}
\newline
%\hspace*{-5ex}
\subfloat[$t=7.5$]{\includegraphics[trim = 10mm 80mm 20mm 85mm, clip, scale = 0.5]{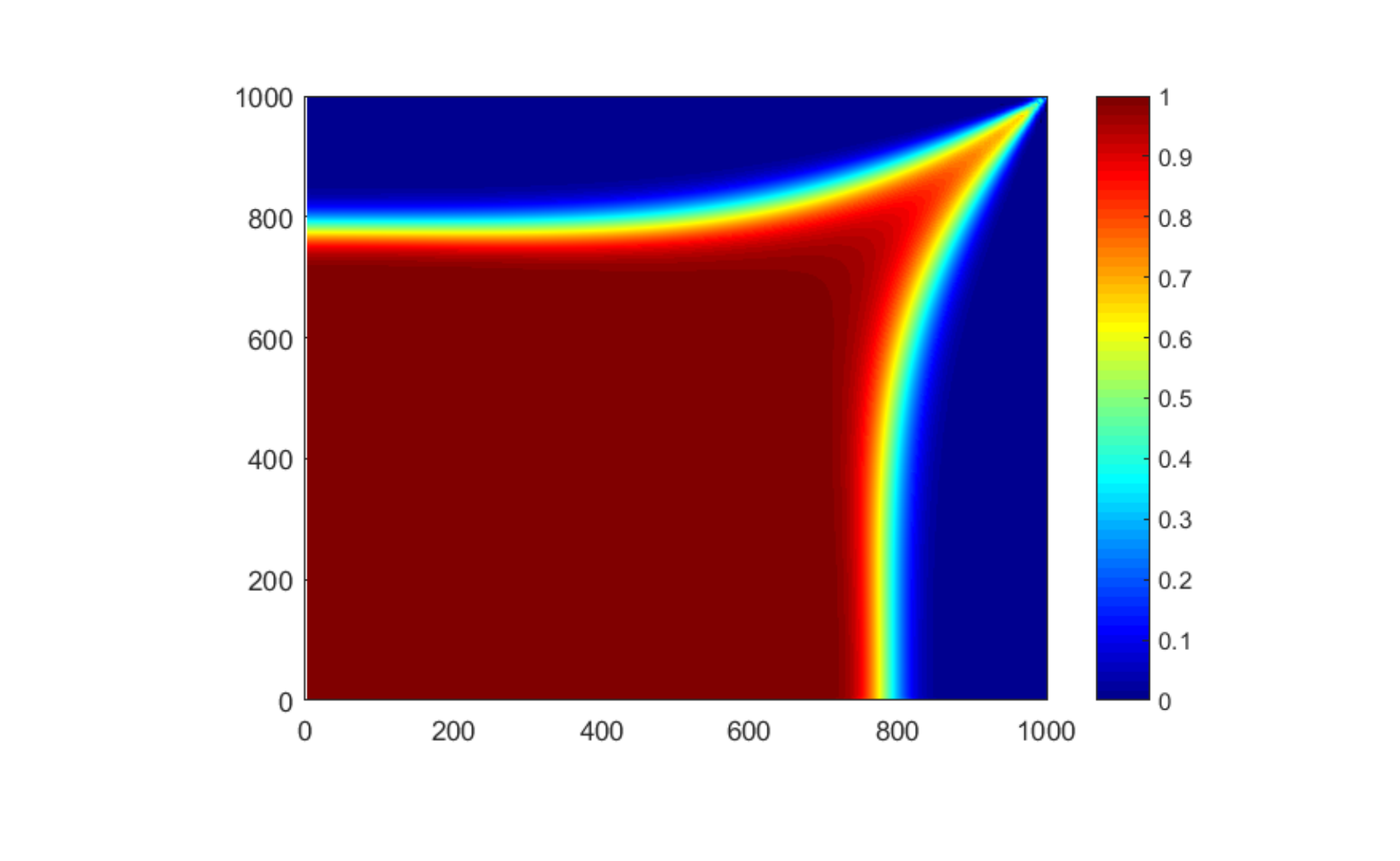}}
\hspace*{-5ex}
\subfloat[$t=10$]{\includegraphics[trim = 10mm 80mm 20mm 85mm, clip, scale = 0.5]{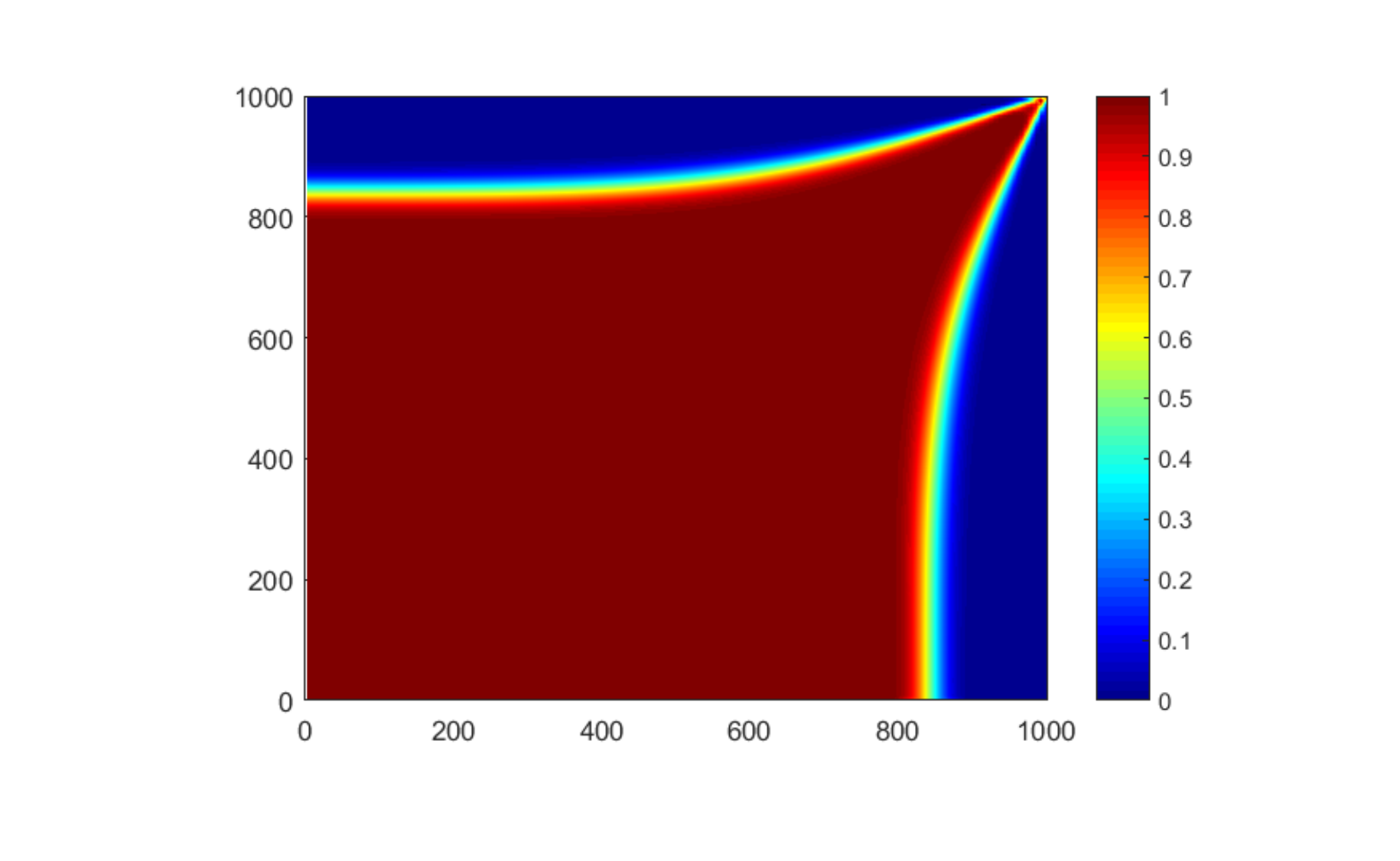}}
\caption{Miscible displacement in a homogeneous medium, with a mesh of 1024 elements, and polynomial order $k=4$.  Snapshots of the solvent profile are displayed at various times.  The concentration travels from the injection well to the production well. }
\label{hom_1}
\end{figure}

Here we test our method in a homogeneous medium $\Omega = [0,1000]^2$.  Initially, the concentration is set to zero.  The dispersion-diffusion tensor coefficients are set as $d_m = 10^{-9}$, $\alpha_t = (1.8) \cdot 10^{-6}$ and $\alpha_l = (1.8) \cdot 10^{-5}$.  Viscosity is same as subsection~\ref{subsec_manu}, and $\bar{c}\equiv 1$.  The homogeneous permeability is $ {  K}   \equiv 10^{-10}$.  We define the source terms such that they are piecewise constant with compact support.  That is, $q^I$ is nonzero on $[0,100]\times [0,100]$ and $q^P$ is nonzero on $[900,1000]\times [900,1000]$.  The non zero constants are determined by the following constraint
$$
\int_\Omega q^I
=
\int_\Omega q^P
=
0.28.
$$
The solvent fluid is injected at the lower left corner, and displaces the fluid mixture to the upper right corner.  Similar test problems can be found in~\cite{li2015numerical}.  A uniform quadrilateral mesh of 1024 elements is used, with discontinuous piecewise quartic basis functions.  The simulation runs to $T = 10$ days, and we provide snapshots at $t=2.5,$ $t=5.0,$ $t=7.5$, and $t=10$.  Our splitting algorithm allows for large timesteps, and in this case we fix $\Delta t =  0.1$ days.  The simulation results are displayed in Fig.~\ref{hom_1}.  As the problem is convection-dominated, localized overshoot and undershoot do occur, but remain bounded.

\begin{figure}[H]
%\hspace*{-5ex}
\subfloat[Complete profile]{\includegraphics[trim = 10mm 80mm 20mm 85mm, clip, scale = 0.50]{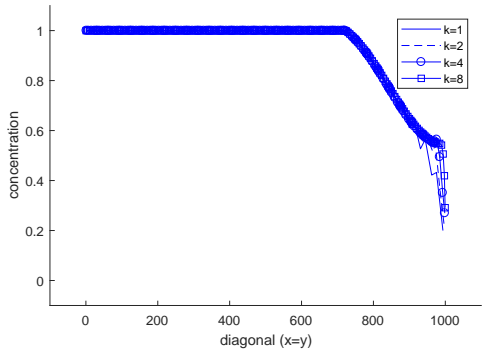}}
\hspace*{-5ex}
\subfloat[Zoom in near $x=y=950$]{\includegraphics[trim = 10mm 80mm 20mm 85mm, clip, scale = 0.50]{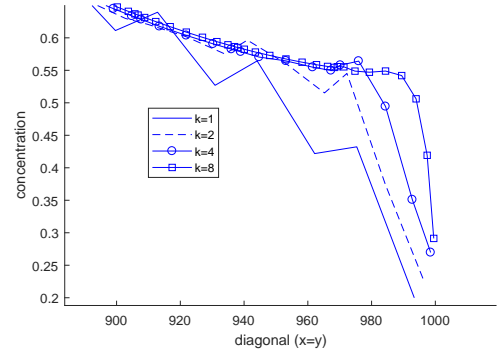}}
\newline
%\hspace*{-5ex}
\subfloat[Zoom in near $x=y=850$]{\includegraphics[trim = 10mm 80mm 20mm 85mm, clip, scale = 0.50]{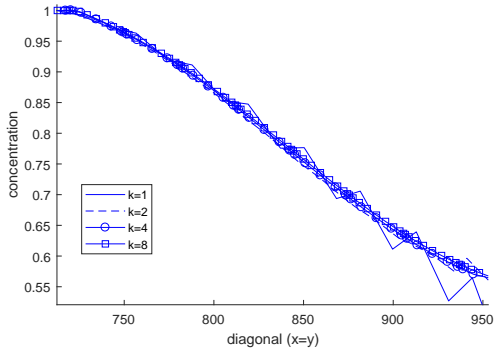}}
\hspace*{-5ex}
\subfloat[Zoom in near $x=y=750$]{\includegraphics[trim = 10mm 80mm 20mm 85mm, clip, scale = 0.50]{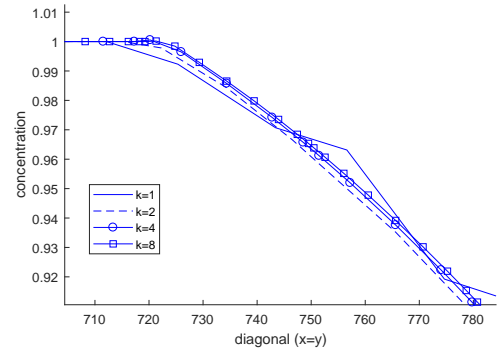}}
%\hspace*{-22ex}
%\subfigure[$h=1/256$]{\includegraphics[trim = 10mm 80mm 20mm 85mm, clip, scale = 0.65]{misc_profile1.pdf}}
\caption{Concentration profile along the line $y=x$.  HDG method on a mesh with 1024 elements, for different polynomial orders.  Smoother concentration profiles are obtained with higher order polynomials.  }
\label{hom_profile}
\end{figure}

\begin{figure}[ht!]
%\hspace*{-5ex}
\subfloat[$k=1$]{\includegraphics[trim = 10mm 80mm 20mm 85mm, clip, scale = 0.5]{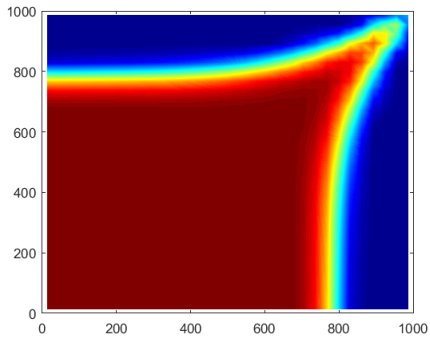}}
\hspace*{-5ex}
\subfloat[$k=2$]{\includegraphics[trim = 10mm 80mm 20mm 85mm, clip, scale = 0.5]{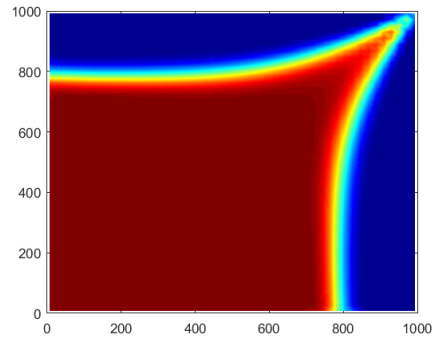}}
\newline
%\hspace*{-5ex}
\subfloat[$k=4$]{\includegraphics[trim = 10mm 80mm 20mm 85mm, clip, scale = 0.5]{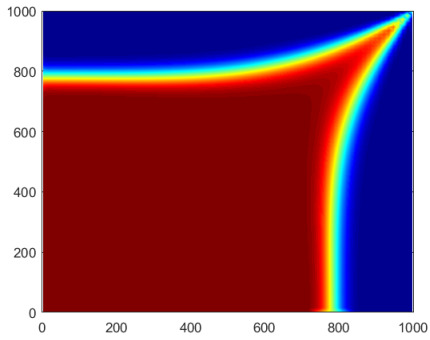}}
\hspace*{-5ex}
\subfloat[$k=8$]{\includegraphics[trim = 10mm 80mm 20mm 85mm, clip, scale = 0.5]{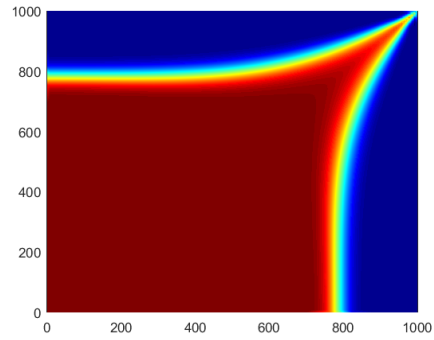}}
\caption{Polynomial order study at $t=7.5$, on a coarse mesh with 256 elements.  As the polynomial order increases, the resolution of the concentration front is sharper.}
\label{hom_2}
\end{figure}

We also study the effect of the polynomial order.  In Fig.~\ref{hom_2}, we show the concentration contours at $t=7.5$ for polynomial orders $k\in \{1,2,4,8\}$ on a mesh with 256 elements.  Features near the concentration front (especially close to the production well) become more defined as the polynomial order is increased.  To get a better sense for the convergence of the method, we examine the profile along the line $y=x$ at $t=7.5$ for polynomial orders varying from piecewise linears to piecewise octics.  Fig.~\ref{hom_profile} shows the profiles.  As the polynomial order is increased, the concentration front is smoother, and the approximations converge.  The lower order approximations are less smooth and more diffusive.
 %High order polynomials reduce dissipation and dispersion, which means that overshoot/undershoot phenomena are less evident in the high order regime.  
 %\clearpage
 
\pagebreak
\clearpage
\pagebreak
\clearpage
\clearpage
\subsection{Permeability lens in 2D}
\label{sec:lens}
\begin{figure}[ht!]
%\hspace*{-5ex}
\subfloat[$t=2.5$]{\includegraphics[trim = 10mm 80mm 20mm 85mm, clip, scale = 0.5]{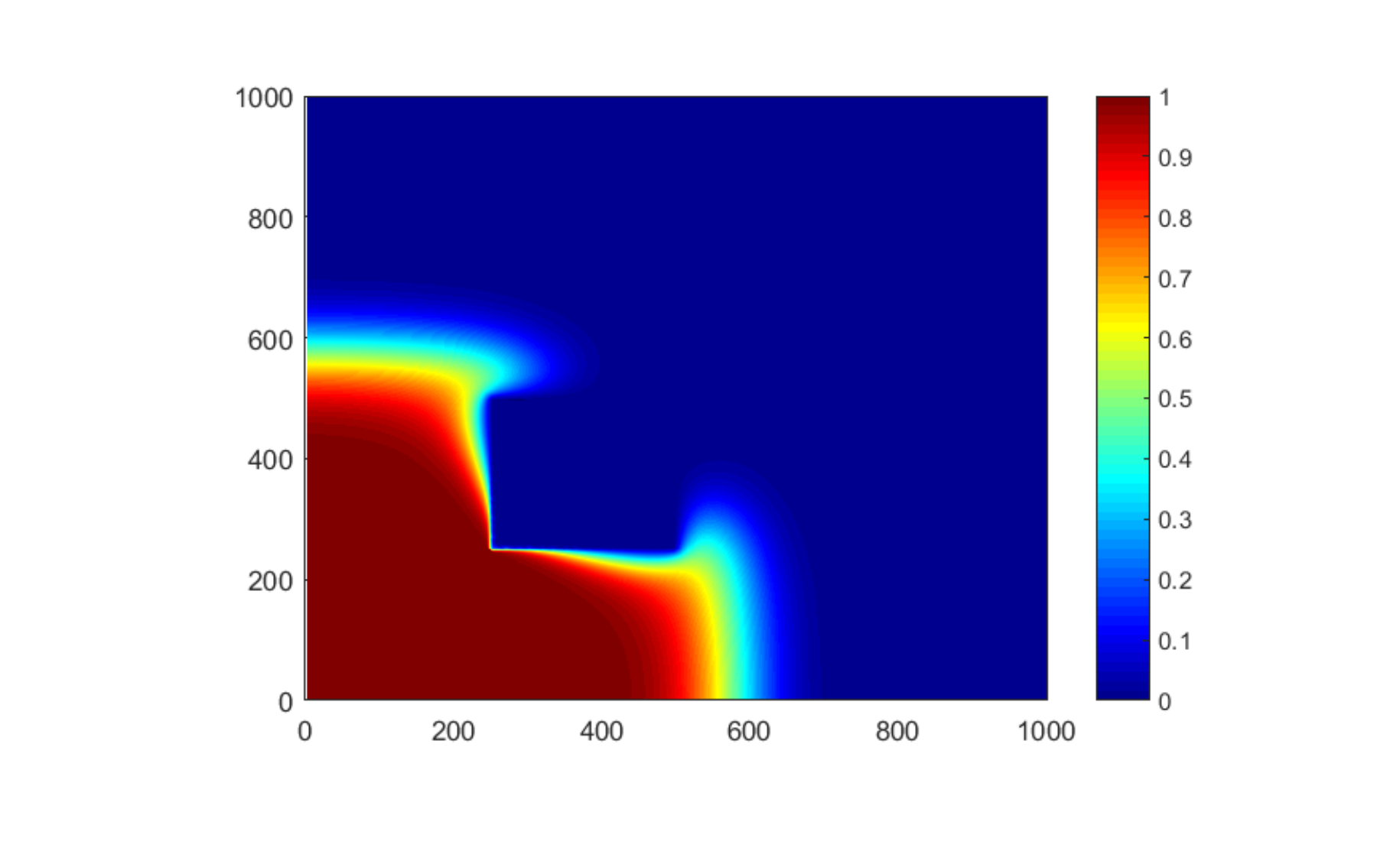}}
\hspace*{-5ex}
\subfloat[$t=5$]{\includegraphics[trim = 10mm 80mm 20mm 85mm, clip, scale = 0.5]{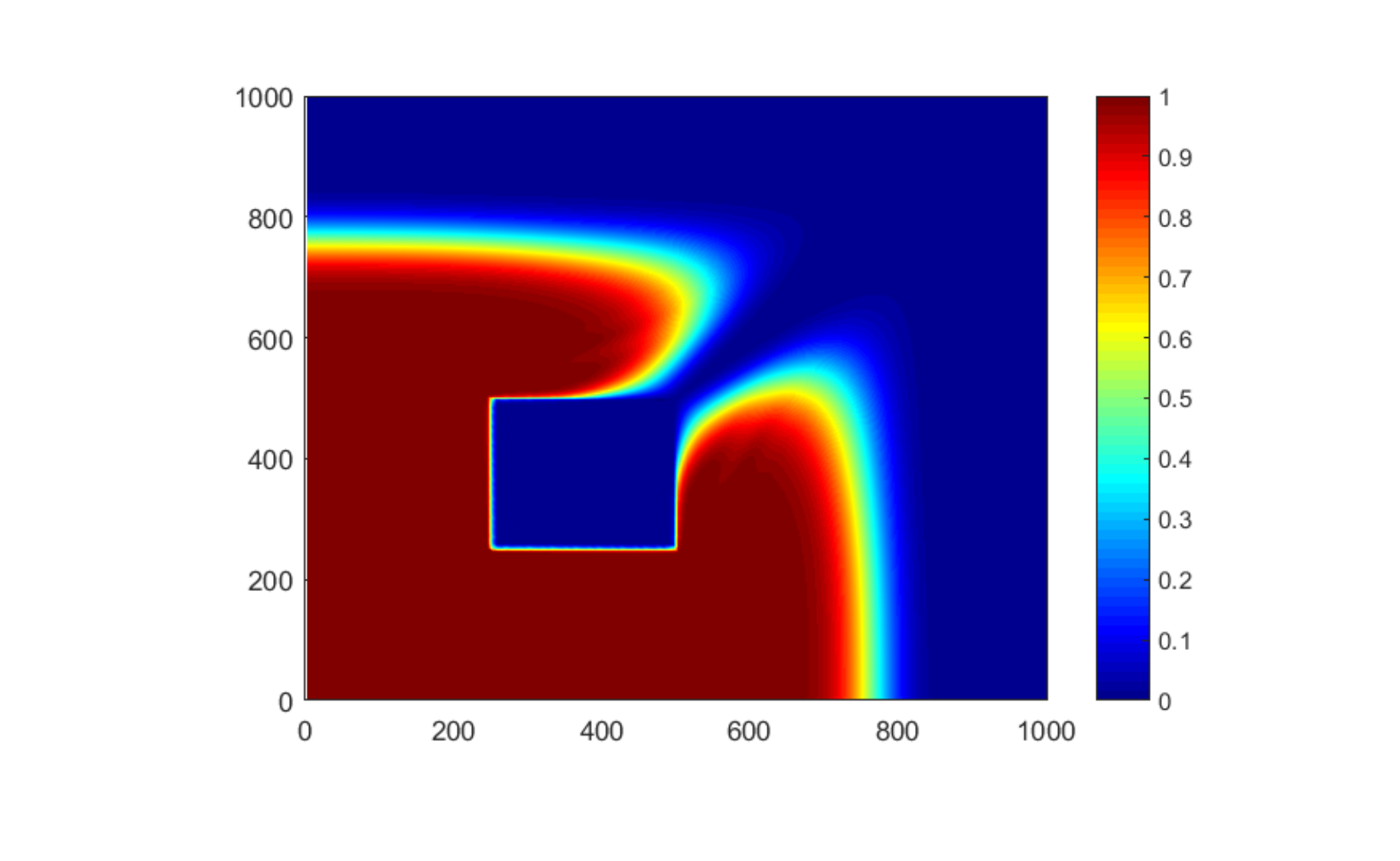}}
\newline
%\hspace*{-5ex}
\subfloat[$t=7.5$]{\includegraphics[trim = 10mm 80mm 20mm 85mm, clip, scale = 0.5]{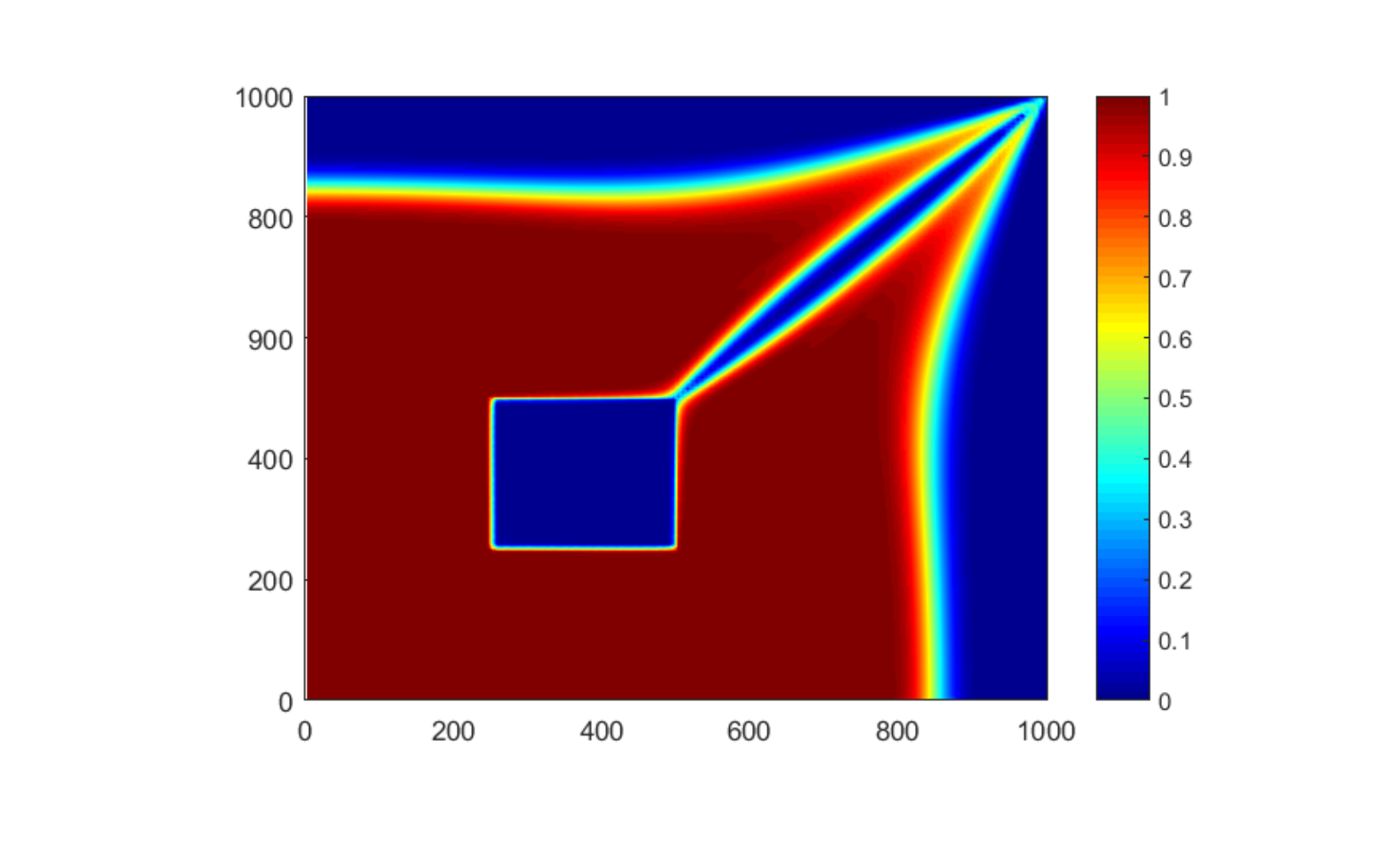}}
\hspace*{-5ex}
\subfloat[$t=10$]{\includegraphics[trim = 10mm 80mm 20mm 85mm, clip, scale = 0.5]{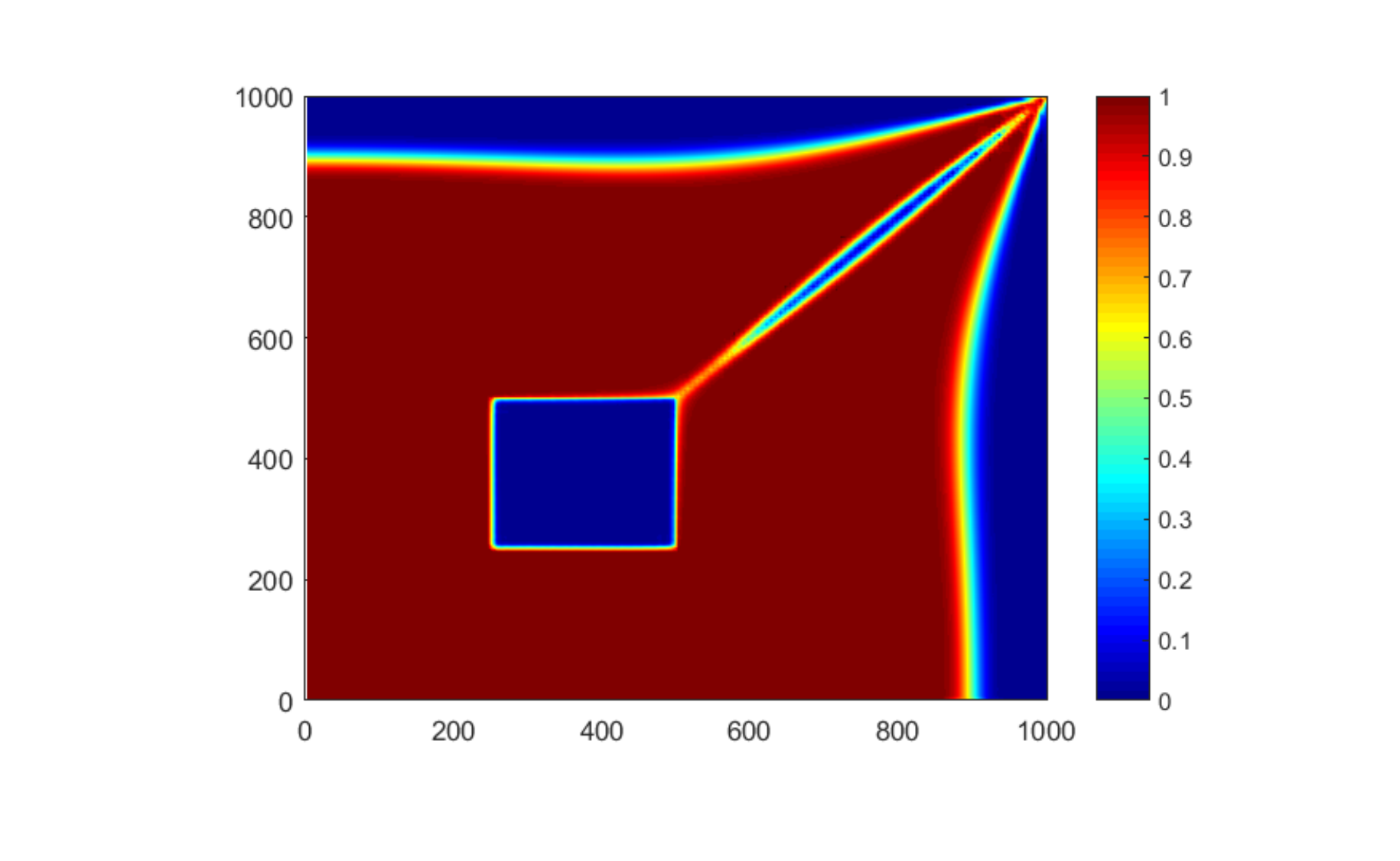}}
\caption{Evolution of concentration for the permeability lens problem.  The HDG method exhibits the correct behavior, since the concentration avoids the region of low permeability.  Mesh with 1024 elements, and polynomial order $k=4$.}
\label{het_lens_1}
\end{figure}

\begin{figure}[ht!]
%\hspace*{-5ex}
\subfloat[$k=1$]{\includegraphics[trim = 10mm 80mm 20mm 85mm, clip, scale = 0.50]{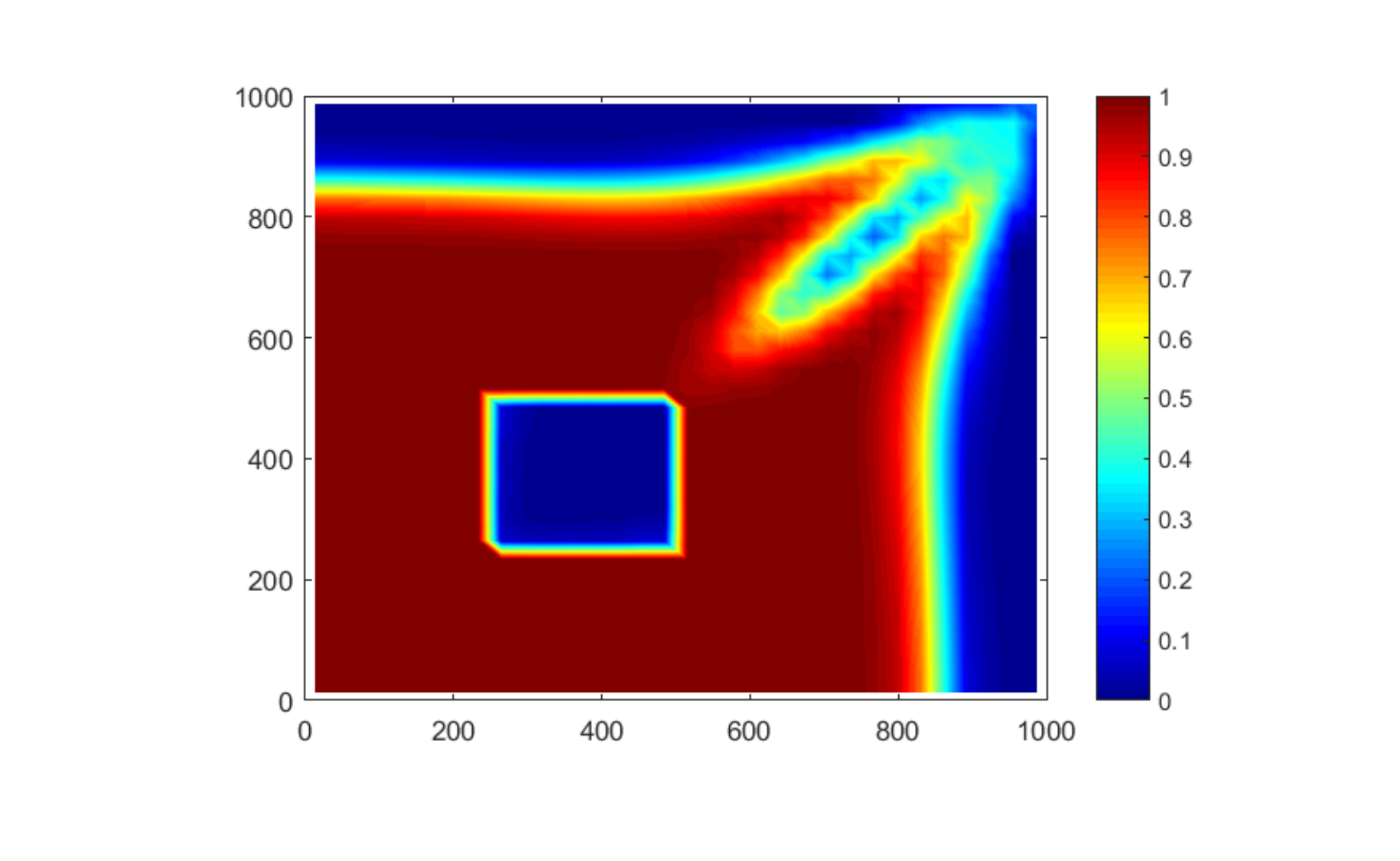}}
\hspace*{-5ex}
\subfloat[$k=2$]{\includegraphics[trim = 10mm 80mm 20mm 85mm, clip, scale = 0.50]{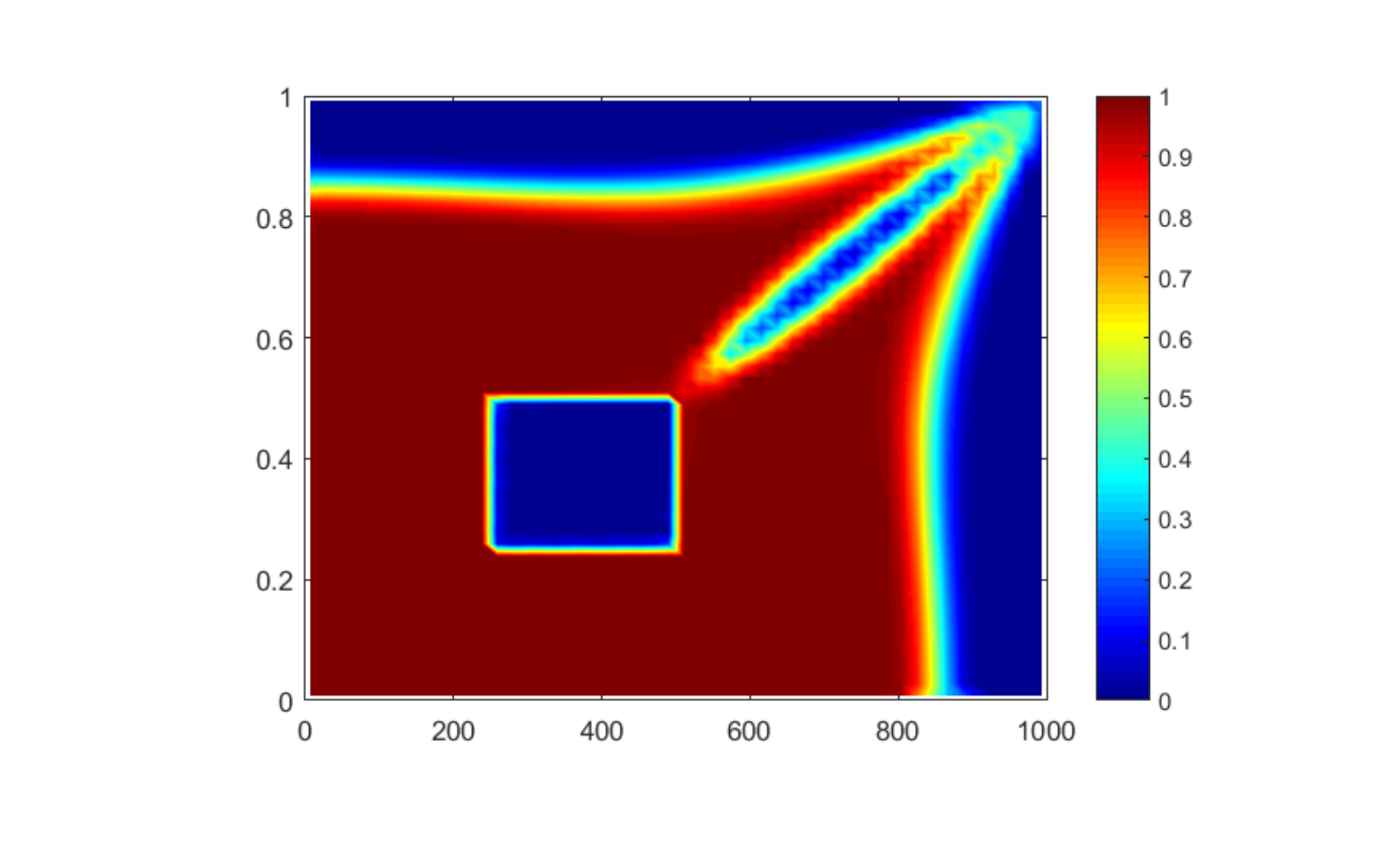}}
\newline
%\hspace*{-5ex}
\subfloat[$k=4$]{\includegraphics[trim = 10mm 80mm 20mm 85mm, clip, scale = 0.50]{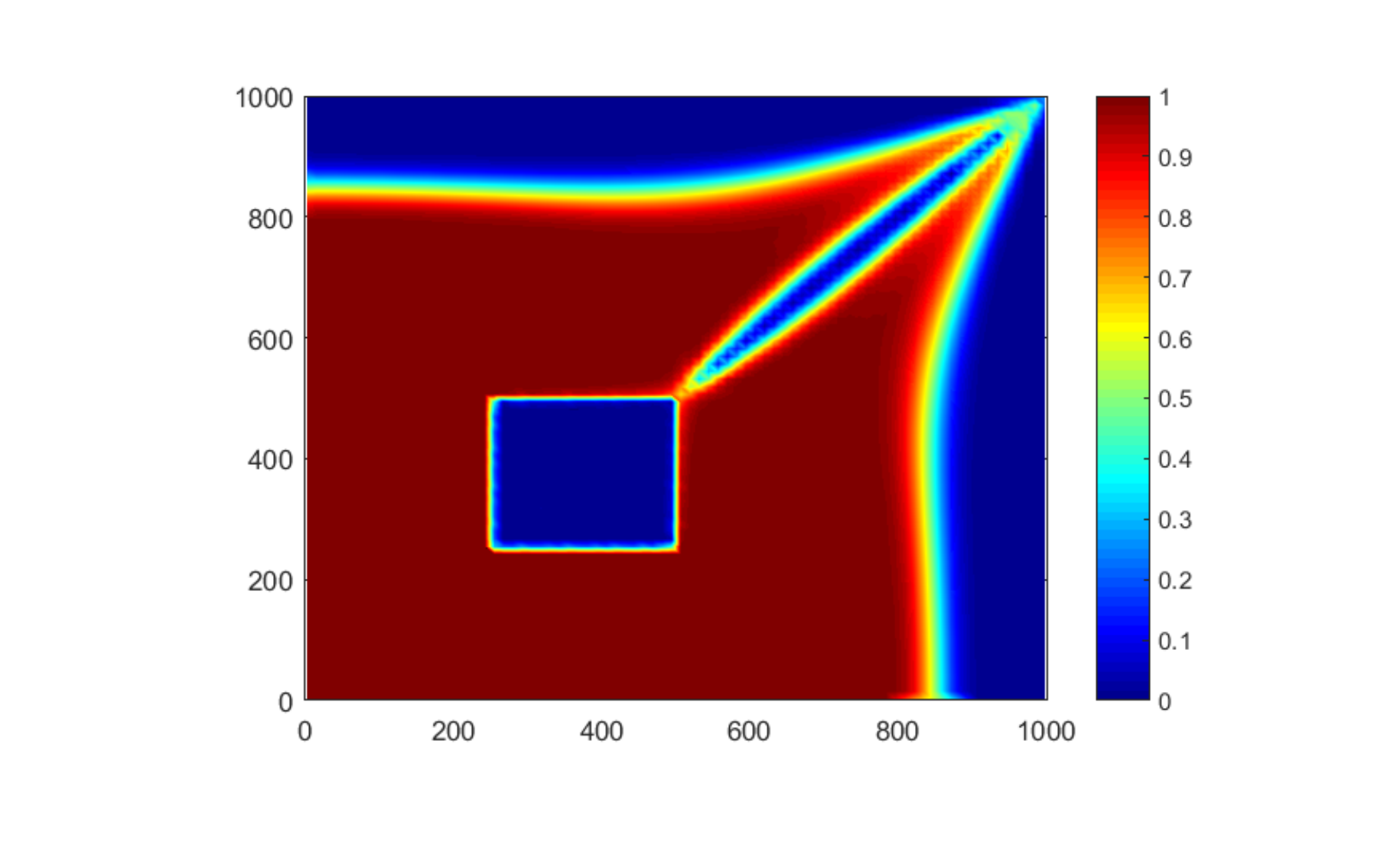}}
\hspace*{-5ex}
\subfloat[$k=8$]{\includegraphics[trim = 10mm 80mm 20mm 85mm, clip, scale = 0.50]{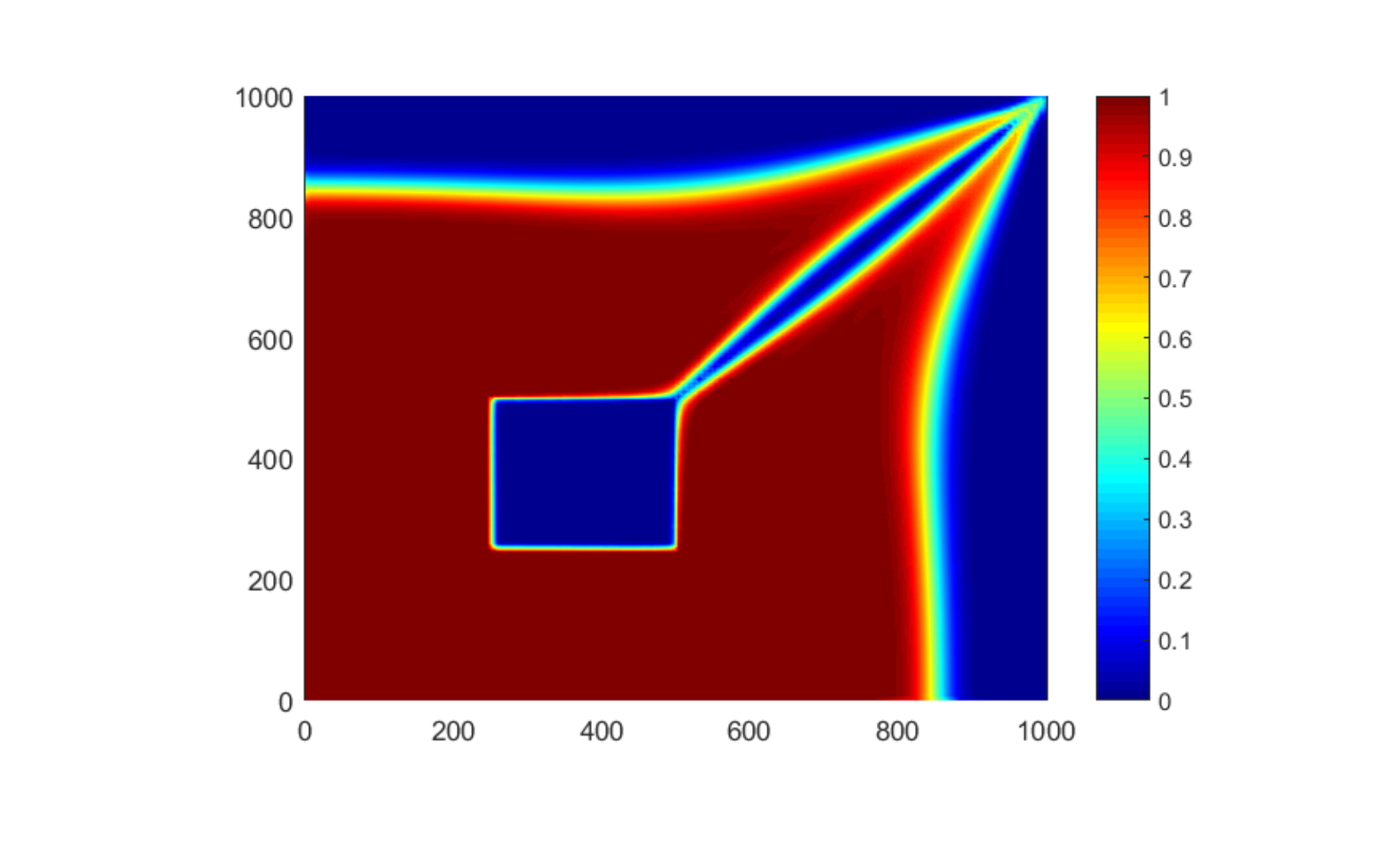}}
\newline
\hspace*{25ex}
\subfloat[$k=16$]{\includegraphics[trim = 10mm 80mm 20mm 85mm, clip, scale = 0.50]{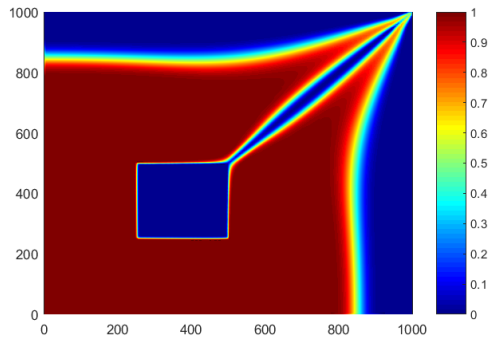}}
\caption{Polynomial order study at $t=7.5$, on a fixed mesh with 256 elements.  As the polynomial order increases, the approximation quality improves significantly.}
\label{het_lens_2}
\end{figure}

In this experiment the domain is $\Omega=[0,1000]^2$, and the permeability is $10^{-10}$ everywhere except the lens $[250,500]\times [250,500]$, where it is 1000 times smaller.  This region of lower permeability acts as an impenetrable area, where the fluid mixture must avoid.  All other parameters are the same as in subsection~\ref{subsec_hom1}.  The concentration at various times is depicted in Fig.~\ref{het_lens_1}, on a mesh with 1024 elements and $k=4$.  As expected, the concentration avoids the region of lower permeability, while still traveling towards the production well.  With high order approximations we are able to resolve the lens boundary with a fine resolution.  %Qualitatively we observe little difference between the HMFE-HDG and HDG-HDG schemes.
\\ \indent
Fig.~\ref{het_lens_2} portrays the effect of increasing the polynomial order on a coarser mesh of 256 elements.  The polynomial degrees vary as: $k=1,2,4,8,16$.  We fix $t=7.5$, and increase the polynomial order in a geometric sequence.  All polynomial orders tested are able to capture the region of low permeability.  Piecewise linears and quadratics give the most diffusive approximations, causing the concentration front to merge faster.  Increasing the polynomial order has the impact of sharpening the borders of the lens.  Further, the approximation is less diffusive, revealing that the two concentration plumes do not fully merge at $t=7.5$.  When using high order approximations, we are able to generate quality simulations on coarser meshes.
%Oscillations are reduced as the polynomial order increases.
 \clearpage
 \clearpage
 
\clearpage
\subsection{Highly heterogeneous media in 2D (SPE Project)}
 \label{sec:spe2d} 
\begin{figure}[ht!]
\centering
    \captionsetup{justification=centering}
    \subfloat[\small{Layer 1 ($64\times 64$ grid, log scale)}]{
        \includegraphics[trim = 40mm 80mm 40mm 90mm, clip, scale = 0.4]{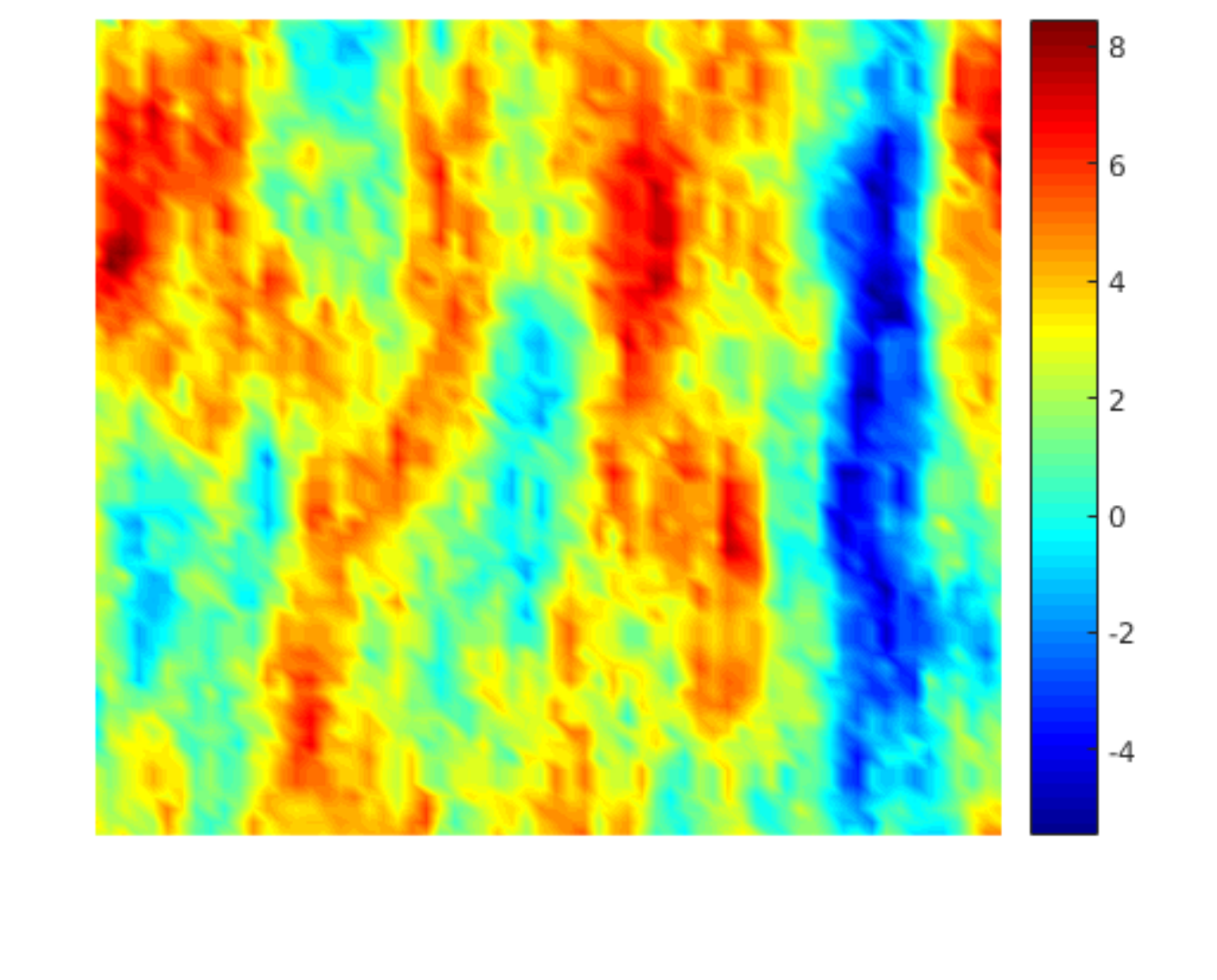}
        \label{Fig:SPEa}
    }
    \subfloat[\small{Layer 44 ($64\times 64$ grid, log scale)}]{
        \includegraphics[trim = 40mm 80mm 40mm 90mm, clip, scale = 0.4]{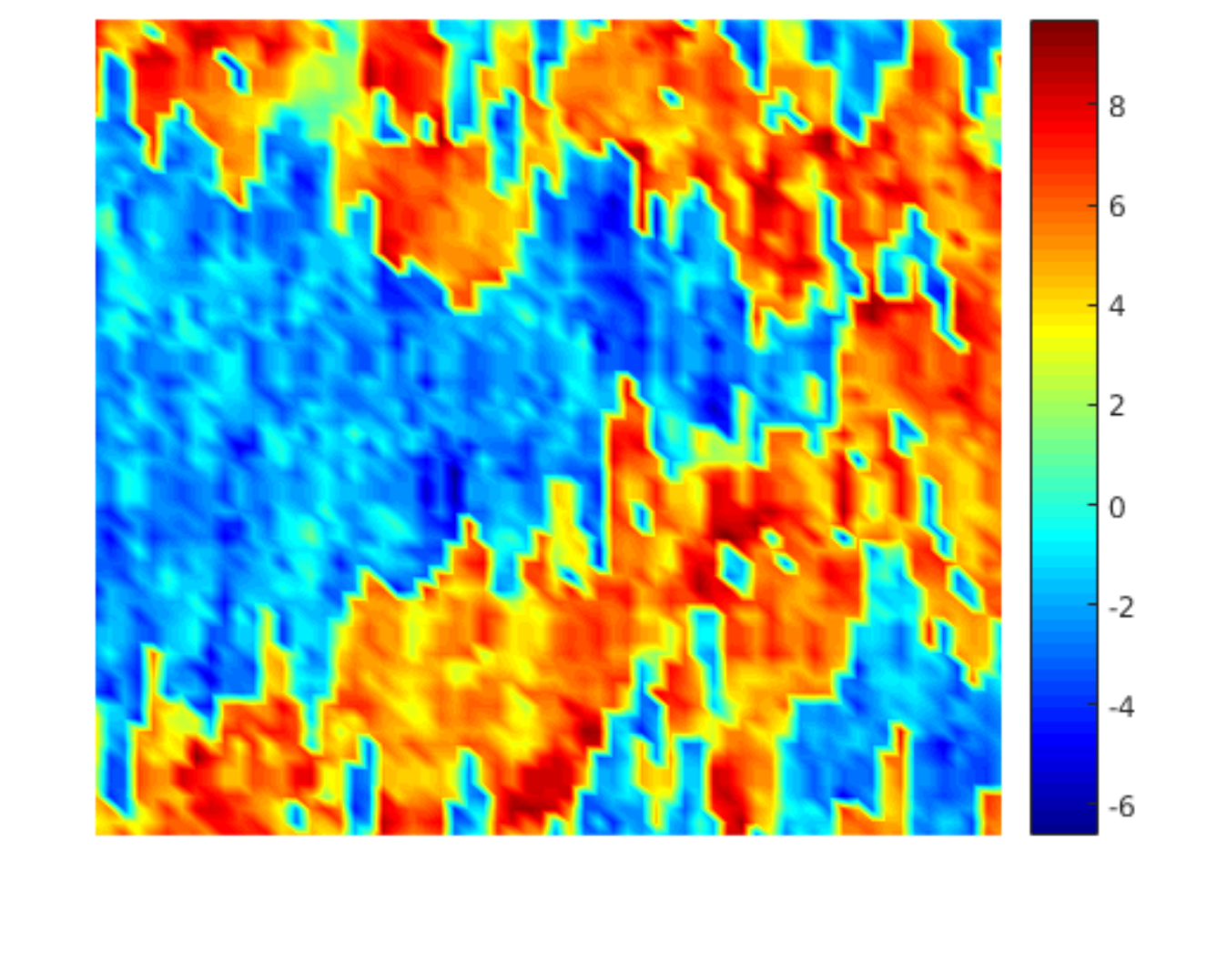}
        \label{Fig:SPEb}        
    }   
    \subfloat[\small{Layer 74 ($64\times 64$ grid, log scale)}]{
        \includegraphics[trim = 40mm 80mm 40mm 90mm, clip, scale = 0.4]{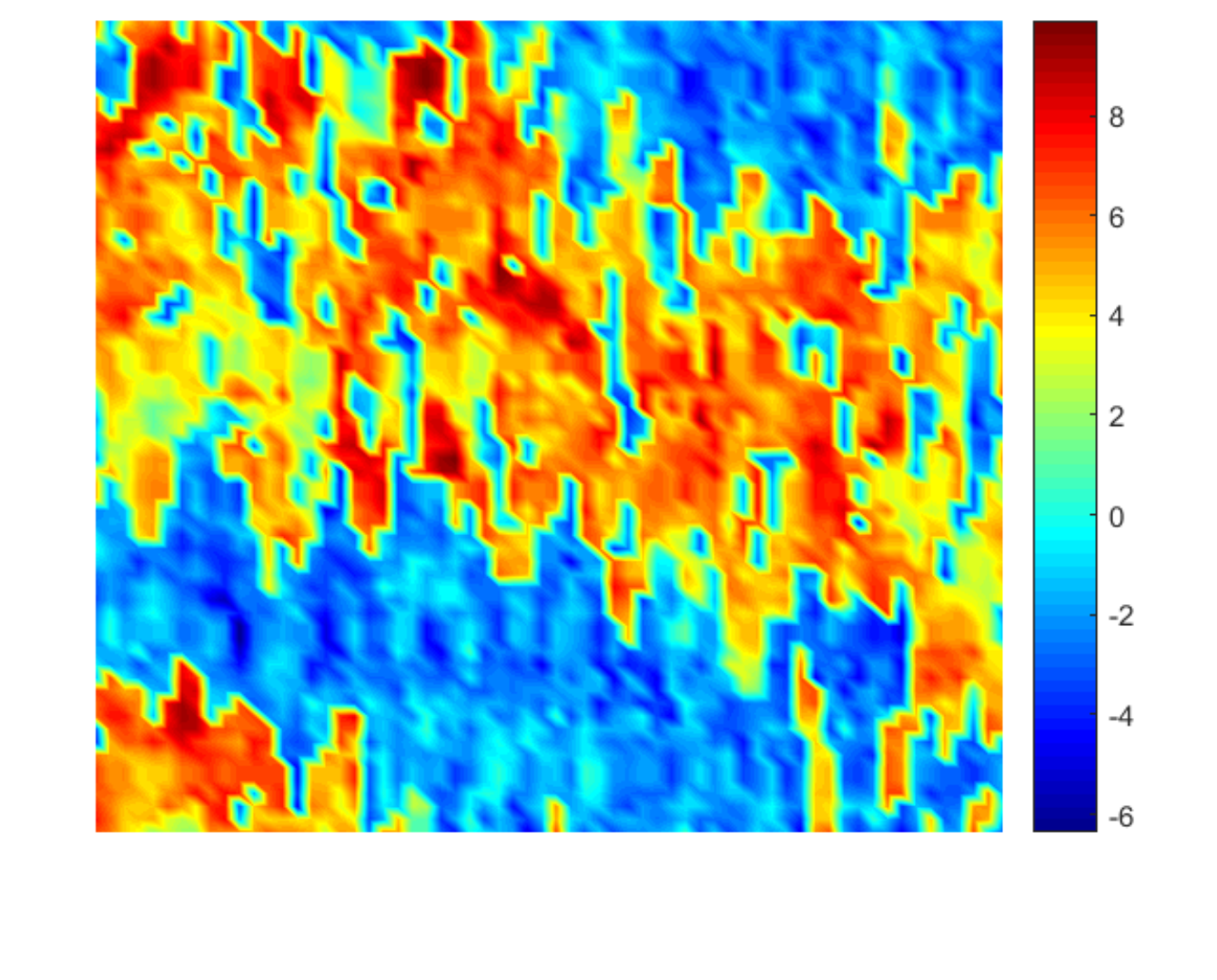}
        \label{Fig:SPEc}        
    }
\caption{Permeability layers.}
\label{Fig:SPE2d_perm}
\end{figure}

 For this numerical experiment, the domain $\Omega=[0,1000]^2$ is heterogeneous, where we take various permeability (horizontal) slices from the SPE10 comparative solution project model 2 \cite{SPE10}.  The porosity is fixed at $20 \%$.  These permeability slices are scaled to a $64 \times 64$ grid, instead of the native $60 \times 220$ grid.  In all experiments we use a mesh with 4096 quadrilateral elements, and discontinuous piecewise quartic basis functions.  All other parameters are the same as in subsection~\ref{subsec_hom1}.% The simulation is driven entirely by the external injection and production wells.  As such, no flow boundary conditions are use for the pressure-velocity system, and no diffusive flux boundary conditions are used for the concentration system.  
\begin{figure}[ht!]
\centering
    \captionsetup{justification=centering}
%%    \subfloat[\small{Layer 1 ($64\times 64$ grid, log scale)}]{
%%        \includegraphics[trim = 40mm 80mm 40mm 90mm, clip, scale = 0.4]{layer_1.pdf}
%%        \label{Fig:SPEa}
%%    }
%%    \subfloat[\small{Layer 44 ($64\times 64$ grid, log scale)}]{
%%        \includegraphics[trim = 40mm 80mm 40mm 90mm, clip, scale = 0.4]{layer_44.pdf}
%%        \label{Fig:SPEb}        
%%    }   
%%    \subfloat[\small{Layer 74 ($64\times 64$ grid, log scale)}]{
%%        \includegraphics[trim = 40mm 80mm 40mm 90mm, clip, scale = 0.4]{layer_74.pdf}
%%        \label{Fig:SPEc}        
%%    }
%%    \\
    \subfloat[\small{Concentration at $t=0.5$ days}]{
        \includegraphics[trim = 40mm 80mm 40mm 90mm, clip, scale = 0.4]{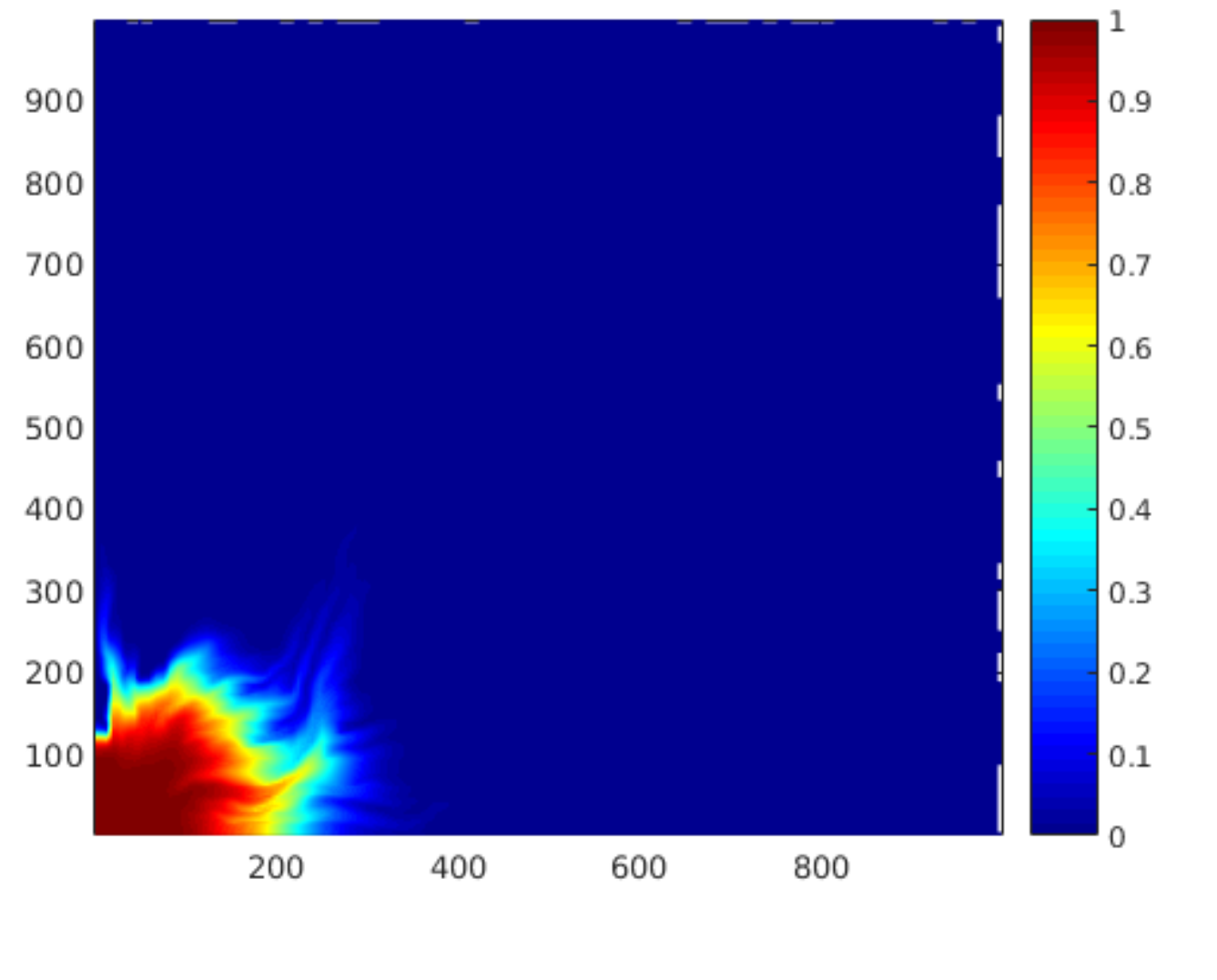}
        \label{Fig:SPEd}         
    }
    \subfloat[\small{Concentration at $t=1.5$ days}]{
        \includegraphics[trim = 40mm 80mm 40mm 90mm, clip, scale = 0.4]{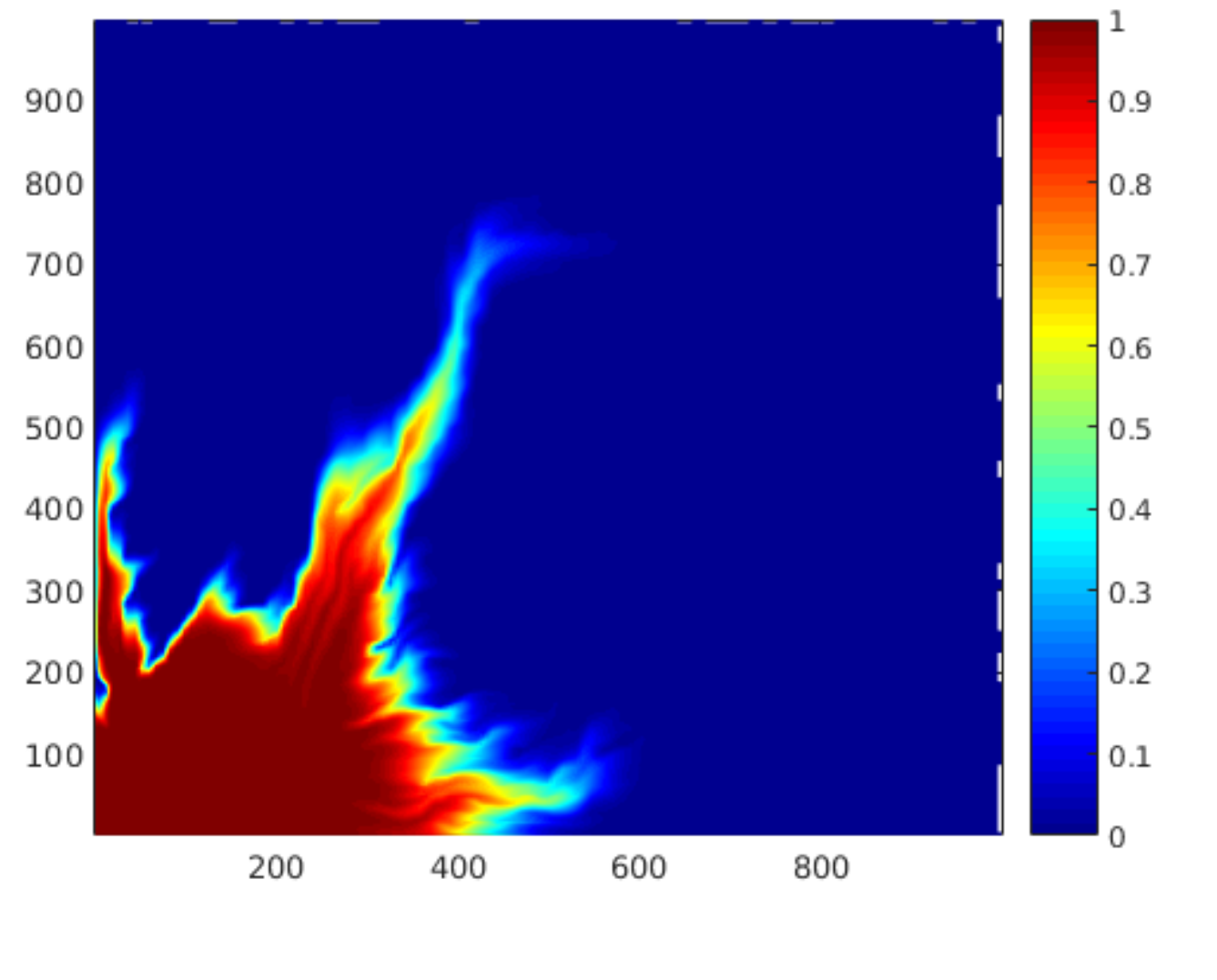}
        \label{Fig:SPEe}         
    }   
    \subfloat[\small{Concentration at $t=2.5$ days}]{
        \includegraphics[trim = 40mm 80mm 40mm 90mm, clip, scale = 0.4]{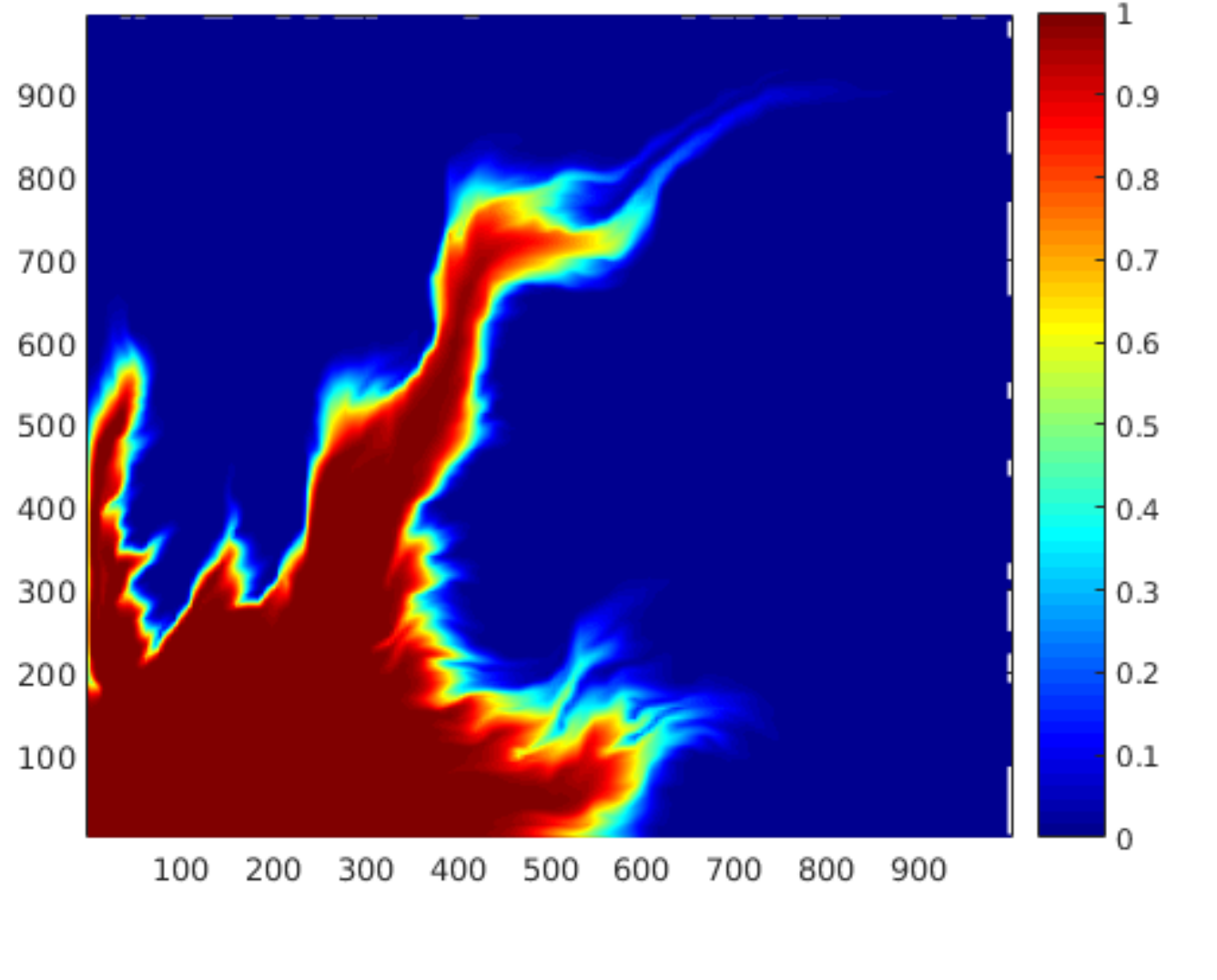}
        \label{Fig:SPEf}         
    }
    \\
    \subfloat[\small{Concentration at $t=0.5$ days}]{
        \includegraphics[trim = 40mm 80mm 40mm 90mm, clip, scale = 0.4]{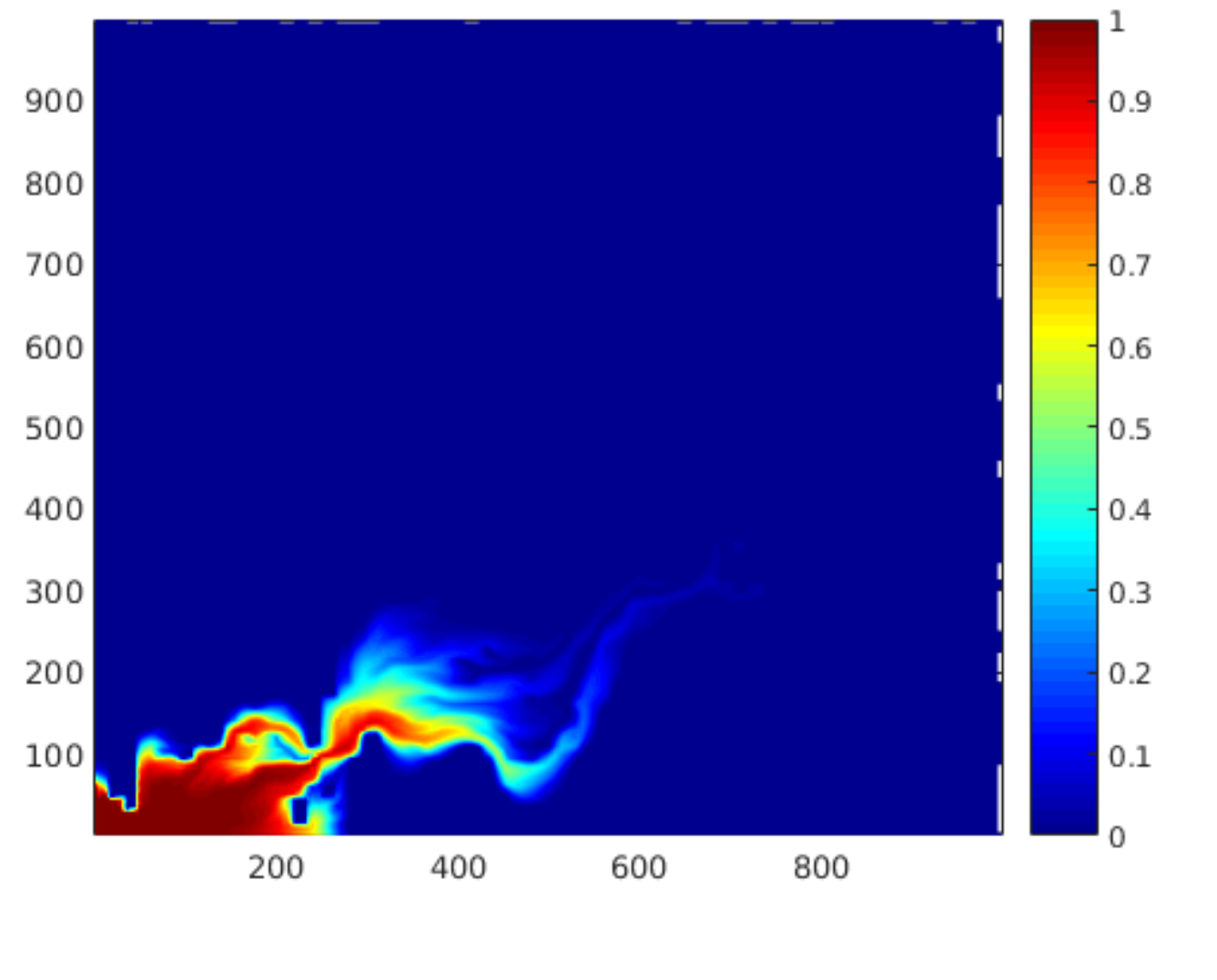}
        \label{Fig:SPEg}         
    }
    \subfloat[\small{Concentration at $t=1.5$ days}]{
        \includegraphics[trim = 40mm 80mm 40mm 90mm, clip, scale = 0.4]{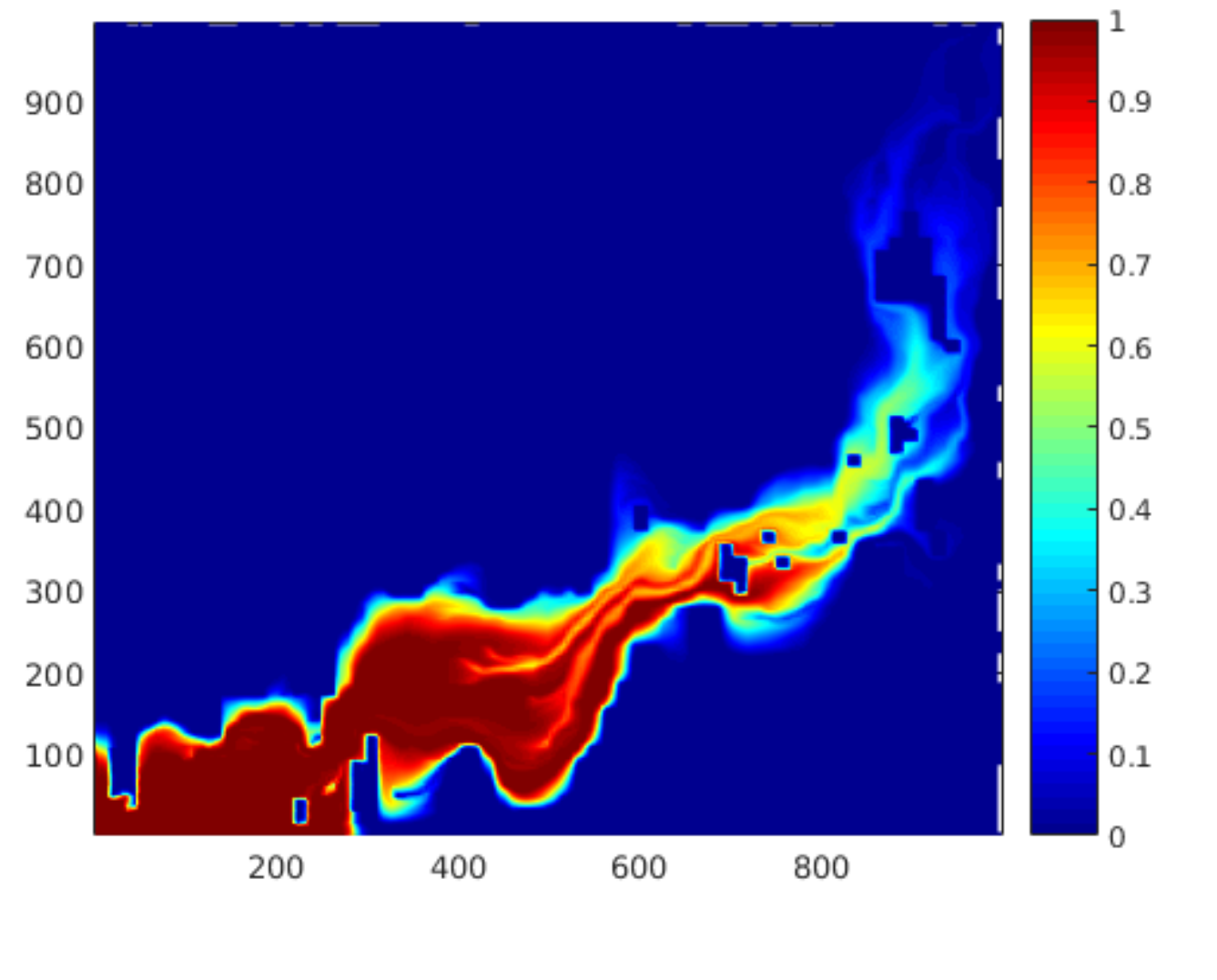}
        \label{Fig:SPEh}         
    }   
    \subfloat[\small{Concentration at $t=2.5$ days}]{
        \includegraphics[trim = 40mm 80mm 40mm 90mm, clip, scale = 0.4]{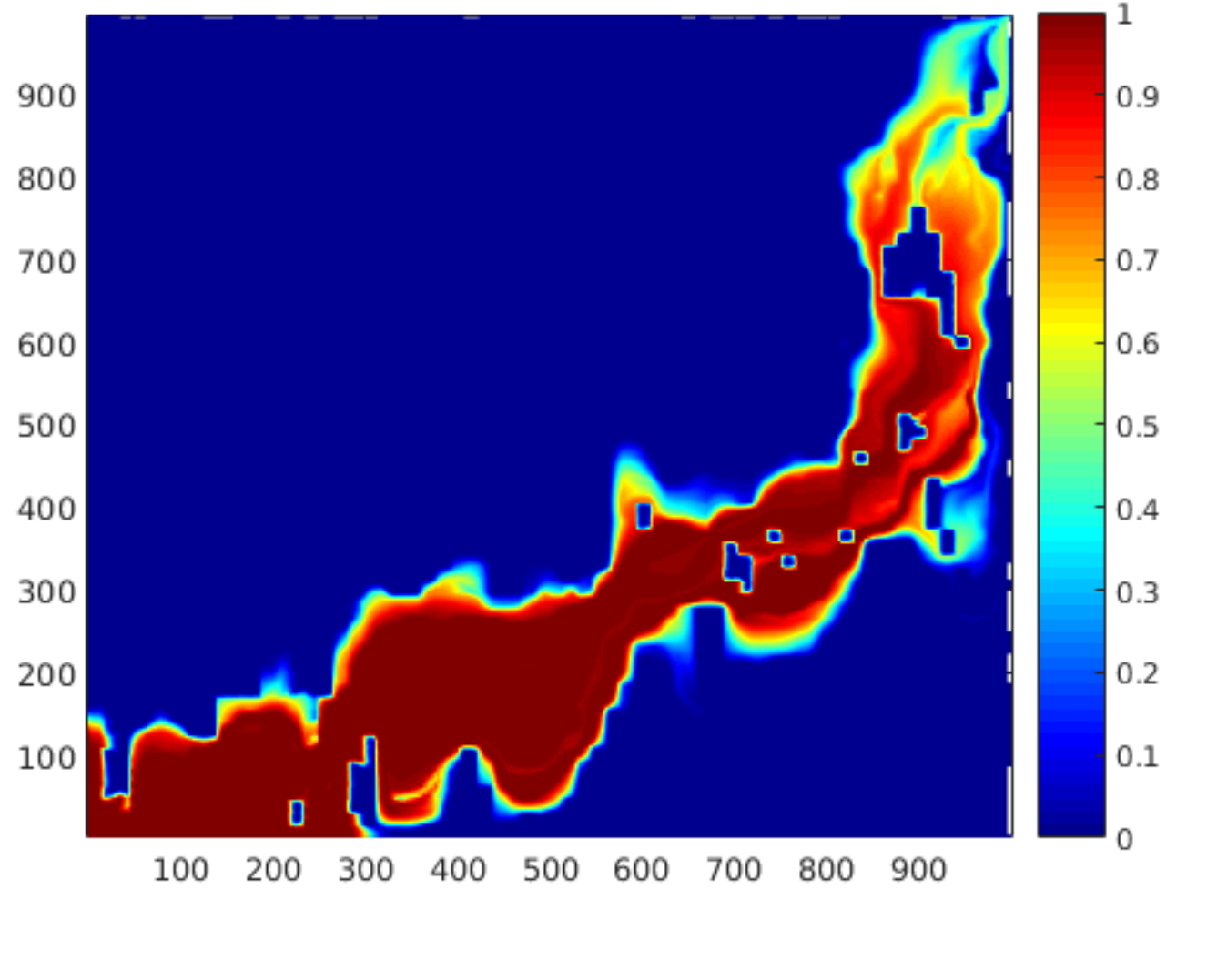}
        \label{Fig:SPEi}         
    }
    \\
    \subfloat[\small{Concentration at $t=0.5$ days}]{
        \includegraphics[trim = 40mm 80mm 40mm 90mm, clip, scale = 0.4]{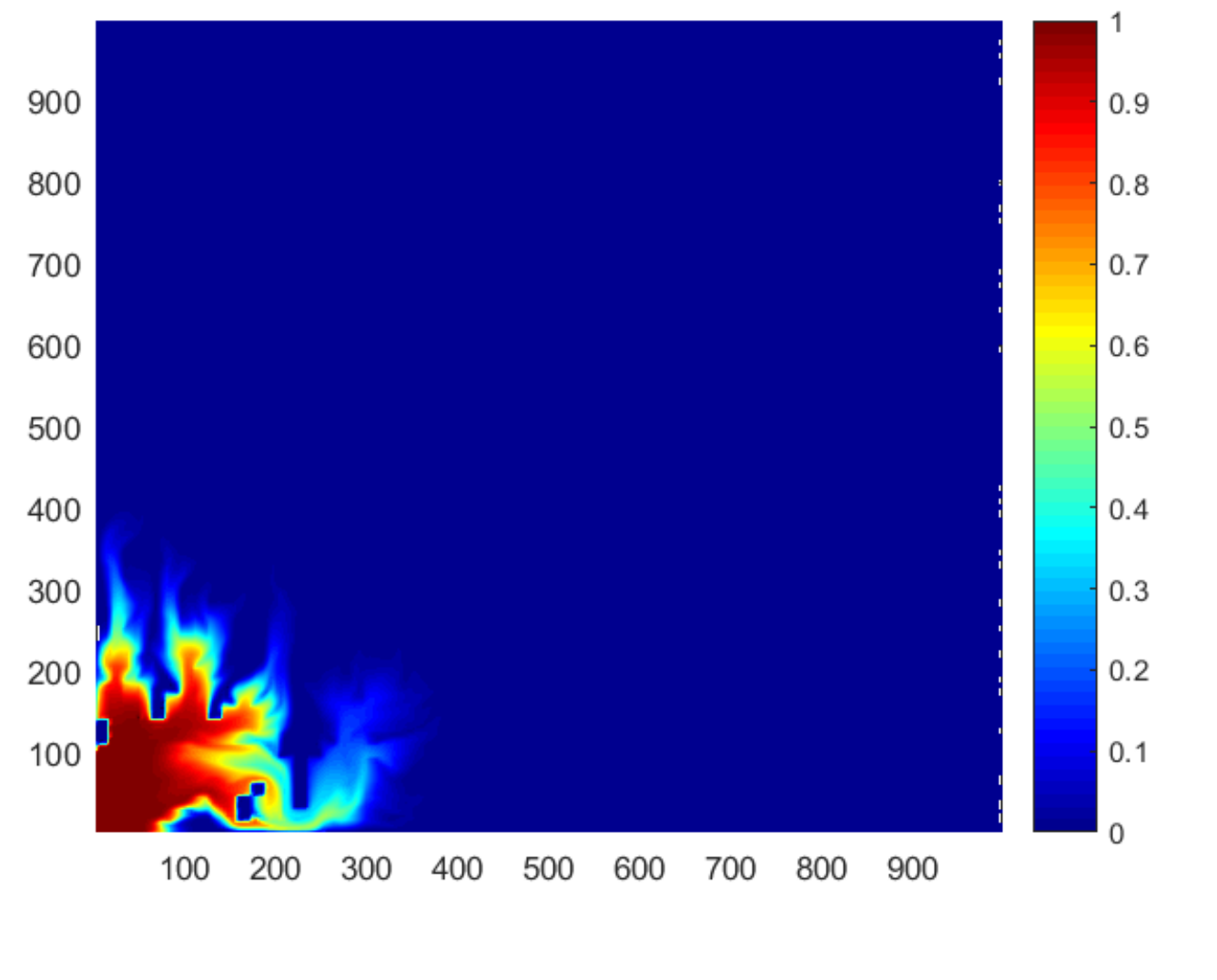}
        \label{Fig:SPEj}         
    }
    \subfloat[\small{Concentration at $t=1.5$ days}]{
        \includegraphics[trim = 40mm 80mm 40mm 90mm, clip, scale = 0.4]{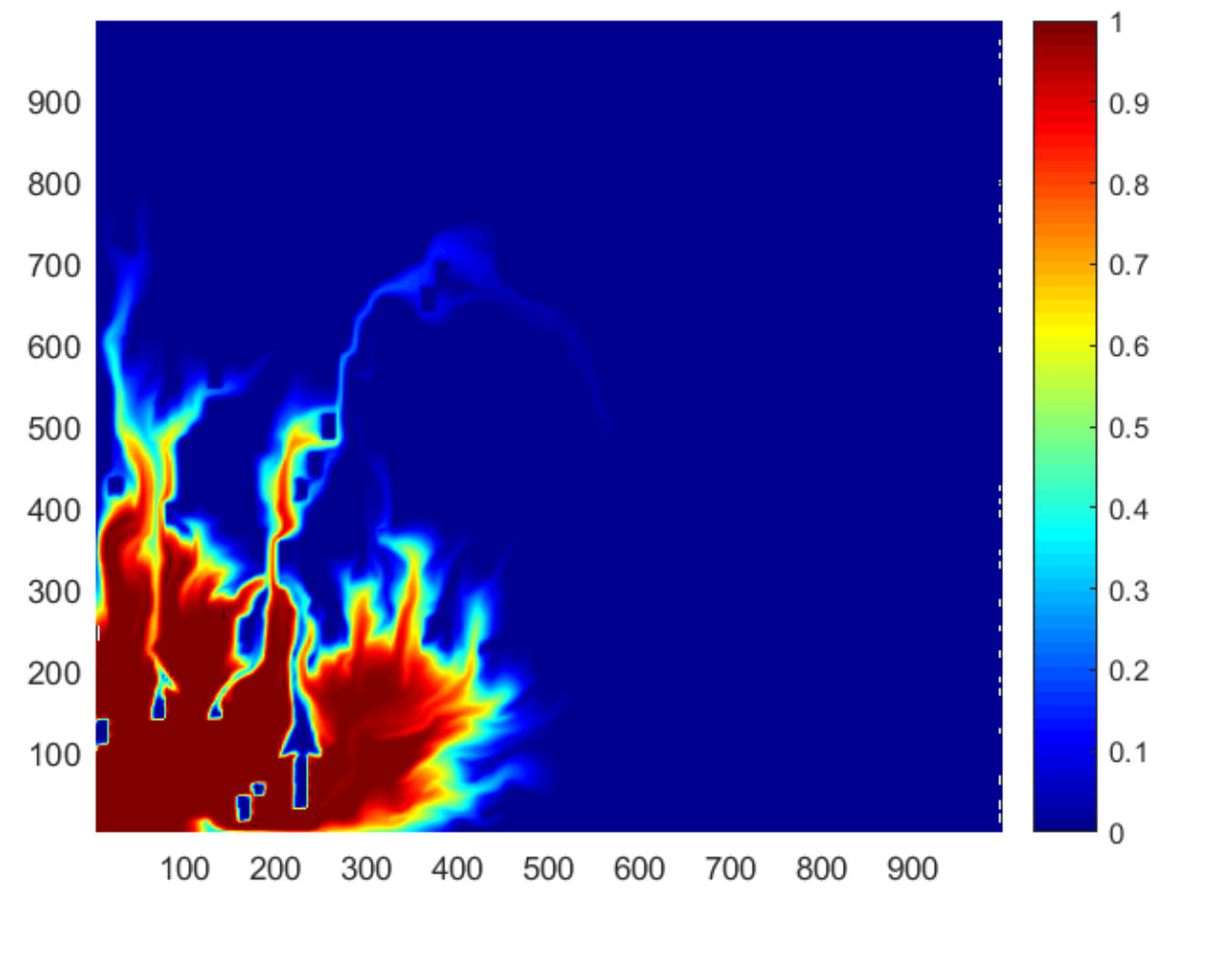}
        \label{Fig:SPEk}         
    }   
    \subfloat[\small{Concentration at $t=2.5$ days}]{
        \includegraphics[trim = 40mm 80mm 40mm 90mm, clip, scale = 0.4]{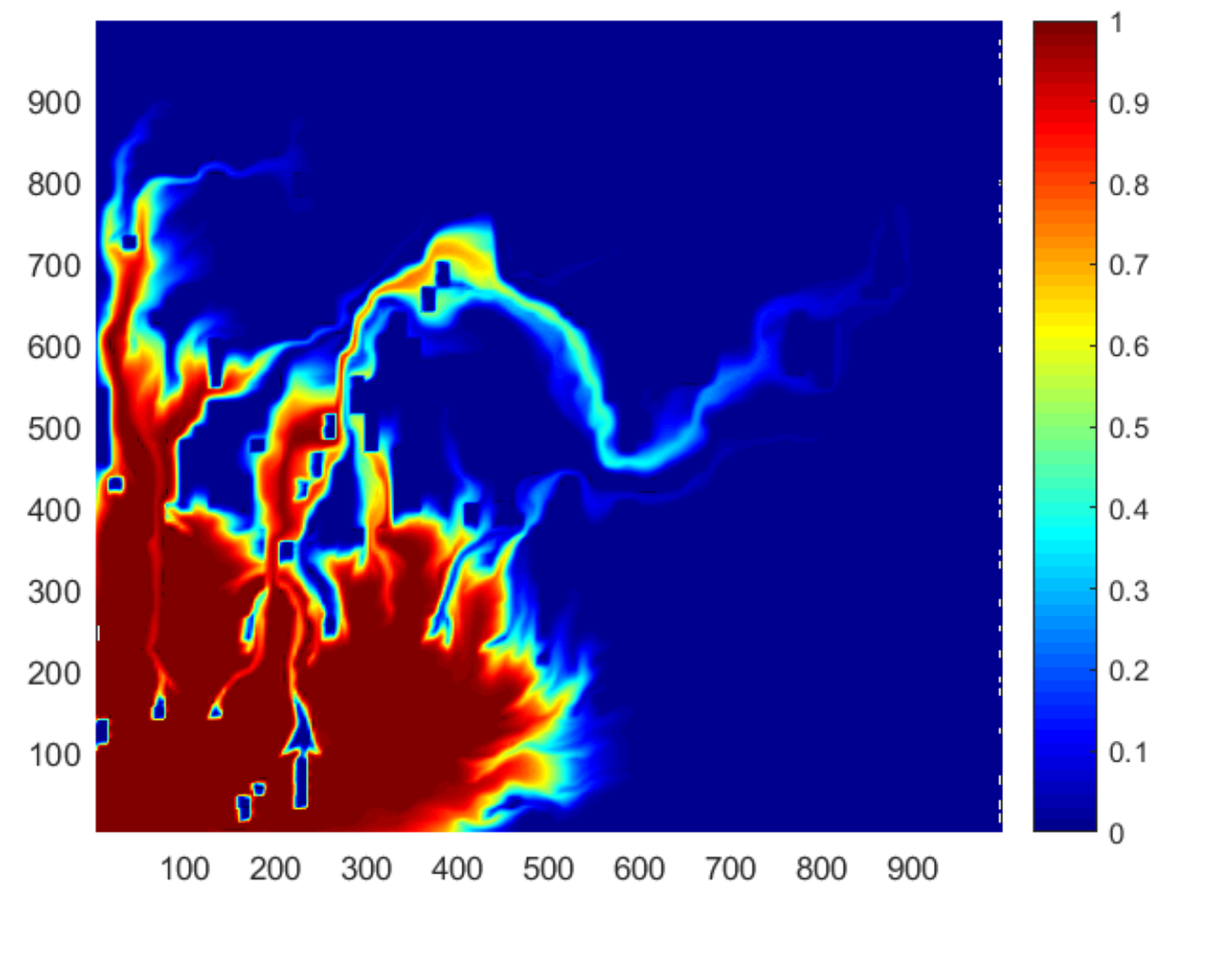}
        \label{Fig:SPEl}         
    }    
\captionsetup{justification=justified}
\caption{Miscible displacement, quarter-five spot problem.  Mesh with 4096 quadrilateral elements, discontinuous piecewise quartic basis functions.}
\label{Fig:SPE}
\end{figure}
Figs.~\ref{Fig:SPEa}, \ref{Fig:SPEb}, and \ref{Fig:SPEc} show the selected permeability layers, which vary over the Talbert and Upper Ness formations.  The concentration profiles are displayed for each of the three permeability layers.  Snapshots of the concentration are given at different times in the simulation.  Layer 1 corresponds to Figs.~\ref{Fig:SPEd}, \ref{Fig:SPEe}, and \ref{Fig:SPEf}.  Layer 44 corresponds to Figs.~\ref{Fig:SPEg}, \ref{Fig:SPEh}, and \ref{Fig:SPEi}.  Layer 74 corresponds to Figs.~\ref{Fig:SPEj}, \ref{Fig:SPEk}, and \ref{Fig:SPEl}.  This example demonstrates that the HDG method is robust for highly heterogeneous porous media.%In all cases we observe that the fluid mixture successfully navigates the highly discontinuous permeability field.  
%\newpage
%\subsection{Homogeneous permeability in 3D}
% \label{sec:3d_hom}
\subsection{Permeability lens in 3D}
 \label{sec:3d_lens}
In this section we validate our HDG method in 3D.  In particular, we examine an analogy of the numerical experiment conducted in subsection~\ref{sec:lens}.  The domain is now $\Omega=[0,1000]^3$.  The permeability lens described in subsection~\ref{sec:lens} is extruded in the $z$-direction, so that the region of lower permeability is a pillar.  Production and injection wells are placed at opposite ends of the domain (e.g. near the origin and the coordinate $(1000,1000,1000)$).  The remaining parameters are the same as in subsection~\ref{sec:lens}.

\begin{figure}[H]
\centering
%\hspace*{-5ex}
    %\captionsetup{justification=centering}
    \subfloat[\small{Top-down view}]{
    	\includegraphics[trim = 10mm 20mm 40mm 40mm, clip,scale = 0.15]{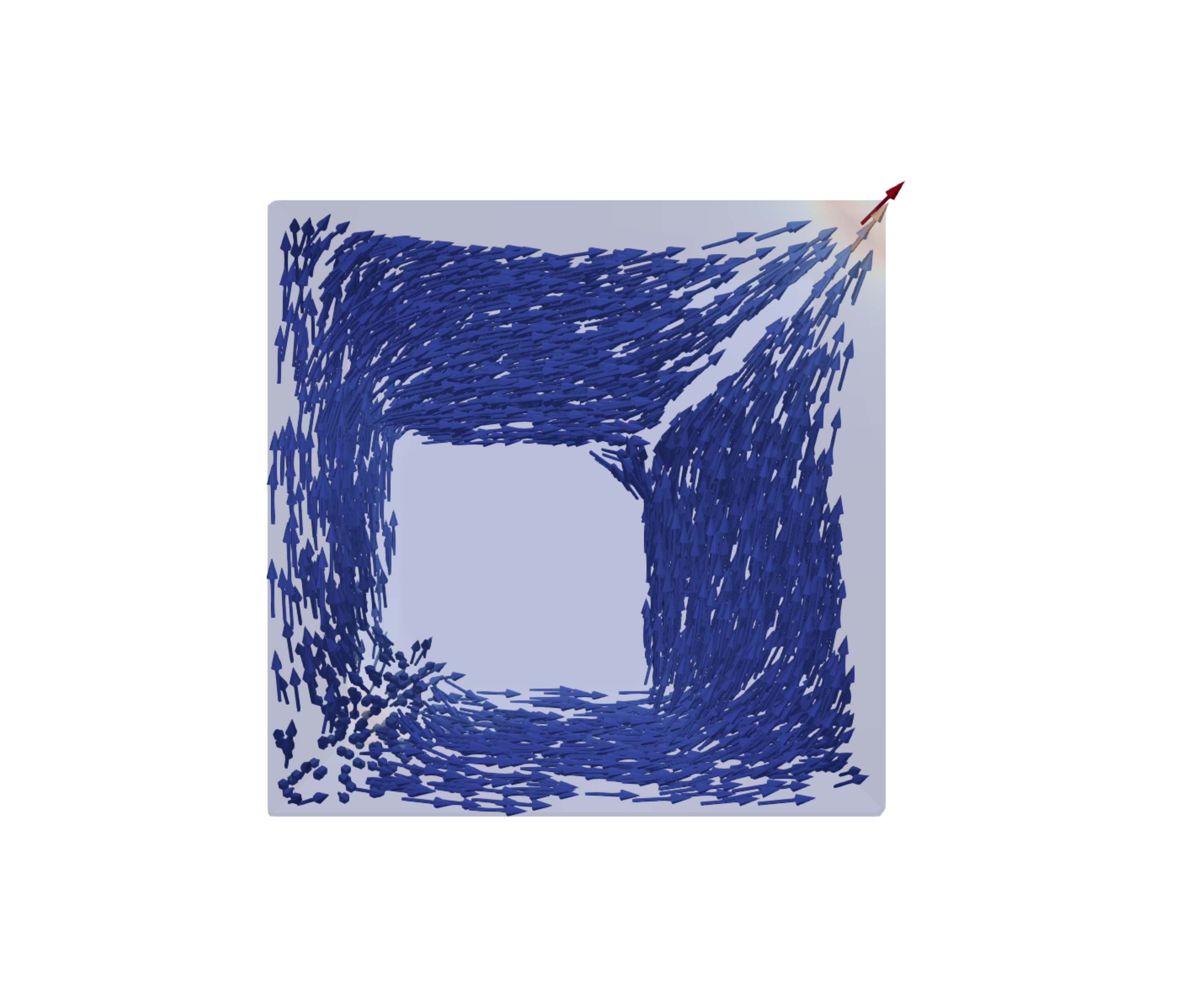}
        \label{Fig:lens_3d1}
    }
    \subfloat[\small{Side view}]{
	    \includegraphics[trim = 10mm 20mm 40mm 40mm, clip,scale = 0.15]{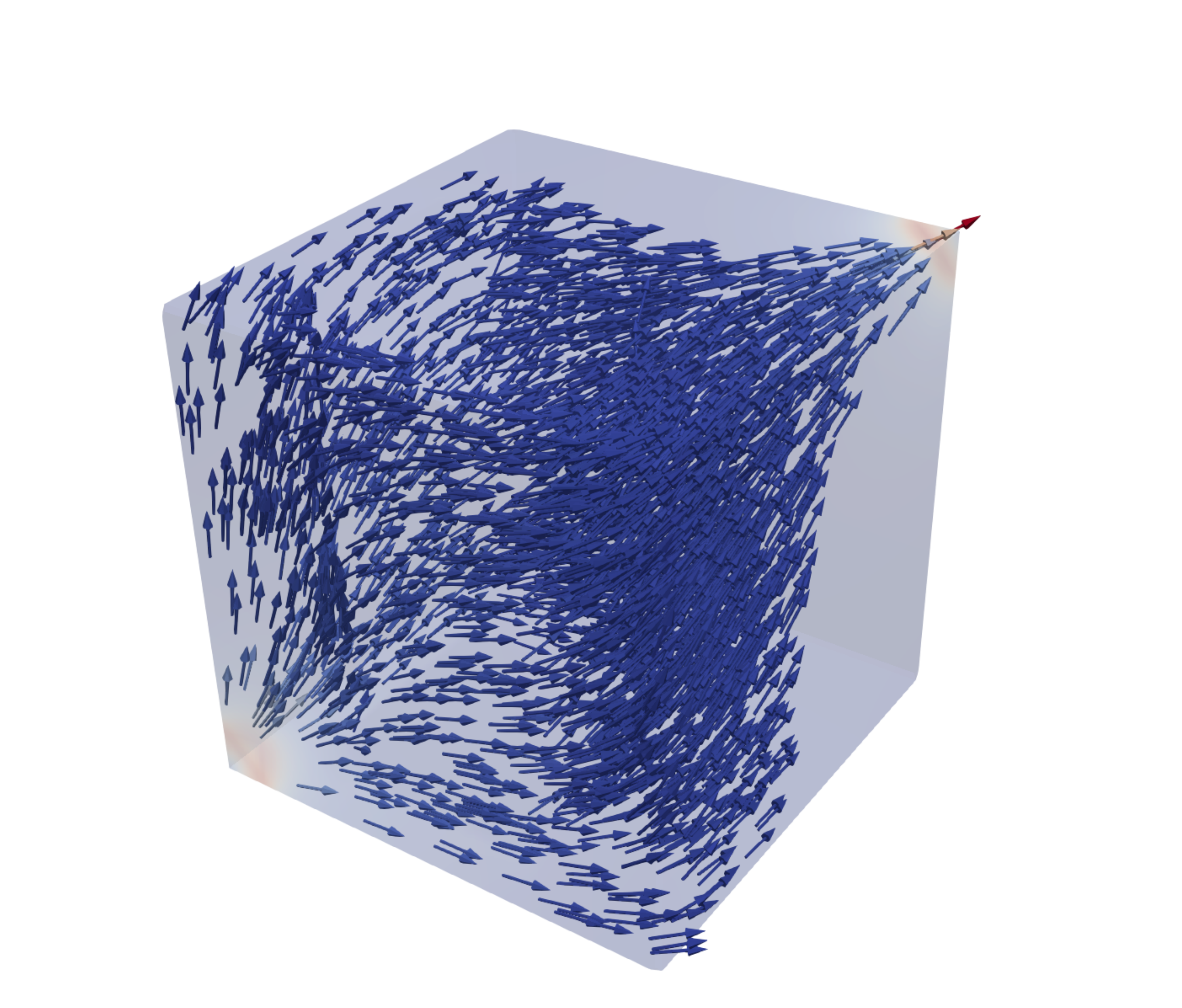}
        \label{Fig:lens_3d2}        
    }
\caption{Miscible displacement in 3D.  Darcy velocity field at $t=7.5$ days (arrows not to scale).  Mesh with 3072 tetrahedral elements, and piecewise quartic basis functions.}
\label{Fig:lens_3d0}
\end{figure}

\begin{figure}[htb!]
	\centering
    \subfloat[\small{$t=2.5$ days}]{
    	\includegraphics[trim = 10mm 10mm 40mm 5mm, clip,scale = 0.45]{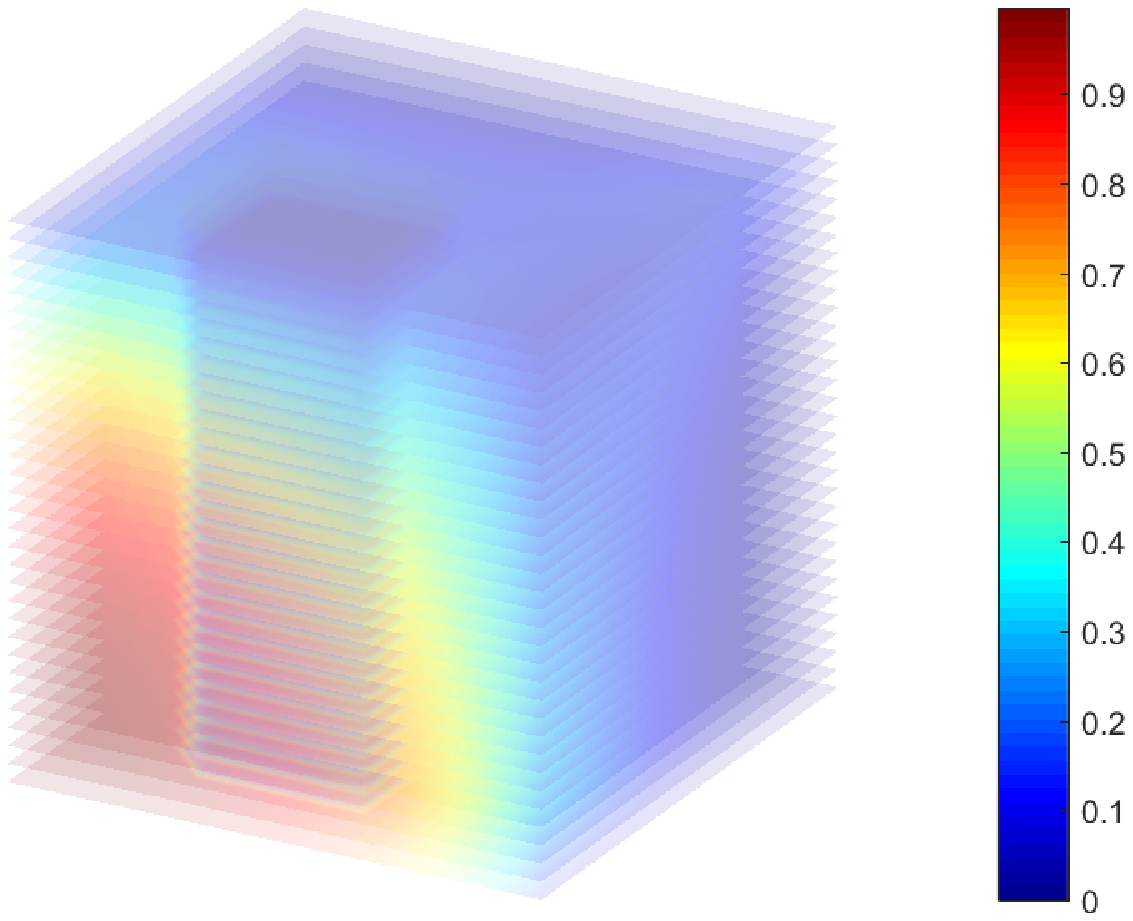}
        %\label{Fig:lens_3d4}
    }  \hspace*{2ex}
    \subfloat[\small{$t=5$ days}]{
	    \includegraphics[trim = 10mm 10mm 40mm 5mm, clip,scale = 0.45]{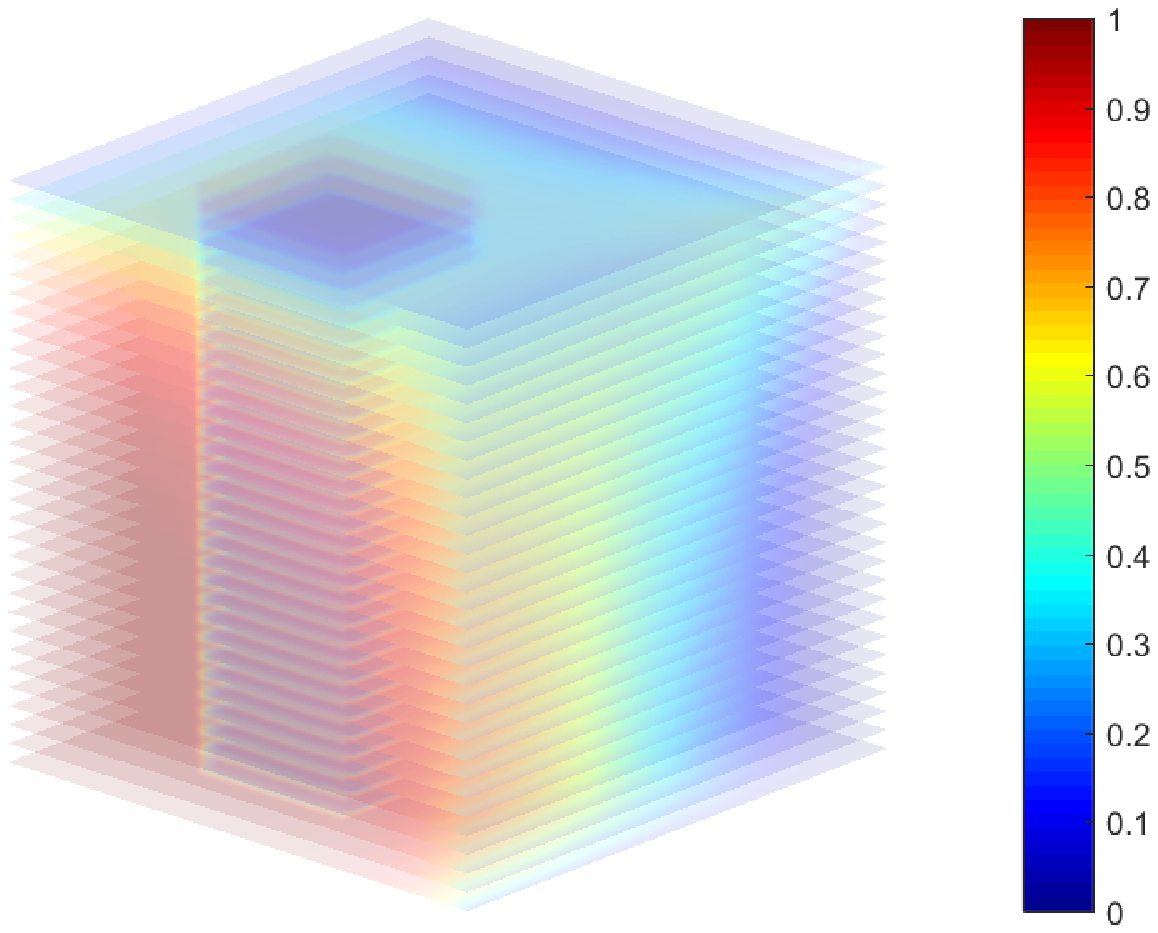}
        %\label{Fig:lens_3d5}        
    }  \hspace*{2ex}
    \subfloat[\small{$t=7.5$ days}]{
	    \includegraphics[trim = 10mm 10mm 40mm 5mm, clip,scale = 0.45]{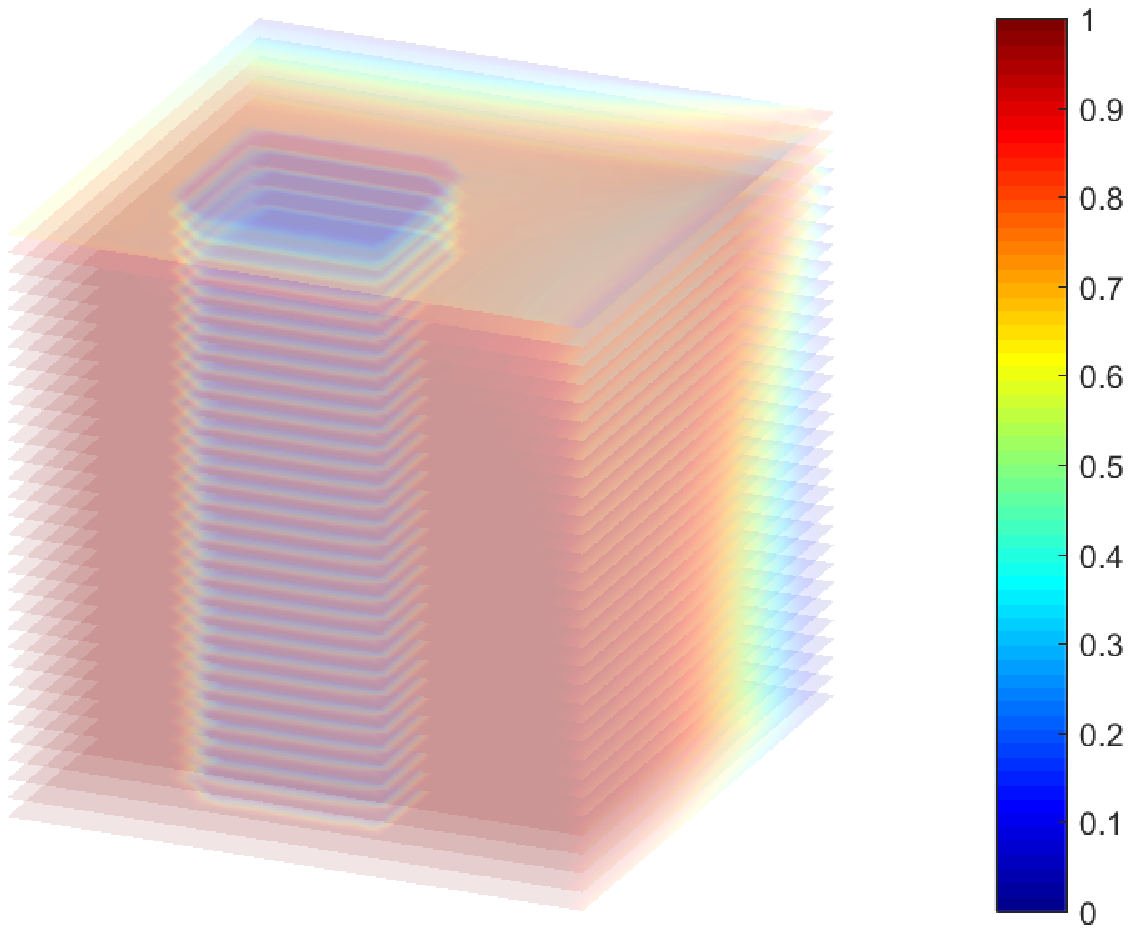}
        %\label{Fig:lens_3d6}        
    }    
\caption{Miscible displacement in 3D.  Concentration snapshots at various times.  The concentration is visualized as volume slices.  Mesh with 3072 tetrahedral elements, and piecewise cubic quartic functions.}
\label{Fig:lens_3d3}
\end{figure}

  We use a structured mesh with 3072 tetrahedral elements, and piecewise quartic basis functions.  Fig.~\ref{Fig:lens_3d0} shows the Darcy velocity field for this simulation.  The flow avoids the low permeability pillar, and travels from the source to the sink, which is clearly visible in Fig.~\ref{Fig:lens_3d1}.  Fig.~\ref{Fig:lens_3d2} plots the flow field from a different angle.
Fig.~\ref{Fig:lens_3d3} plots snapshots of the concentration at different times.  The volumetric slices show that the fluid mixture is navigating around the region of lower permeability.  These plots agree with the Darcy velocity field displayed in Fig.~\ref{Fig:lens_3d0}.  This is the expected outcome for a quarter-five spot simulation with a permeability contrast in 3D.
           
\subsection{Highly heterogeneous media in 3D (SPE Project)}
 \label{sec:spe3d}
\begin{figure}[H]
\centering 
%\hspace*{-5ex}
    %\captionsetup{justification=centering}
    \subfloat[\small{$t=1$ days}]{
    	\includegraphics[trim = 10mm 30mm 40mm 0mm, clip,scale = 0.45]{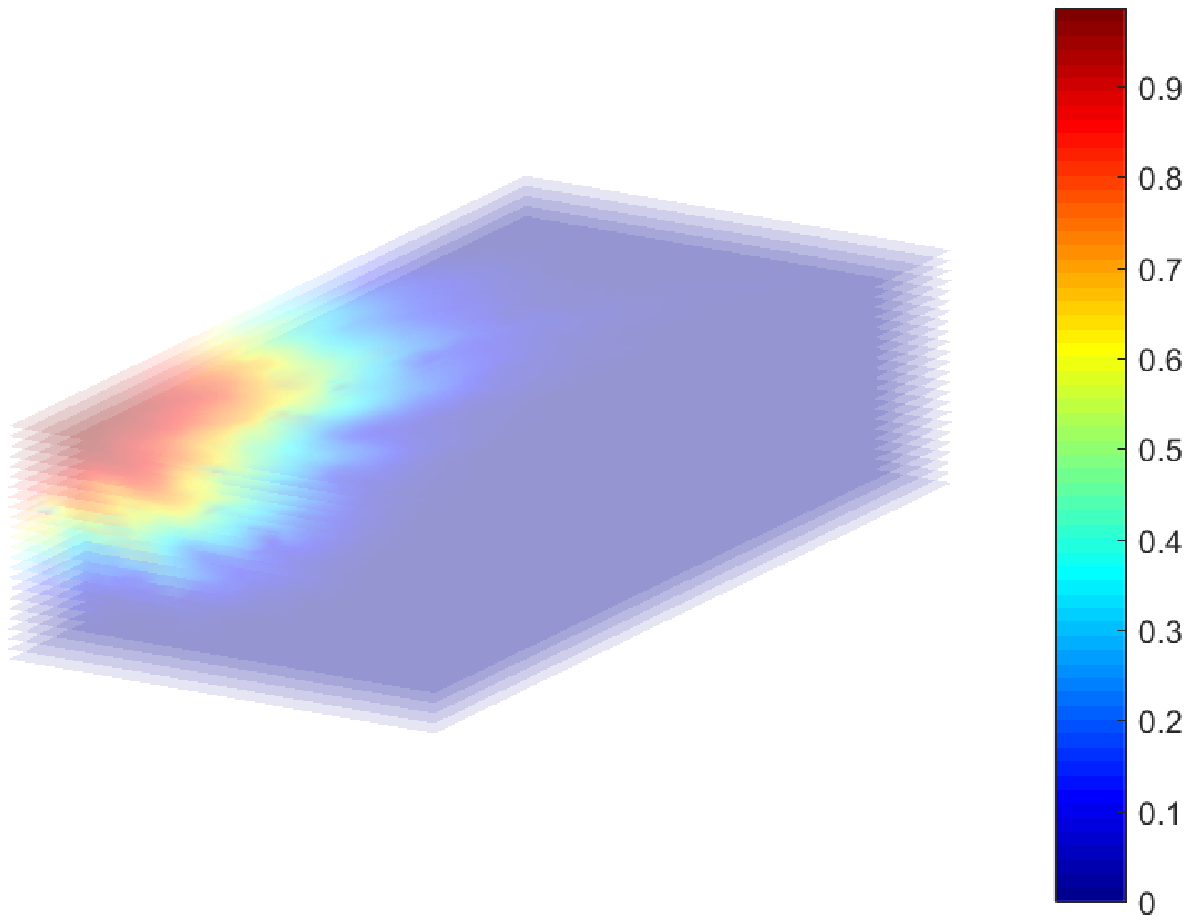}
        \label{Fig:SPE3da}
    }  \hspace*{2ex}
    \subfloat[\small{$t=5$ days}]{
	    \includegraphics[trim = 10mm 30mm 40mm 0mm, clip,scale = 0.45]{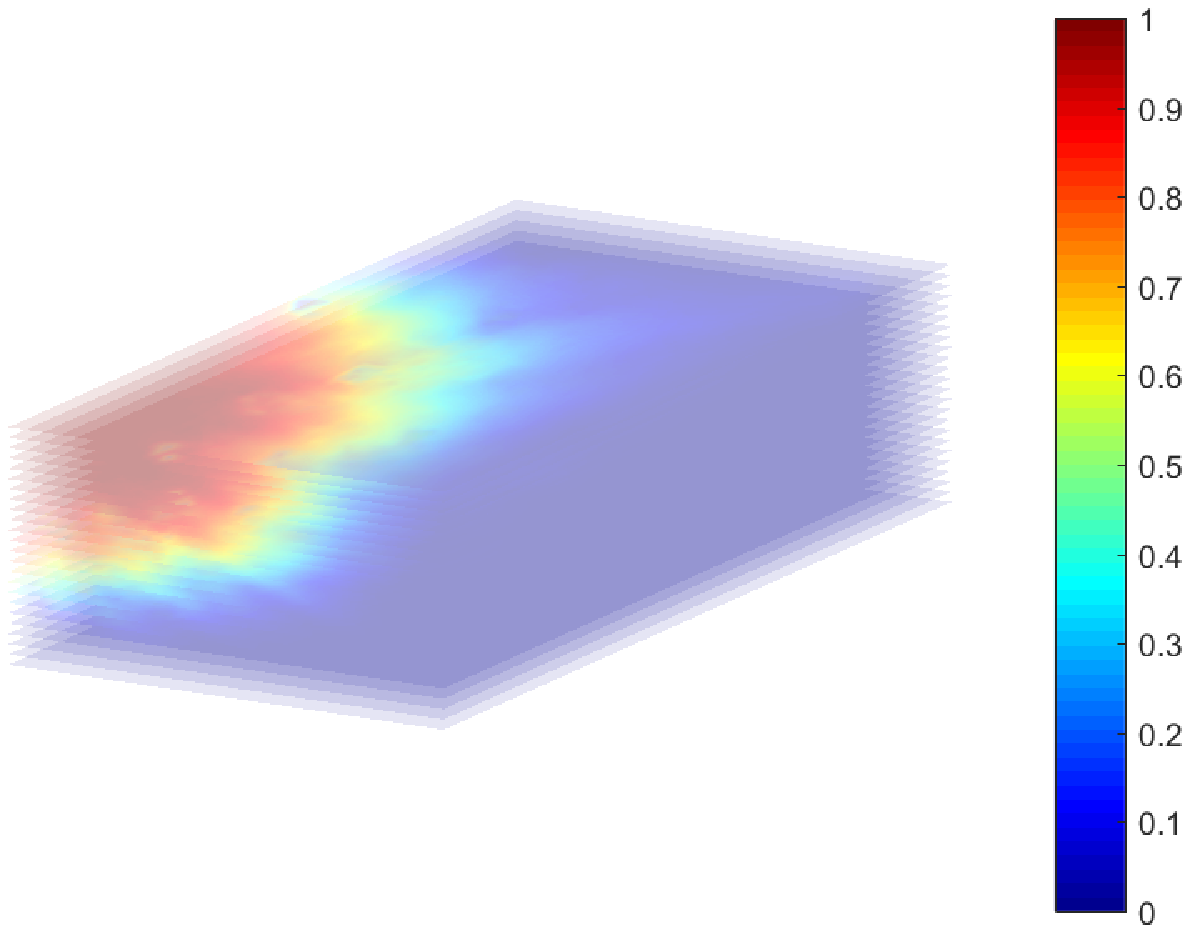}
        \label{Fig:SPE3db}        
    }  \hspace*{2ex}
    \subfloat[\small{$t=8$ days}]{
		\includegraphics[trim = 10mm 30mm 40mm 0mm, clip,scale = 0.45]{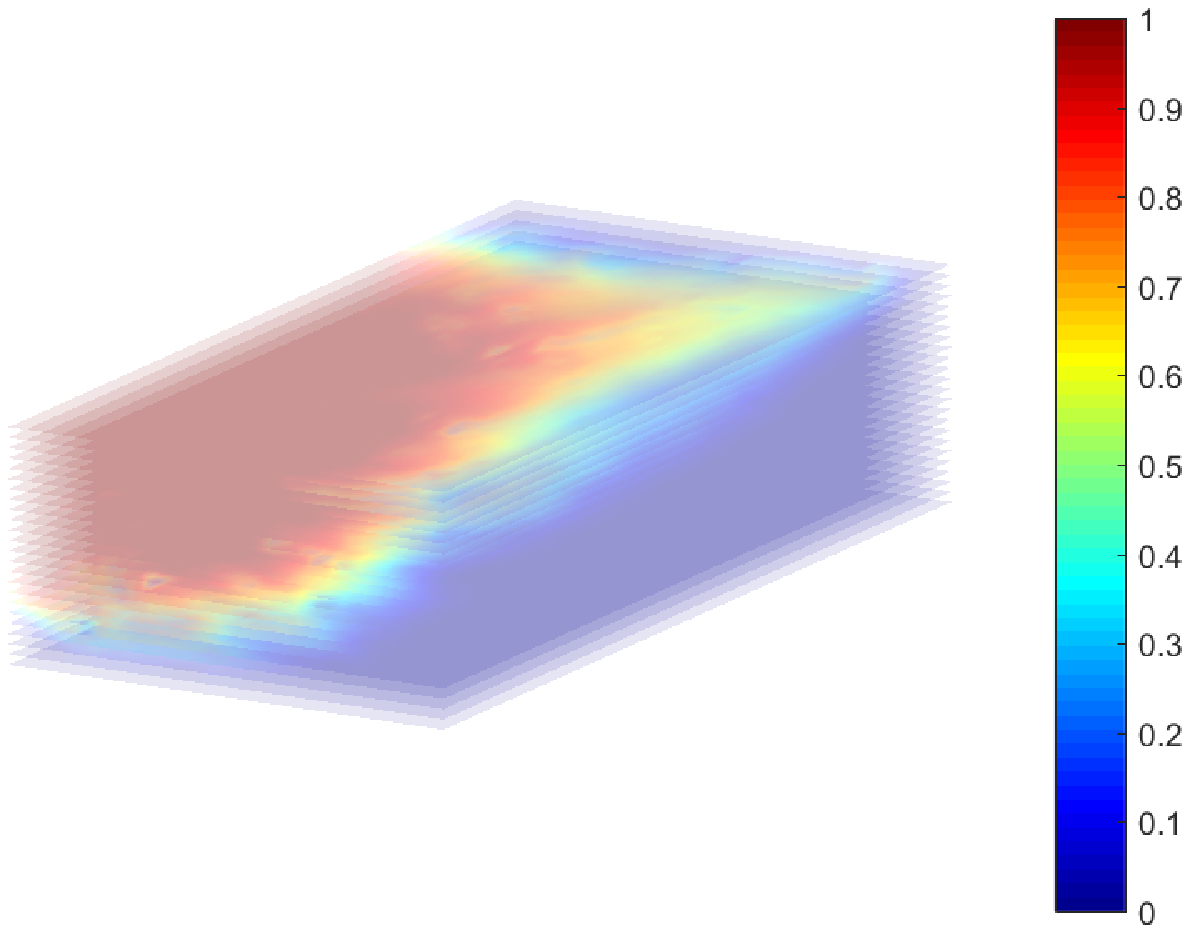}    
        \label{Fig:SPE3dc}        
    }
    \\
    \subfloat[\small{Permeability (log scale)}]{
		\includegraphics[scale = 0.45]{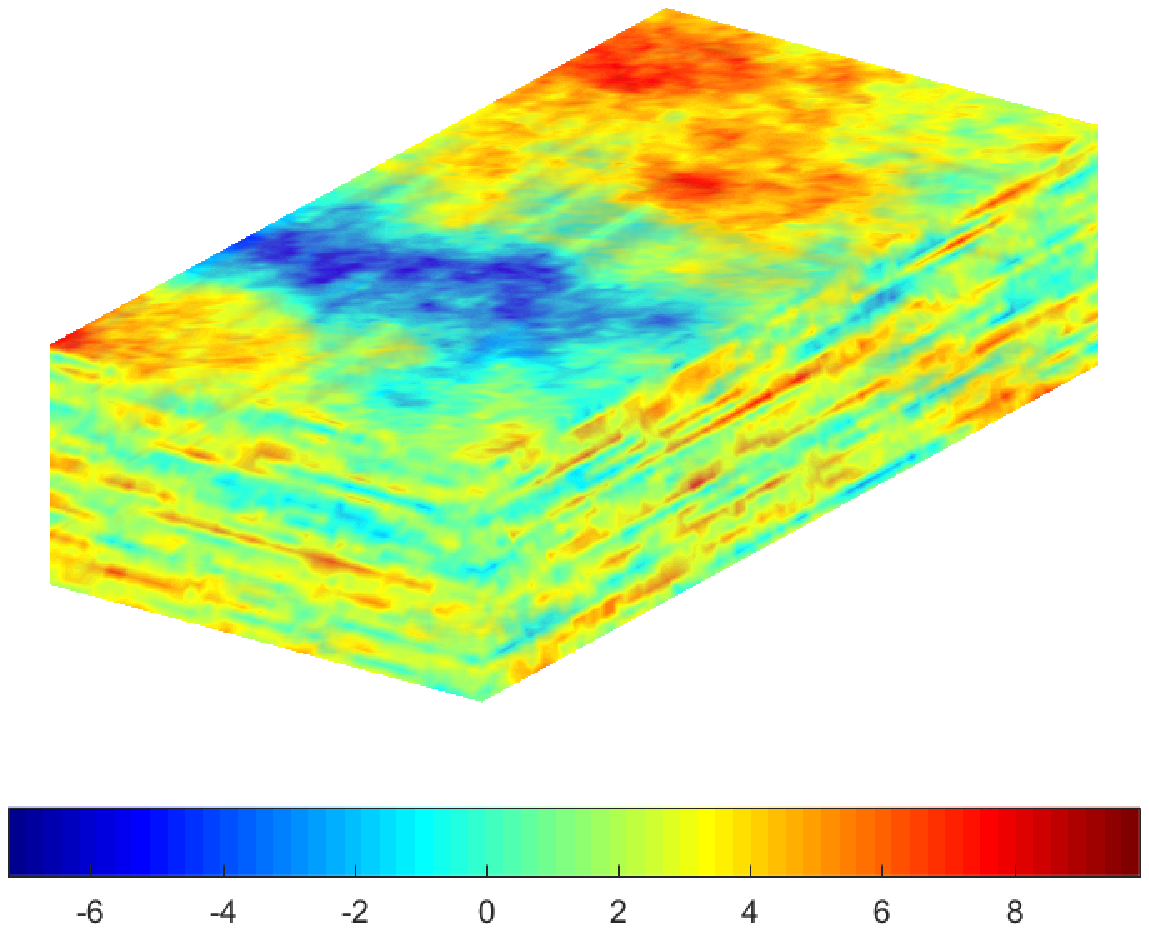}    
        \label{Fig:SPE3dd}        
    }
%\captionsetup{justification=justified}
\caption{Miscible displacement in 3D.  The concentration snapshots are visualized as volume slices, and permeability field ($k_x=k_y$).}
\label{Fig:SPE3d}
\end{figure}

Here we consider another 3D experiment.  The domain is now $\Omega=[0,50]\times[0,100]\times[0,25]$.  For the permeability (see Fig.~\ref{Fig:SPE3dd}), we take a $32\times 64\times 16$ sample from the SPE10 comparative solution project model 2 \cite{SPE10}.
  A quarter-five spot problem is set up as follows: we place an injection well at the coordinate $(0,0,25)$ and a production well at $(50,100,25)$, with $\int_\Omega q^I =\int_\Omega q^P = 1.5$.  The mesh consists of 174080 tetrahedral elements, and we use piecewise cubic basis functions.  All other parameters are the same as in subsection~\ref{subsec_hom1}.  The concentration at various times is displayed in Figs.~\ref{Fig:SPE3da},~\ref{Fig:SPE3db}, and~\ref{Fig:SPE3dc}.  It is evident that the concentration  exhibits the expected behavior, since it travels from the injection well to the production well.  This example illustrates that the HDG method for miscible displacement remains robust for highly heterogeneous porous media in 3D.
 
\section{Conclusion}
We presented a new discretization for the miscible displacement problem.  The algorithm splits the system Darcy system and the transport equations.  The subsequent linear PDEs are solved sequentially in an implicit fashion.  We observe numerically that no slope limiters are needed.  A key feature of our approach is that we use high order hybridized finite element discretizations in space.  This allows us to retain all favorable aspects DG methods, with the added bonus of increased accuracy, superconvergent postprocessing, and significantly fewer degrees of freedom for high order polynomials~\cite{CockburnDG08,cockburn2016static}.

\indent  Accurate simulations are obtained using a HDG method for both the Darcy system the transport system.  This is possible due to HDG being compatible discretizations \cite{dawson2004compatible}; so local and global mass conservation is retained.  For the HDG method, given sufficiently smooth solutions, the concentration, pressure, gradient of concentration, and velocity all converge at the rate of $k+1$ in the $L^2$ norm~\cite{nguyen2009implicit}.  Classical primal DG methods do not share this property, and in some cases they are not compatible discretizations.  In this situation, they require a $H(\text{div})$ flux reconstruction for the velocity variable, which often lowers accuracy.

%\indent  The reduction in degrees of freedom (and total nonzeros in discretization operators) is of great importance, as we use high order accurate approximations for the simulation of complex flow-transport systems.  Furthermore, since our algorithm decouples flow and transport, at every time step, multiple linear systems are to be solved.  In the most simple situation only two linear solves are needed per time step, one for obtaining the pressure and velocity and the other for concentration.  In some situations one could lag the pressure/velocity update and use it for multiple time steps before obtaining a new profile.  However, iterative coupling and high order time stepping may be utilized to enhance the solution, which increases the number of linear solves that are needed per time step.   Efficient linear solvers are required for this class of problems~\cite{fabien1}, as the dominant cost occurs during this phase of the simulation.
 %Hybridization will allow for unprecedented large scale simulations in porous media using high order MFE and DG discretizations.  %Its relevance will become more apparent for 3D problems, and multicomponent and multiphase problems.
  
Our framework does not satisfy a discrete maximum principle in general, unlike other approaches~\cite{chang2017variational}.  Overshoot and undershoot do occur (under 10\%), which is to be expected in the high order regime for convection-dominated problems.  Increasing the polynomial order reduces overshoot/undershoot phenomena as well as dissipation and dispersion.  Another benefit of high order approximations is that they give us the opportunity to use coarser meshes than those typically used for low order discretizations.  The hybridization technology allows us to consider polynomial orders that are computationally intractable for traditional DG methods; in this paper we utilize polynomial orders up to $k=16$.%That is, we can obtain similar or better accuracy on a coarse mesh with high order approximations than low order approximations on a fine mesh.  The hybridization technology allows us to consider polynomial orders that are computationally intractable for traditional DG methods; in this paper we utilize polynomial orders up to $k=16$.

%, can capture unstable flow (viscous finger-type effect), and is able to admit unstructured meshes.
We have shown through several 2D and 3D numerical experiments that the HDG method for the miscible displacement problem is high order accurate, and robust for realistic heterogeneous media.  Additionally, it is capable of static condensation, which significantly reduces the total number of degrees of freedom for high polynomial orders, compared to classical DG methods.   

\newpage
\clearpage

\bibliographystyle{elsarticle-num}
\bibliography{references_misc}

\begin{thebibliography}{10}
\expandafter\ifx\csname url\endcsname\relax
  \def\url#1{\texttt{#1}}\fi
\expandafter\ifx\csname urlprefix\endcsname\relax\def\urlprefix{URL }\fi
\expandafter\ifx\csname href\endcsname\relax
  \def\href#1#2{#2} \def\path#1{#1}\fi

\bibitem{lantz1970rigorous}
R.~Lantz, et~al., Rigorous calculation of miscible displacement using
  immiscible reservoir simulators, Society of Petroleum Engineers Journal
  10~(02) (1970) 192--202.

\bibitem{killough1987fifth}
J.~Killough, C.~Kossack, et~al., Fifth comparative solution project: evaluation
  of miscible flood simulators, in: SPE Symposium on Reservoir Simulation,
  Society of Petroleum Engineers, 1987.

\bibitem{todd1972development}
M.~Todd, W.~Longstaff, et~al., The development, testing, and application of a
  numerical simulator for predicting miscible flood performance, Journal of
  Petroleum Technology 24~(07) (1972) 874--882.

\bibitem{homsy1987viscous}
G.~M. Homsy, Viscous fingering in porous media, Annual review of fluid
  mechanics 19~(1) (1987) 271--311.

\bibitem{douglas1983approximation}
J.~Douglas~Jr, R.~E. Ewing, M.~F. Wheeler, The approximation of the pressure by
  a mixed method in the simulation of miscible displacement, RAIRO-Analyse
  num{\'e}rique 17~(1) (1983) 17--33.

\bibitem{li2015high}
J.~Li, B.~Riviere, High order discontinuous {G}alerkin method for simulating
  miscible flooding in porous media, Computational Geosciences 19~(6) (2015)
  1251.

\bibitem{dawson2004compatible}
C.~Dawson, S.~Sun, M.~F. Wheeler, Compatible algorithms for coupled flow and
  transport, Computer Methods in Applied Mechanics and Engineering 193~(23)
  (2004) 2565--2580.

\bibitem{cockburn2004characterization}
B.~Cockburn, J.~Gopalakrishnan, A characterization of hybridized mixed methods
  for second order elliptic problems, SIAM Journal on Numerical Analysis 42~(1)
  (2004) 283--301.

\bibitem{arnold1985mixed}
D.~N. Arnold, F.~Brezzi, Mixed and nonconforming finite element methods:
  implementation, postprocessing and error estimates, RAIRO-Mod{\'e}lisation
  math{\'e}matique et analyse num{\'e}rique 19~(1) (1985) 7--32.

\bibitem{zhang2017combined}
J.~Zhang, J.~Zhu, R.~Zhang, D.~Yang, A.~F. Loula, A combined discontinuous
  {G}alerkin finite element method for miscible displacement problem, Journal
  of Computational and Applied Mathematics 309 (2017) 44--55.

\bibitem{cockburn2000development}
B.~Cockburn, G.~E. Karniadakis, C.-W. Shu, The development of discontinuous
  {G}alerkin methods, in: {D}iscontinuous {G}alerkin {M}ethods, Springer, 2000,
  pp. 3--50.

\bibitem{CockburnDGRS09}
B.~Cockburn, J.~G. B.~Dong, M.~Restelli, R.~Sacco, A hybridizable discontinuous
  {G}alerkin method for steady-state convection-diffusion-reaction problems,
  {SIAM} J. Scientific Computing 31~(5) (2009) 3827--3846.

\bibitem{CockburnGL09}
B.~Cockburn, J.~Gopalakrishnan, R.~D. Lazarov, Unified hybridization of
  discontinuous {G}alerkin, mixed, and continuous {G}alerkin methods for second
  order elliptic problems, {SIAM} J. Numerical Analysis 47~(2) (2009)
  1319--1365.

\bibitem{CockburnDG08}
B.~Cockburn, B.~Dong, J.~Guzm{\'{a}}n, A superconvergent {LDG}-hybridizable
  {G}alerkin method for second-order elliptic problems, Math. Comput. 77~(264)
  (2008) 1887--1916.

\bibitem{anderson2018arbitrary}
D.~Anderson, J.~Droniou, An arbitrary-order scheme on generic meshes for
  miscible displacements in porous media, SIAM Journal on Scientific Computing
  40~(4) (2018) B1020--B1054.

\bibitem{cockburn2016bridging}
B.~Cockburn, D.~A. Di~Pietro, A.~Ern, Bridging the hybrid high-order and
  hybridizable discontinuous {G}alerkin methods, ESAIM: Mathematical Modelling
  and Numerical Analysis 50~(3) (2016) 635--650.

\bibitem{huerta2013efficiency}
A.~Huerta, A.~Angeloski, X.~Roca, J.~Peraire, Efficiency of high-order elements
  for continuous and discontinuous galerkin methods, International Journal for
  numerical methods in Engineering 96~(9) (2013) 529--560.

\bibitem{samii2016parallel}
A.~Samii, C.~Michoski, C.~Dawson, A parallel and adaptive hybridized
  discontinuous {G}alerkin method for anisotropic nonhomogeneous diffusion,
  Computer Methods in Applied Mechanics and Engineering 304 (2016) 118--139.

\bibitem{fabien1}
M.~S. {Fabien}, M.~G. {Knepley}, R.~T. {Mills}, B.~M. {Rivi{\`e}re},
  {Heterogeneous computing for a hybridizable discontinuous {G}alerkin
  geometric multigrid method}, ArXiv e-prints\href
  {http://arxiv.org/abs/1705.09907} {\path{arXiv:1705.09907}}.

\bibitem{fabien2}
M.~S. Fabien, B.~M. Knepley, Matthew G.~Rivi{\`e}re, A hybridizable
  discontinuous {G}alerkin method for two-phase flow in heterogeneous porous
  media, International Journal for Numerical Methods in Engineering\href
  {http://dx.doi.org/10.1002/nme.5919} {\path{doi:10.1002/nme.5919}}.

\bibitem{riviere2002discontinuous}
B.~Rivi{\`e}re, M.~F. Wheeler, Discontinuous {G}alerkin methods for flow and
  transport problems in porous media, International Journal for Numerical
  Methods in Biomedical Engineering 18~(1) (2002) 63--68.

\bibitem{ern2008discontinuous}
A.~Ern, A.~F. Stephansen, P.~Zunino, A discontinuous {G}alerkin method with
  weighted averages for advection--diffusion equations with locally small and
  anisotropic diffusivity, IMA Journal of Numerical Analysis.

\bibitem{bastian2003superconvergence}
P.~Bastian, B.~Rivi{\`e}re, Superconvergence and {H} (div) projection for
  discontinuous {G}alerkin methods, International journal for numerical methods
  in fluids 42~(10) (2003) 1043--1057.

\bibitem{ern2007accurate}
A.~Ern, S.~Nicaise, M.~Vohral{\'\i}k, An accurate {H} (div) flux reconstruction
  for discontinuous {G}alerkin approximations of elliptic problems, Comptes
  Rendus Mathematique 345~(12) (2007) 709--712.

\bibitem{koval1963method}
E.~Koval, et~al., A method for predicting the performance of unstable miscible
  displacement in heterogeneous media, Society of Petroleum Engineers Journal
  3~(02) (1963) 145--154.

\bibitem{berrut2004barycentric}
J.-P. Berrut, L.~N. Trefethen, Barycentric {L}agrange interpolation, SIAM
  review 46~(3) (2004) 501--517.

\bibitem{nguyen2009implicit}
N.~C. Nguyen, J.~Peraire, B.~Cockburn, An implicit high-order hybridizable
  discontinuous {G}alerkin method for linear convection--diffusion equations,
  Journal of Computational Physics 228~(9) (2009) 3232--3254.

\bibitem{chen2015robust}
H.~Chen, J.~Li, W.~Qiu, Robust a posteriori error estimates for {HDG} method
  for convection--diffusion equations, IMA Journal of Numerical Analysis 36~(1)
  (2015) 437--462.

\bibitem{fu2015analysis}
G.~Fu, W.~Qiu, W.~Zhang, An analysis of {HDG} methods for convection-dominated
  diffusion problems, ESAIM: Mathematical Modelling and Numerical Analysis
  49~(1) (2015) 225--256.

\bibitem{qiu2016hdg}
W.~Qiu, K.~Shi, An {HDG} method for convection diffusion equation, Journal of
  Scientific Computing 66~(1) (2016) 346--357.

\bibitem{chen2014analysis}
Y.~Chen, B.~Cockburn, Analysis of variable-degree {HDG} methods for
  convection-diffusion equations. part ii: Semimatching nonconforming meshes,
  Mathematics of Computation 83~(285) (2014) 87--111.

\bibitem{cockburn2016static}
B.~Cockburn, Static condensation, hybridization, and the devising of the {HDG}
  methods, in: Building bridges: connections and challenges in modern
  approaches to numerical partial differential equations, Springer, 2016, pp.
  129--177.

\bibitem{bui2016construction}
T.~Bui-Thanh, Construction and analysis of {HDG} methods for linearized shallow
  water equations, SIAM Journal on Scientific Computing 38~(6) (2016)
  A3696--A3719.

\bibitem{li2015numerical}
J.~Li, B.~Riviere, Numerical solutions of the incompressible miscible
  displacement equations in heterogeneous media, Computer Methods in Applied
  Mechanics and Engineering 292 (2015) 107--121.

\bibitem{SPE10}
{SPE Comparative Solution Project} model 2,
  \url{http://www.spe.org/web/csp/datasets/set02.htm}, accessed: 2017-05-27.

\bibitem{chang2017variational}
J.~Chang, K.~Nakshatrala, Variational inequality approach to enforcing the
  non-negative constraint for advection--diffusion equations, Computer Methods
  in Applied Mechanics and Engineering 320 (2017) 287--334.

\end{thebibliography}






% \end{thebibliography}

\end{document}